\documentclass[a4paper,onecolumn,11pt,accepted=2023-05-09]{quantumarticle}
\pdfoutput=1
\usepackage[utf8]{inputenc}
\usepackage{amsmath}
\usepackage{bbold}
\usepackage{color}
\usepackage{graphicx}
\usepackage{subcaption}
\usepackage{scalerel,stackengine}
\usepackage{braket}
\usepackage{leftidx}
\usepackage[title]{appendix}
\usepackage{tikz-cd}
\usepackage{dsfont}
\usepackage{cite}
\usepackage{soul}
\usepackage{hyperref}    
\usepackage{authblk}
\usepackage{hyperref}

\def\be{\begin{equation}}
\def\ee{\end{equation}}
\def\ba{\begin{eqnarray}}
\def\ea{\end{eqnarray}}

\newcommand\nn{\nonumber}
\newcommand\q{\quad}
\stackMath
\newcommand\reallywidehat[1]{%
\savestack{\tmpbox}{\stretchto{%
  \scaleto{%
    \scalerel*[\widthof{\ensuremath{#1}}]{\kern.1pt\mathchar"0362\kern.1pt}%
    {\rule{0ex}{\textheight}}
  }{\textheight}%
}{2.4ex}}%
\stackon[-6.9pt]{#1}{\tmpbox}%
}

\newcommand{\cc}{\mathcal C}

\newcommand{\cg}{\mathcal G}
\newcommand{\ch}{\mathcal H}

\newcommand{\cp}{\mathcal P}
\newcommand{\cq}{\mathcal Q}
\newcommand{\calr}{\mathcal R}
\newcommand{\cs}{\mathcal S}
\newcommand{\ct}{\mathcal T}

\def\f{\frac}

\def\p{\partial}

\usepackage{bbm}

\def\q{{\quad}}

 \newcommand{\pah}[1]{{ #1}}
  \newcommand{\av}[1]{{ #1}}
\newcommand{\fla}[1]{{ #1}}

\usepackage{geometry}
\title{Switching quantum reference frames in the N-body problem and the absence of global relational perspectives}

\author{Augustin Vanrietvelde}
\orcid{0000-0001-9022-8655}
\affiliation{Laboratoire Méthodes Formelles, Inria Saclay,  France}
\affiliation{Institute for Quantum Optics and Quantum Information, Austrian Academy of Sciences, Vienna}
\affiliation{Quantum Group, Department of Computer Science, University of Oxford}
\affiliation{Department of Physics, Imperial College London}

\author{Philipp A.\ H\"ohn}
\orcid{0000-0001-7108-1184}
\email{philipp.hoehn@oist.jp}
\thanks{corresponding author}
\affiliation{Okinawa Institute of Science and Technology Graduate University, Onna, Okinawa}
\affiliation{Institute for Quantum Optics and Quantum Information, Austrian Academy of Sciences, Vienna}
\affiliation{Department of Physics and Astronomy, University College London, London}
\affiliation{Faculty of Physics, University of Vienna}

\author{Flaminia Giacomini}
\orcid{0000-0002-2475-2296}
\affiliation{Institute for Theoretical Physics, ETH Z{\"u}rich, Wolfgang-Pauli-Str. 27, Z{\"u}rich, Switzerland}
\affiliation{Institute for Quantum Optics and Quantum Information, Austrian Academy of Sciences, Vienna}
\affiliation{Faculty of Physics, University of Vienna}
\affiliation{Perimeter Institute for Theoretical Physics, Waterloo, Ontario, Canada}

\date{}

\begin{document}

\maketitle

\begin{abstract}
{\small Given the importance of quantum reference systems to both quantum and gravitational physics, it is pertinent to develop a systematic method for switching between the descriptions of physics relative to different choices of quantum reference systems. Here, we expand on a unifying approach, begun in \cite{vanrietvelde2018change}, which blends the operational language of \cite{Giacomini:2017zju} with a gravity-inspired symmetry principle. The latter enforces physics to be relational and leads, thanks to gauge related redundancies, to a perspective-neutral structure which contains all frame choices at once and via which frame perspectives can be consistently switched. Formulated in the language of constrained systems, the perspective-neutral structure turns out to be the constraint surface classically and the gauge invariant Hilbert space in the Dirac quantized theory. By contrast, a perspective relative to a specific frame corresponds to a gauge choice and involves a symmetry reduction procedure to an associated  reduced phase and Hilbert space. Quantum reference frame switches thereby amount to a particular gauge transformation.}

{\small Here, we show that they take the form of `quantum coordinate changes'. We illustrate this method in a general mechanical particle model, namely the relational $N$-body problem in three-dimensional space with rotational and translational symmetry. This model is particularly interesting because it features an analog of the generic Gribov problem so that globally valid gauge fixing conditions, and thereby relational frame perspectives, are absent. We will show that the constraint surface is topologically non-trivial and foliated by three-, five- and six-dimensional gauge orbits, where the lower dimensional orbits are a set of measure zero. In consequence, the $N$-body problem also does not admit globally valid canonically conjugate pairs of Dirac observables. These challenges notwithstanding, we exhibit how one can construct the quantum reference frame transformations for the three-body problem. Our construction also sheds new light on the generic inequivalence of Dirac and reduced quantization through its interplay with quantum frame perspectives.}
\end{abstract}

\section{Introduction}

Reference frames appear ubiquitously in both quantum and gravitational physics. A concrete description of some physical situation is usually given from the perspective of some appropriate choice of reference frame. In most cases, such a frame is considered as an idealized external system that can be used as a vantage point but does not itself back-react on the remaining physical systems. 

Reference frames, however, are always physical systems and treating them fundamentally as being quantum in nature is inevitable in quantum gravity \cite{DeWitt:1967yk,Rovelli:2004tv,Rovelli:1990pi,Rovelli:1990ph,Kuchar:1991qf,Isham:1992ms,Brown:1994py,Dittrich:2004cb,Dittrich:2005kc,Tambornino:2011vg,Thiemann:2007zz,Bojowald:2010xp,Bojowald:2010qw,Hohn:2011us, Dittrich:2016hvj,Dittrich:2015vfa} and also in quantum information and foundations, for example once considering that measurements are carried out with physical systems \cite{Aharonov:1967zza, PhysRev.158.1237,PhysRevD.30.368,Bartlett:2007zz,bartlett2009quantum, gour2008resource, Palmer:2013zza,bartlett2006degradation,smith2016quantum, poulin2007dynamics,PhysRevLett.111.020504, loveridge2017relativity, pienaar2016relational, angelo2011physics, Hoehn:2014vua, guerin2018observer, oreshkov2012quantum,Hardy:2018kbp}. In fact, every  physical measurement \emph{is} relational: they relate some physical property of the measured systems to a physical property of the systems comprising the measurement device. For example, in a Stern-Gerlach magnet, the spin of the measured particles is related to the (coherent) spin of the atoms composing the magnet. 
Treating reference systems fundamentally as quantum objects is ultimately a consequence of the universality of quantum theory, according to which all physical systems are subject to its laws.

Both in quantum gravity and quantum foundations it therefore becomes crucial to clarify how to describe physics relative to quantum reference systems and how the descriptions relative to different such choices are related. Exploiting a fruitful interplay of ideas from both fields, we began developing a unifying method for transforming between quantum reference systems in \cite{vanrietvelde2018change} that  aims at ultimately encompassing both quantum and gravitational physics. The key ingredients of this method are a gravity inspired (gauge) symmetry principle and the relational and operational approach to quantum reference frames recently put forward in \cite{Giacomini:2017zju}.

The symmetry principle is, in fact, inspired by Mach's principle, which was a pivotal motivation for the development of general relativity and is closely related to the diffeomorphism symmetry of the latter. Mach's principle essentially states that all physics is relational \cite{Barbour295,mercati2018shape,Rovelli:2004tv}. In particular, what inertial frames are is not determined with respect to an absolute space (as in Newtonian physics), but by the remaining dynamical content of the universe. We likewise employ the symmetry principle to enforce physical observables to be relational in the systems we consider and to generate an inherent redundancy in their description. Exploiting this redundancy allows us to develop, as proposed in \cite{Hoehn:2017gst}, a perspective-neutral meta-structure, which contains, so to speak, all frame perspectives at once and via which they are changed. 

This symmetry principle is implemented using the language of constrained Hamiltonian systems \cite{Dirac,Henneaux:1992ig}, which also underlies the canonical formulation of general relativity and quantum gravity \cite{Rovelli:2004tv,Thiemann:2007zz}. In our approach, the perspective-neutral structure corresponds to the constraint surface classically and to the gauge invariant physical Hilbert space in the Dirac quantized theory. Taking the perspective of a specific frame is closely related to imposing a gauge that fixes the redundancies in the description and changing from one frame perspective to another amounts to a symmetry transformation via the perspective-neutral structure. A specific perspective relative to one choice of frame is classically encoded in a gauge-fixed reduced phase space and to a symmetry reduced Hilbert space in the quantum theory. The latter will be the quantum analog of a gauge-fixed reduced phase space, but not necessarily equivalent to a quantization of it. This gives rise to a systematic method for changing between the perspectives relative to different choices of quantum reference systems and can be applied to both temporal and spatial quantum reference systems and in both quantum and gravitational physics. Compellingly, by always passing from one frame perspective through the perspective-neutral Hilbert space in order to map to a new frame perspective, the quantum reference frame transformations will take the form of `quantum coordinate transformations'.\footnote{Notice that some authors distinguish between quantum reference frames, i.e., where the reference frame is treated as a physical system and has a quantum state associated to it, and quantum coordinates~\cite{Hardy:2018kbp, hardy2020implementation, zych2018relativity}, where the reference frame transformation is generalised but reference frames are not treated as physical systems. Here, we do not make this distinction, and use both interchangeably.}

Our method extends the approach to changes of temporal quantum reference systems, i.e.\ relational `clocks',  developed in \cite{Bojowald:2010xp,Bojowald:2010qw,Hohn:2011us} for models of quantum gravity and cosmology, which precisely uses the above concepts and tools at a semiclassical level. Altogether, our approach can also be viewed as an expansion of the view, advocated in \cite{Rovelli:2013fga}, that gauge related redundancies are not just a mathematical artifact, but physically crucial for a completely relational description of the world.

As already indicated, our operational language, ultimately originating in \cite{Giacomini:2017zju}, also sheds new light on the relation between Dirac and reduced quantization -- i.e.\ on the relation between quantizing first, then solving the constraints, and solving the constraints first, then quantizing -- which has been the topic of a lively debate in the literature \cite{Ashtekar:1982wv,guillemin1982geometric, tian1998analytic,hochs2008guillemin,gotay1986constraints,kucha1986covariant,Ashtekar:1991hf,Schleich:1990gd,Kunstatter:1991ds,Hajicek:1990eu,Romano:1989zb,Dittrich:2016hvj,Dittrich:2015vfa,Loll:1990rx}. While the agreement is that the two methods are generally not equivalent, the debate also revolves around the question when one or the other would be the correct one to apply. In our approach, the Dirac quantized theory provides the perspective-neutral structure from which internal perspectives are constructed via `quantum coordinate maps'. These are the quantum analog of classical phase space reductions by gauge fixing and map to symmetry reduced Hilbert spaces; they proceed as follows:
\begin{enumerate} 
\item Trivialize the symmetry generating quantum constraints. That is, transform them in such a way that they only apply to (and fix) the subsystem one would like to use as a reference and, accordingly, whose degrees of freedom one considers as the redundant ones in this specific description.
\item Condition on the corresponding classical gauge fixing conditions.
\end{enumerate}
The symmetry reduced quantum theories so constructed -- constituting the quantum frame perspectives -- will generally be unitarily \emph{inequivalent} to the quantization of the classically reduced theories, i.e.\ the direct quantization of the classical frame perspectives. This manifests the generic inequivalence between Dirac and reduced quantization; for example, we will see that the Hamiltonians will differ. However, this is not a problem for our framework of quantum frame covariance. As we shall argue, it is the Dirac quantized theory which should be given primacy as this is the one treating all degrees of freedom on an equal footing and thereby translating the redundancy into the quantum theory that enables one to define a notion of quantum covariance in the first place. In reduced quantization one simply generally lacks the structure to relate different frame perspectives and they will in general not be unitarily equivalent. In other words, both classically and in the quantum theory, we give primacy to the quantization of the perspective-neutral structure over the quantization of the internal perspectives; it is the perspective-neutral structure which links all the internal frame perspectives.

It should be emphasized that it \emph{is} possible to render Dirac and reduced quantization equivalent in the class of models considered here by invoking constructions exhibited in \cite{kucha1986covariant,Kunstatter:1991ds}. But this involves a non-standard factor ordering of the physical Hamiltonian which essentially amounts to somewhat artificially adapting Dirac to reduced quantization so that the two methods produce equivalent results. \pah{This leads to a serious drawback, however: the so modified Dirac quantized theory is then equivalent to the quantization of only one of the classical frame perspectives and explicitly depends on this choice.} Specifically, the modification of the physical Hamiltonian then depends on the choice of perspective, with different choices leading to inequivalent results, manifesting that the direct quantizations of the different classical frame perspectives are unitarily inequivalent. In particular, \emph{constructing the quantum frame perspectives either by directly quantizing the classical frame perspectives or by quantum reduction from the perspective-neutral Hilbert space are in general two physically inequivalent procedures}. We comment further on this below and in Appendix~\ref{HamiltonianDiscrepancy}. 

In \cite{vanrietvelde2018change}, we have introduced the general conceptual underpinning of this approach and illustrated it in a simple $N$-particle model in one-dimensional space with a single {\it linear} translation generator constraint. Accordingly, globally valid gauge fixing conditions and thus globally valid relational perspectives on the physics were possible in this simple model. Using our new method, we recovered  in \cite{vanrietvelde2018change} some of the quantum frame transformations of \cite{Giacomini:2017zju}, which were constructed using a different approach, thereby embedding them in a perspective-neutral framework. These transformations were also employed to show in \cite{vanrietvelde2018change} how entanglement and classicality of an observed system depend on the quantum frame perspective.

Here, our aim is to generalize our approach, begun in \cite{vanrietvelde2018change}, by applying it to a significantly more complicated model, namely the relational $N$-body problem in three-dimensional space which now features not only translational, but also rotational invariance. This is a very general mechanical particle model and it will serve to substantiate both our conceptual and technical line of argumentation, proving the capability of our approach.

The most crucial difference to the model in \cite{vanrietvelde2018change} is the {\it absence of globally valid internal perspectives}, i.e.\ any internal perspective will fail to fully describe all physical situations. This is analogous to the absence of global coordinates on generic spacetime manifolds and links with the conceptual discussion in \cite{vanrietvelde2018change}. The origin of this property in the present model is the additional rotational symmetry. The gauge orbits of the rotation group in the constraint surface are compact (and thus closed) and this results in the impossibility of finding globally valid gauge fixing conditions. This feature is a mechanical analog of the Gribov problem in gauge field theories, and it is also the reason why globally valid pairs of canonically conjugate gauge invariant (i.e.\ Dirac) observables are absent. In consequence, it is only possible to develop non-global descriptions of the physics relative to a particular reference frame in both the classical and quantum theory. 
This is closely related to the global problem of time in general relativistic systems \cite{Kuchar:1991qf,Isham:1992ms,Bojowald:2010xp,Bojowald:2010qw,Hohn:2011us,Dittrich:2016hvj,Dittrich:2015vfa}. Concretely, this means that generic relational clocks (temporal quantum reference systems) in these systems feature turning points -- and, accordingly, can start running `backwards' -- and thus cannot be used as globally valid temporal references. Therefore, non-global relational descriptions appear, and should be expected, in most interesting physical scenarios, as argued also in \cite{vanrietvelde2018change}. The below is still a fairly benign illustration of this Gribov problem. 

It must be emphasized, however, that this is not a problem for the physics per se, but only of descriptions of it. In fact, the perspective-neutral structures in both the classical and quantum theory can be formulated globally without any issues. Challenges only arise when defining specific perspectives and so much of our attention here will be devoted to exhibiting how to describe physics relative to non-global relational perspectives and how to nevertheless switch between them via a perspective-neutral structure that itself can be formulated globally. \av{Moreover, one should keep in mind that in this work, the internal perspectives only fail on very specific, highly degenerate configurations (typically, when all particles are aligned); and moreover that they all fail on the same configurations \pah{for the $N=3$ case which we focus on}.\footnote{However, the second point will not be true anymore for $N>3$; see the discussion at the very end of Section \ref{quantum3D}.} The first of these features shows that the situation is less severe than, for example, in the case of Schwarzschild coordinates, which leave out a consequent chunk of spacetime; the second puts it in contrast with the case of spherical coordinates, in which the failure happens over an arbitrary axis that bears no relationship with the physical configuration.}

The rest of the article is arranged as follows. In sec.\ \ref{sec_classical}, we introduce the classical $N$-body problem in three-dimensional Newtonian space, but subject to a global rotational and translational invariance, which renders the spatial physics fully relational and leads to a redundancy in the description. Here, we discuss all the technical challenges that arise due to the appearance of the (mechanical analog of the) Gribov problem and how one can nevertheless choose a frame perspective. In sec.\ \ref{quantum3D}, we quantize the three-body problem in both the Dirac and reduced method. Our main ambition here is to construct explicitly the quantum symmetry reduction, i.e.\ quantum coordinate maps, from the Dirac to the  reduced quantum theories in the particular frame perspectives and to compare the result to the quantization of the classical frame perspectives. We finally then use this quantum reduction method to also construct the quantum reference frame transformation that takes \fla{one from a} quantum frame perspective to another. Importantly, this transformation proceeds by inverting the quantum reduction, mapping back to the perspective-neutral Hilbert space and then performing the forward quantum reduction to the desired perspective, hence a `quantum coordinate transformation'. Despite the absence of global classical relational perspectives, these quantum frame changes are unitary for regular Hilbert space states because the pathological configurations comprise a set of measure zero; for distributional states, the situation differs somewhat. We illustrate the transformation on example states which show the quantum frame dependence of entanglement, before concluding in sec.\ \ref{sec_conc}. For better readability, we have moved most technical details to various appendices and for better orientation, we also provide here a table of contents.

\tableofcontents

\section{Internal perspectives in the classical $N$-body problem}\label{sec_classical}

\subsection{A toy model for Mach's principle in 3D space}\label{sec_mach3d}

We employ a toy model\footnote{We thank T.\ Koslowski for suggesting this model.} for Mach's principle in a system of $N$ interacting particles in three-dimensional space with translation and rotation invariance. This gauge symmetry implies that the position, orientation and motion of the particles with respect to the Newtonian background space has no physical significance and that the gauge invariant information is purely relational. As such, our system bears some resemblance to the Barbour-Bertotti and related models \cite{Barbour295,mercati2018shape,Barbour:2014bga,Barbour:2015sba}, except that here we will not have dilation or Hamiltonian constraints.

For simplicity and in analogy to \cite{vanrietvelde2018change}, we restrict to unit mass particles and a Euclidean phase space $\cq=\mathbb{R}^{3N}$ so that our phase space, coordinatized by particle positions $\vec{q}_i$ and momenta $\vec{p}_i$, is $T^*\cq=\mathbb{R}^{6N}$. 
We take an $N$ particle system whose Lagrangian, described in Appendix \ref{app_lagr}, features the desired gauge symmetries and upon Legendre transformation leads to the following six (primary) constraints on phase space:\footnote{Henceforth, the Einstein convention holds and repeated spatial indices are implicitly summed over.}
\begin{subequations}\label{constraints}
\be\label{constraintP}
P^a = \sum_i p_i^a \approx 0\,,
\end{equation}
\be\label{constraintR}
R^a = \sum_i \epsilon^{abc}q_i^b p_i^c \approx 0\,,
\ee
\end{subequations}
where $a,b,c\in\{x,y,z\}$ denote spatial components. These constraints are independent\footnote{In the $N=2$ case, the system (\ref{constraints}) is not independent and there are only five independent constraints, which induces some peculiarities. In the rest of the paper, unless stated otherwise, we will assume $N  \geq 3$.} (almost everywhere, see below) and first-class, clearly constituting the generators of the Euclidean group:
\ba\label{classalgebra}
\{P^a,P^b\} = 0\,,\q\q\q\q
\{R^a,R^b\} = \epsilon^{abc} R^c \approx 0\,,\q\q\q\q
\{P^a,R^b\} = - \epsilon^{abc} P^c \approx 0\,.
\ea
They are conserved so that no secondary constraints arise. The total Hamiltonian following from the Lagrangian of Appendix \ref{app_lagr} is
\begin{equation}\label{H3D}
H_{\rm tot} = \frac{1}{2}\,\sum_{i=1}^N\, \vec{p_i}^2 + V(\{|\vec{q}_i - \vec{q}_j|\}_{i,j=1}^N) +  \lambda^a P^a + \mu^a R^a\,,
\end{equation}
where the $\lambda^a$ and $\mu^a$ are six a priori arbitrary Lagrange multipliers, accounting for gauge freedom, and the potential $V$ depends only on the relative absolute distances between the particles.

Consequently, the six constraints generate gauge transformations on the constraint surface; clearly, the $P^a$ induce global translations, namely infinitesimally
\begin{equation}\label{gaugeTransfoP}
\begin{cases}
q_i^a \to q_i^a + \{q_i^a, P^b\} \,\varepsilon^b = q_i^a + \varepsilon^a\,, \\
p_i^a \to p_i^a + \{p_i^a, P^b\}\, \varepsilon^b = p_i^a\,,
\end{cases}
\end{equation}
while the $R^a$ are generators of global rotations, infinitesimally given by 
\begin{equation}\label{gaugeTransfoR}
\begin{cases}
q_i^a \to q_i^a + \{q_i^a, R^b\}\, \varepsilon^b = q_i^a + \epsilon^{abc}\, \varepsilon^b \,q_i^c\,,\\
p_i^a \to p_i^a + \{p_i^a, R^b\}\, \varepsilon^b = p_i^a + \epsilon^{abc}\, \varepsilon^b \,p_i^c\,.\\
\end{cases}
\end{equation}

A subtlety arises: while the six constraints (\ref{constraints}) are independent on generic points of the constraint surface $\cc$ in phase space (so generically, the gauge orbits will be six-dimensional), there exist special pathological points on $\cc$ (thus a priori allowed), where the constraints become partially dependent. 
These correspond to $N$ particle collisions and situations when all particles are collinear. As shown in Appendix \ref{app_colcol}, in these cases, only three and five constraints are independent, respectively, and the corresponding configurations reside in lower dimensional gauge orbits, so that the constraint surface is, in fact, foliated by six-, five- and three-dimensional gauge orbits. This is a direct consequence of the compactness of the pure rotation orbits and the (complete or partial) rotational invariance of these special configurations. Given that total collisions and collinearity are a set of measure zero, almost all gauge orbits are six-dimensional. However, the lower dimensional ones are topologically relevant and have physical consequences: globally valid gauge conditions will not be possible because on the lower dimensional orbits one cannot fully fix the rotational gauge freedom as some rotations will act trivially (see Appendix \ref{app_colcol} and the discussion in sec.\ \ref{classicalGaugeFix3D}).\footnote{Also, as a manifestation of the Gribov problem, other gauge fixing surfaces might miss the lower dimensional orbits entirely.} Since we will, again, interpret choices of internal perspectives as gauge fixings, this is the origin of why there will be no globally valid internal perspectives in this model. 

This implies significant repercussions for the reduced phase space, which is the set of all gauge orbits (every gauge orbit is the equivalence class corresponding to one physical state). Mathematically it is the quotient $\cp_{\rm red}=\cc/\!\!\sim$, where $\sim$ identifies points if they lie in the same orbit, and so will be topologically non-trivial because of the differing dimensions of the gauge orbits. 

Clearly, this also has consequences for gauge invariance and, specifically, gauge invariant functions, i.e.\ Dirac observables. Before we discuss these consequences, we recall that any function can be expressed in terms of our basic phase space variables $\vec{q}_i,\vec{p}_j$ and inquire what the most general form of a Dirac observable is. Given that the full set of rotations only leaves inner products invariant, any Dirac observable must be a function of inner products of our basic variable vectors. These inner products must also be translation invariant and so any Dirac observable must be a function of
\ba
(\vec{q}_i-\vec{q}_j)\cdot(\vec{q}_k-\vec{q}_l),\q\q\q\q  (\vec{q}_i-\vec{q}_j)\cdot\vec{p}_k\,,\q\q\q\q\vec{p}_i\cdot\vec{p}_j\,,\q\q\q i,j,k,l=1,\dots,N\,,\label{indirac3d}
\ea
which comprise all rotation and translation invariant inner products quadratic in our basic variables.\footnote{Note that also all Dirac observables which involve cross products will be functions of these combinations.} In particular, the unconstrained part of $H_{\rm tot}$ is obviously also gauge invariant.

\begin{figure}
    \centering
    \begin{subfigure}[b]{0.3\textwidth}
        \includegraphics[width=\textwidth]{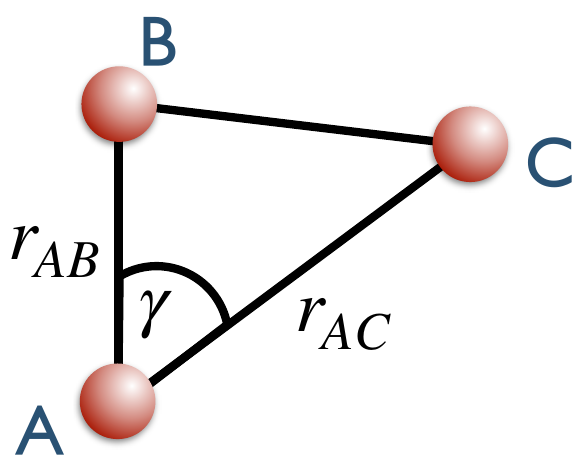}
        \caption{}
        \label{fig:1a}
    \end{subfigure}
 ~  \q\q\q\q\q\q\q
    \begin{subfigure}[b]{0.3\textwidth}
        \includegraphics[width=\textwidth]{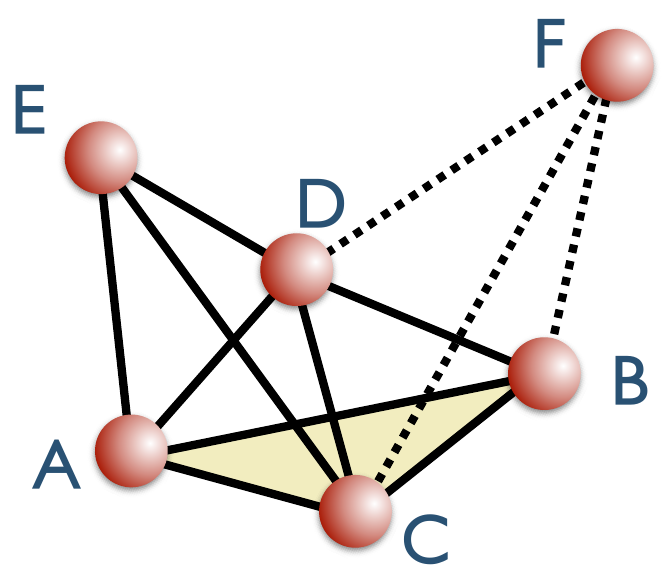}
        \caption{}
        \label{fig:1b}
    \end{subfigure}
    \caption{\small (a) Illustration of the three-body problem and gauge invariant configuration degrees of freedom. Our three configuration Dirac observables from the (almost everywhere) conjugate pairs in (\ref{conjDirac3d}) are functions of these: $\rho_{BA},\rho_{CA}$ are simply the logs of the relative distances $r_{AB}=|\vec{q}_A-\vec{q}_B|,r_{AC}=|\vec{q}_A-\vec{q}_C|$ between $A$ and $B,C$, respectively, while $u$ is (minus) the cotangent of the relative angle between $B$ and $C$ as seen from $A$. (b) Illustration of a generic situation in the six-body problem as a triangulation. In order to fully localize particle $F$ relative to particles $A$ to $E$, three relative distances between $F$ and the rest are sufficient and not all relative distances are independent. }\label{fig:1}
\end{figure}

Again, there is a redundancy among the elementary Dirac observables (\ref{indirac3d}). Indeed, the $N$ particles have $3N$ coordinates and the symmetry consists of three global translations and three global rotations so there can only be (at most) $3N-6$ independent gauge invariant configuration degrees of freedom. Similarly, there are $3N$ momentum coordinates and six constraints (\ref{constraints}) that can be solved for momenta so there can likewise only be (at most) $3N-6$ independent gauge invariant momentum degrees of freedom. For intuition of these statements, it is helpful to visualize the $N$ particle motion and its gauge invariant information geometrically: a generic $N$ particle configuration (i.e., no total collision or collinearity) corresponds to a triangulation with $N$ vertices in 3D Euclidean space (see fig.\ \ref{fig:1}) whose $3N-6$ edges are labeled by {\it relative distances} 
$\big|\vec{q}_i - \vec{q}_j\big|$ 
from among the (square roots of the) configuration observables in (\ref{indirac3d}). In this manner, we argue geometrically in Appendix \ref{app_generic} that for generic $N$ particle configurations there are indeed $3N-6$ independent configuration and $3N-6$ independent momentum Dirac observables.

But what about $N$ particle collisions and total collinearity where we saw that the symmetry generators become dependent? We prove in Appendix \ref{app_absence} that (i) none of the elementary Dirac observables (\ref{indirac3d}) are independent on the $N$ particle collisions residing in three-dimensional gauge orbits, and (ii) only $2(N-1)$ of them are independent on totally collinear $N$ particle configurations residing in five-dimensional gauge orbits. 

As we also show in Appendix \ref{app_absence}, this has a severe ramification: {\it canonically conjugate or affine \footnote{ \pah{ Two phase space variables $q,\pi$ form an affine (rather than canonical) pair if $\{q,\pi\}=q$. That is, $\pi$ is the generator of dilatations (rather than translations) of $q$.}} pairs of Dirac observables do not exist globally on $\cc$.} Accordingly, there are no canonically (or affinely) conjugate coordinates covering all of the reduced phase space, manifesting its topologically non-trivial nature alluded to above. A priori this will complicate also the reduced quantization of the physics in any of the internal perspectives, as we can not simply apply canonical or affine quantization methods \cite{isham2} to the reduced phase space as a whole without restricting the dynamics and introducing boundary conditions. In this sense, also the internal perspectives in the quantum theory will be non-global.

Let us now construct canonically conjugate Dirac observables that are `as global as possible'. Constructing sets of independent (non-global) canonically conjugate Dirac observables for the $N$-body problem depends on $N$ and so here we shall only provide such a set for the three-body problem. Labeling the three particles as before by $A,B,C$, we assembled a convenient set of $3N-6=3$ (almost everywhere) independent and canonically conjugate Dirac observable pairs for our purposes as follows
\ba
\rho_{BA}&:=&\ln r_{BA}= \ln|\vec{q}_B-\vec{q}_A|\,,\q\q\q\q\q\q\q\q p_{\rho_{BA}}:=\vec{p}_B\cdot\left(\vec{q}_B-\vec{q}_A\right)\,,\nn\\
\rho_{CA}&:=&\ln r_{CA}=\ln |\vec{q}_C-\vec{q}_A|\,,\q\q\q\q\q\q\q\q p_{\rho_{CA}}:=\vec{p}_C\cdot\left(\vec{q}_C-\vec{q}_A\right)\,,\label{conjDirac3d}\\
u&:=&-\cot\gamma=-\f{(\vec{q}_B-\vec{q}_A)\cdot(\vec{q}_C-\vec{q}_A)}{\big|(\vec{q}_B-\vec{q}_A)\times(\vec{q}_C-\vec{q}_A)\big|}\,,\nn\\ p_{u}&:=&\sin\gamma\,\,\vec{p}_C\cdot\left(\cos\gamma\,(\vec{q}_C-\vec{q}_A)-\f{r_{CA}}{r_{BA}}\,(\vec{q}_B-\vec{q}_A)\right).\nn
\ea
We choose this set as it has a direct geometrical interpretation: the configuration Dirac observables are the logarithms of the relative distances $r_{BA}, r_{CA}\in[0,\infty)$ of $B$ and $C$ from $A$ and (minus) the cotangent of the angle $\gamma\in[0,\pi]$ in the triangle between $B$ and $C$ at $A$ (see fig.\ \ref{fig:1a}). However, in contrast to $r_{BA},r_{CA},\gamma$, our geometric variables $\rho_{BA},\rho_{CA},u$ each take value in all of $\mathbb{R}$ which will be more convenient for quantization. (We choose $u=-\cot\gamma$ so that $u$ goes from $-\infty$ to $+\infty$ as $\gamma$ runs from $0$ to $\pi$.) The momentum Dirac observables are projections of the particle momenta in the appropriate directions. The same construction can be carried out, of course, for permutations of $A,B,C$ with the obvious meaning. 

Clearly, this set will not define global coordinates on the reduced phase space as $\rho_{BA},\rho_{CA},p_u$ become singular when $A$ collides with either $B$ or $C$ and $u$ is undefined in those cases and when the three particles are collinear. (These configurations lie in the boundary of the gauge invariant configuration space and are a set of measure zero \cite{littlejohn1995}.) For other non-global constructions of canonically conjugate Dirac observables for the three-body problem, e.g., in terms of Jacobi (mass-weighted difference) vectors, see \cite{littlejohn1995,littlejohn1997,Barbour:2015sba}. 

Our treatment for these illnesses is to dynamically restrict the model: we shall henceforth assume (but not explicitly write) that the potential $V(\{|\vec{q}_i - \vec{q}_j|\}_{i,j=A,B,C})$ becomes infinitely repulsive when any of the relative distances vanishes. For example, we might imagine three electrons and a Coulomb potential. This rules out any particle collisions (for finite energies) and so the three-dimensional gauge orbits will no longer become dynamically accessible. Note that the pairs $(\rho_{BA},p_{\rho_{BA}})$ and $(\rho_{CA},p_{\rho_{CA}})$ will then indeed be canonically conjugate on the dynamically accessible configurations and also $p_u$ will remain finite. Only $u$ will, strictly speaking, remain undefined when $A,B,C$ are collinear, although $\gamma$ is well-defined then. Since these configurations also remain a set of measure zero in the boundary of the gauge invariant configuration space \cite{littlejohn1995}, we shall simply interpret the divergence of $u$ for collinearity as `infinite distance' in the variable $u$. In this sense, after dynamical restriction, our variables (\ref{conjDirac3d}) are `as good as it gets' and canonically conjugate on the dynamically accessible regions of $\cc$.

In summary, owing to the gauge symmetry, the $N$ particle system has no physically meaningful absolute position, orientation and motion in space, yet \textit{relative} position, orientation and motion among the particles is gauge invariant. Building up on our discussion in \cite{vanrietvelde2018change}, we interpret the structures described here and, in particular, all the structure on the constraint surface $\cc$ as a perspective-neutral meta-structure which, thanks to its inherent redundancy, contains all perspectives at once. Choosing a specific internal perspective of a particle frame will again amount to a gauge fixing, as we shall discuss shortly. 

We noted that there cannot be a globally valid internal perspective on account of the topological non-trivialities of the gauge orbits. However, it must be emphasized that all ensuing challenges only arise when collecting the structures necessary for moving into a specific internal perspective. The perspective-neutral structure has no problems per se as $\cc$ can be consistently and globally described in terms of the original phase space variables $\vec{q}_i,\vec{p}_j$ and, while Dirac observables are necessary for gauge invariance, global canonically conjugate pairs of them are not required.

\subsection{Choosing an internal perspective = choosing a gauge }\label{classicalGaugeFix3D}

In order to completely fix the gauge, we have to impose six gauge fixing conditions as we have six independent first class constraints on generic points of $\cc$. Altogether, as already noted, we then have
\begin{equation}\label{numberDegrees}
F = 6N - 12
\end{equation}
independent gauge invariant phase space degrees of freedom away from pathological configurations -- the dimension of the reduced phase space.\footnote{Except for $N=2$, where $F=2$.}

Suppose we want to describe the physics from the internal perspective of particle $A$. In complete analogy to the purely translation invariant case of \cite{vanrietvelde2018change}, we define $A$ to be the origin from which all distances are measured:
\begin{equation}\label{constraintsChi}
\chi^a = q_A^a \overset{!}{=} 0\,,\q\q\q a=x,y,z\,,
\end{equation}
\fla{where the symbol $\overset{!}{=}$ means that we demand the equality \pah{to hold}.} However, we also need to fix the rotational gauge symmetry. Before we move on, we already note that {\it there can be no globally valid gauge fixing conditions} that will pierce every gauge orbit. This follows from pure dimension counting. Indeed, a set of gauge fixing conditions that also completely fixes the rotational symmetry for generic configurations must define a $(6N-12)$-dimensional gauge fixing surface $\cg$ within the $(6N-6)$-dimensional constraint surface $\cc$. Locally, within $\cc$, $\cg$ will be described by six independent conditions. We noted that the gauge orbits in which the total collisions reside are three-dimensional (see Appendix \ref{app_colcol}) and so these will locally, within $\cc$, be described by $6N-9$ independent conditions. Clearly, it is impossible to satisfy the $6+6N-9=6N-3>6N-6$ independent conditions simultaneously within  $\cc$.  The analogous state of affairs holds for the five-dimensional orbits in which totally collinear configurations reside. Hence, any gauge fixing surface that fixes the rotational symmetry for generic configurations will necessarily miss the lower-dimensional gauge orbits. This is an incarnation of the Gribov problem.  {\it Without a global gauge fixing surface, we will also not have a globally valid internal perspective.}

\begin{figure}
    \centering
        \includegraphics[width=.4\textwidth]{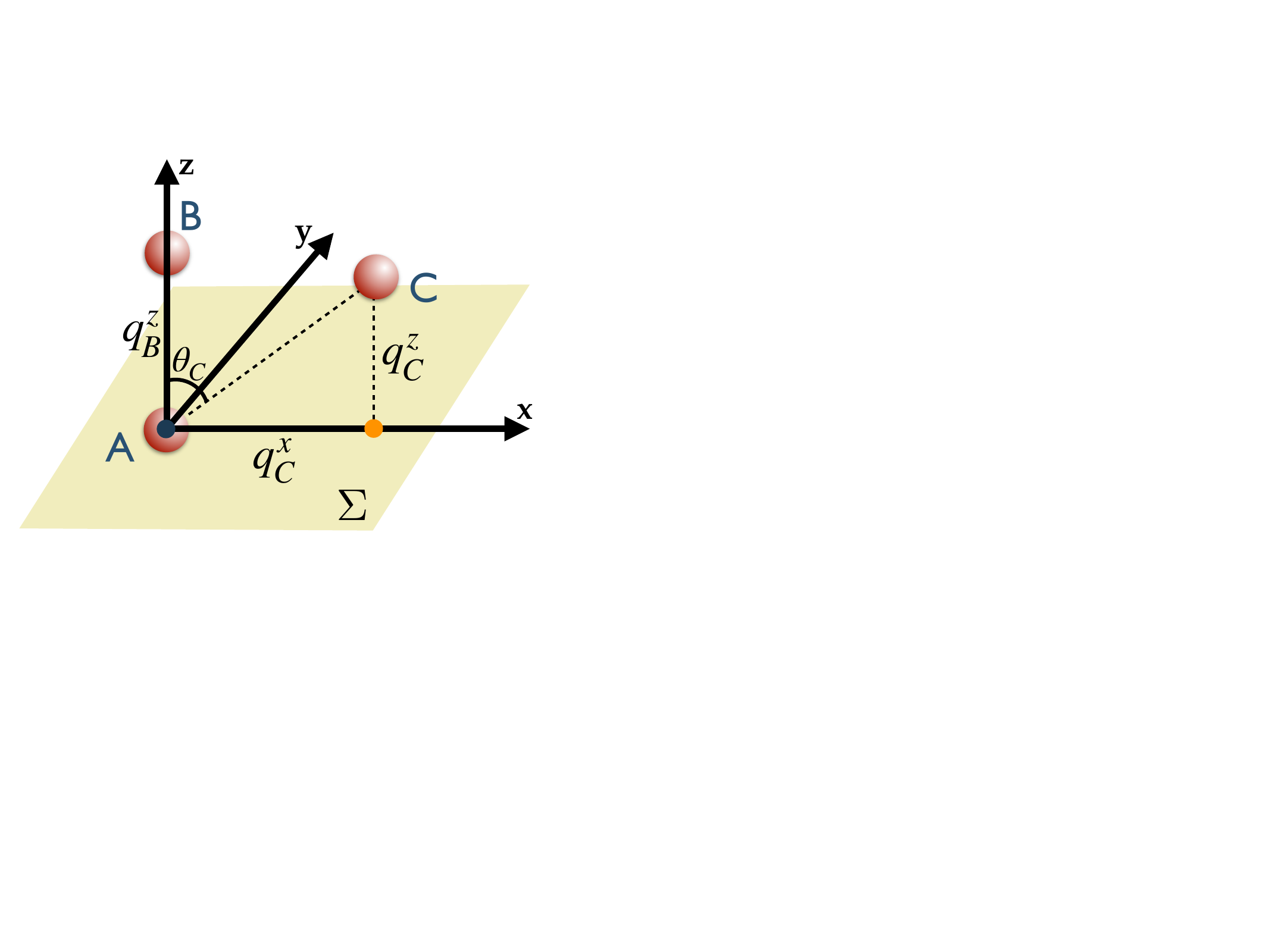}
          \caption{\small Illustration of the gauge-fixing procedure (\ref{constraintsChi}--\ref{constraintDiscrete12}) for the three-body problem. We choose $A$ as the origin from which to measure all distances, $B$ to define the $z$-axis and $C$ to define the $x$-$z$-plane. This procedure works in the same way for $N>3$ particles by choosing a three-body subsystem. }\label{fig:2}
\end{figure}

We proceed by focusing on generic configurations and accept that we will have to miss the lower-dimensional orbits. In order to complete the choice of reference frame, we now also need to define the three axes of space as seen from particle $A$, thereby fixing the three rotational gauge degrees of freedom. This has to be done in terms of the other particles as these provide the only physically meaningful reference for $A$. We have illustrated our gauge-fixing procedure for better visualization in fig.\ \ref{fig:2}.
First, picking another particle $B$, we can get two gauge conditions by fixing \textit{the direction in which $A$ sees $B$}, i.e., the direction (but not the norm) of $\vec{q}_{BA} = \vec{q}_{B} - \vec{q}_{A} \approx \vec{q}_{B}$. Choosing it as $A$'s $z$-axis imposes
\begin{subequations}\label{constraintPhi12}
\begin{equation}\label{constraintPhi1}
\phi_1 =  q_{BA}^y \overset{!}{=} 0\,,
\end{equation}

\begin{equation}\label{constraintPhi2}
\phi_2 =   q_{BA}^x  \overset{!}{=} 0\,.
\end{equation}
\end{subequations}
There is then only one continuous gauge freedom left, corresponding to the rotation of direction $z$ around the origin. To fix it, we consider a third particle $C$, and fix it to lie in the $(x,z)$ plane:
\begin{equation}\label{constraintPhi3}
\phi_3 = q_{CA}^y \overset{!}{=}0\,.
\end{equation}

In fact, we have only blocked all \textit{continuous} gauge transformations, but there remain some allowed \textit{discrete} gauge transformations, namely, the rotations of angle $\pi$ around one of the axes, whose effect is to invert the two other axes. Indeed, those transformations leave invariant the gauge-fixing conditions (\ref{constraintsChi}--\ref{constraintPhi3}). This is a consequence of the fact that the rotational orbits are compact so that the hyperplane defined by (\ref{constraintsChi}--\ref{constraintPhi3}) pierces them multiple times.
To get rid of this residual gauge symmetry, it is necessary and sufficient to fix the orientation of two of the axes. This can be done with conditions, which fix the positive part of the $z$ and $x$ axes through:
\begin{subequations}\label{constraintDiscrete12}
\begin{equation}\label{constraintDiscrete1}
q_{BA}^z > 0\,,
\end{equation}
\begin{equation}\label{constraintDiscrete2}
q_{CA}^x  > 0\,.
\end{equation}
\end{subequations}
With those conditions, $B$ is now constrained to lie on the \textit{semi}-axis $[Az)$, and $C$ to lie on the \textit{half}-plane spanned by the axis $(Az)$ and the semi-axis $[Ax)$, see fig.\ \ref{fig:2}.
Indeed, it is clear from the previous discussion and that in Appendix \ref{app_colcol} that we could not fix the gauge completely, using $B$ and $C$ as material references, if at least one of them were coincident with $A$ or the three particles were collinear. 
Altogether, we then have a system of 12 second-class constraints (given by (\ref{constraints}), (\ref{constraintsChi}--\ref{constraintDiscrete12})), that is, a completely gauge-fixed system, as proven in Appendix \ref{proofGaugeFix}.

In summary, a complete gauge fixation is given by the choice of three particles $A$,$B$,$C$\footnote{The most general case of gauge choice would be given by introducing gauge parameters, which allow for example to fix $A$ to lie not at the origin, but at any position in space - the same goes for the directions of $B$ and $C$. We will not consider in detail this formal possibility which does not add to the physical meaning of this analysis.} out of the $N$ particles, provided they are not collinear. We can denote such a gauge choice as [$A$,$B$,$C$].
Physically, such a gauge choice corresponds to an operational definition of the axes with which $A$ assesses positions of other particles in space: $B$ provides with the first axis of reference $z$, and $C$ defines the plane $(Axz)$. This information is then sufficient for $A$ to build a non-ambiguous set of orthogonal axes $(x,y,z)$. One can also see this procedure as an operational construction of spherical coordinates: the origin is defined as being particle $A$; the zenith direction $z$, to which the polar angle $\theta$ is relative, is defined as the direction of particle $B$; and the plane $(Axz)$, to which the azimuth angle $\phi$ is relative, is defined as the one in which $C$ is lying.

We need to check that the gauge choice is consistent with the equations of motion
\begin{subequations}\label{EOM3D}
\begin{equation}\label{EOMq3D}
\dot{q}_i^a = \frac{\partial H_{\rm tot}}{\partial p_i^a} = p_i^a + \lambda^a + \epsilon^{abc} \mu^b q_i^c\,,
\end{equation}
\begin{equation}\label{EOMp3D}
\dot{p}_i^a = - \frac{\partial H_{\rm tot}}{\partial q_i^a} = \epsilon^{abc} \mu^b p_i^c-\f{\p\,V}{\p q_i^a}\,.
\end{equation}
\end{subequations}
The conservation of our gauge conditions fixes the vectors $\vec{\lambda}$ and $\vec{\mu}$ as defined in (\ref{H3D}), 
\begin{subequations}\label{LagrangeMultipliers}
\begin{equation}\label{}
\vec{\lambda} = - \vec{p}_A\,,
\end{equation}
\begin{equation}\label{}
\mu^y = - \frac{p_{BA}^x}{q_B^z}\,,
\end{equation}
\begin{equation}\label{}
\mu^x =  \frac{p_{BA}^y}{q_B^z}\,,
\end{equation}
\begin{equation}\label{}
\mu^z = -\frac{1}{q_C^x} \Big( p_{CA}^y -  \frac{q_C^z}{q_B^z} p_{BA}^y \Big)\,,
\end{equation}
\end{subequations}
where we write $p_{ij} = p_i - p_j$. Putting (\ref{LagrangeMultipliers}) into (\ref{EOM3D}) then gives us the equations of motion for all remaining particles as seen from $A$'s frame. Note that the validity of the latter requires (\ref{constraintDiscrete12}) to hold; $q^z_B>0$ is implied by our dynamical restriction of infinite repulsion on collisions. However, $q^x_C=0$ is not dynamically ruled out and happens on the measure zero set on the boundary of the gauge invariant configuration space when $u\rightarrow\pm\infty$.

Let us now limit ourselves to $N=3$ particles for clarity. The canonically conjugate Dirac observables (\ref{conjDirac3d}) take the following form in our gauge (see fig.\ \ref{fig:2} for illustration):
\ba \label{DobsClass}
\rho_{BA}&\equiv&\rho_B:=\ln\,r_B= \ln\,{q}^z_B\,,\q\q\q\q\q\q\q\,\,\,\,\, p_{\rho_{BA}}\equiv p^{\rho}_B:=q^z_B\,{p}^z_B\,,\nn\\
\rho_{CA}&\equiv&\rho_C:=\ln\,r_C= \ln\sqrt{(q^x_C)^2+(q^z_C)^2}\,,\q\q\, p_{\rho_{CA}}\equiv p^\rho_C:=p_C^x\,q_C^x+p_C^z\,q_C^z\,,\label{redcoord3d}\\
u&\equiv&u_C:=-\cot\theta_C=-\f{q_C^z}{q^x_{C}}\,,
\nn\\p_{u}&\equiv& p^u_C:=\sin^2\theta_C\,\left(q_C^z\,p_C^x-q_C^x\,p_C^z\right)=\sin^2\theta_C\,R_C^y\,.\nn
\ea
Hence, we may interpret $q^z_B,r_C$ as the distances of $B$ and $C$ as seen from $A$, while $\theta_C$ becomes the angle which $A$ sees between $B$ and $C$. Note also that $C$'s angular momentum around $y$ is essentially conjugate to the angle between $B$ and $C$.

After having gauge fixed our model, we need to replace the Poisson by Dirac brackets \cite{Dirac,Henneaux:1992ig}
\begin{equation}
 \{F,G\}_D := \{F,G\} - \{F,\Lambda_\alpha\} (C^{-1})^{\alpha\beta} \{\Lambda_\beta,G\}\,,
\end{equation}
where $\Lambda_\alpha$, $\alpha=1,\ldots,12$, runs over the 12 second class constraints (\ref{constraints}) and (\ref{constraintsChi}--\ref{constraintDiscrete12}) and $C^{-1}$ is given in (\ref{inverseConstraintsMatrix}) in Appendix \ref{proofGaugeFix}, to construct the bracket structure for our gauge-fixed reduced phase space. The Dirac brackets of the gauge-fixed $q_A^x,q_A^y,q_A^z,q_B^x,q_B^y,q_C^y$ and their conjugated momenta vanish and we can now drop these variables (the latter being solved for by the constraints, see Appendix \ref{appSolv}). By contrast, the $(q_B^z,q_C^z,q_C^x) \in \mathbb{R}_+ \times \mathbb{R} \times \mathbb{R}_+$ (due to (\ref{constraintDiscrete12})\footnote{The gauge invariant configuration space of the three-body problem is homeomorphic to half of $\mathbb{R}^3$ with all collisional and collinear configurations residing in the boundary \cite{littlejohn1995}.}) and their $p_B^z,p_C^x,p_C^z$ remain canonically conjugate also with respect to $\{.,.\}_D$, so we could use them as canonical coordinates on the dynamically accessible region of the reduced phase space. However, for quantization it will be more convenient to use the radial and angle coordinates $(\rho_{BA},\rho_{CA},u) \in \mathbb{R}^3$ and their respective momenta in (\ref{redcoord3d}) which remain canonically conjugate in $\{.,.\}_D$ there too.

Before we continue, let us make an important remark. Denote the reduced phase space, which we have obtained after gauge fixing to $A$'s perspective, by $\cp_{BC|A}$ as it encodes the physics of $B$ and $C$ relative to $A$. Strictly speaking, $\cp_{BC|A}$ is {\it not} actually equivalent to the abstract -- and, in fact, perspective-neutral -- reduced phase space $\cp_{\rm red}=\cc/\!\!\sim$, alluded to in sec.\ \ref{sec_mach3d}. Indeed, $\cp_{BC|A}$ is equivalent to the intersection $\cc\cap\cg_{BC|A}$ (and can be canonically embedded as such into $\cc$), where $\cg_{BC|A}$ is the gauge fixing surface defined by the gauge conditions (\ref{constraintsChi}--\ref{constraintDiscrete12})). As argued above, $\cg_{BC|A}$ entirely misses the lower-dimensional gauge orbits for dimensional reasons and so does not intersect each gauge orbit once and only once. Hence, total collisions and total colinearity are contained in $\cp_{\rm red}$, but not in $\cp_{BC|A}$. The non-equivalence, of course, only concerns a set of measure zero, from which we have also dynamically ruled out the total collisions. \av{The configurations left out in $\cp_{BC|A}$ \pah{for $N=3$} do not depend on our choice of perspective; indeed, gauge-fixing to $B$'s perspective (i.e.\ considering $\cp_{AC|B}$) or to $C$'s perspective ($\cp_{AB|C}$) would leave out exactly the same configurations. This is why the internal perspectives will still be equivalent with one another, even though they are not fully equivalent with the perspective-neutral $\cp_{\rm red}$.} \pah{However, the situation changes for $N>3$, where gauge choices relative to distinct particles can fail on different configurations.} We have also argued above that there are no global gauge fixing conditions, due to the Gribov problem, and so one cannot obtain a gauge fixed phase space that is fully equivalent to $\cp_{\rm red}$. We can thus take $\cp_{BC|A}$ (or any similarly gauge fixed phase space) as a best possible description of the perspective-neutral $\cp_{\rm red}$ relative to a choice of reference frame. We shall work with $\cp_{BC|A}$ and refer to it as the reduced phase space in $A$ perspective.

Using the constraint solutions for the redundant momenta (see Appendix \ref{appSolv}), we can now express our Hamiltonian (\ref{H3D}) in terms of the surviving degrees of freedom on $\cp_{BC|A}$:

\be\label{HamiltonRed3D}
H_{BC|A} = (p_B^z)^2 + (p_C^z)^2 + (p_C^x)^2 + p_B^z \,p_C^z + \f{1}{(q_B^z)^2} (R_C^y)^2 - \f{p_C^x}{q_B^z}  R_C^y  + V(q_B^z,q_C^z,q_C^x)\,.
\ee
This is the Hamiltonian for the dynamics of particles $B$ and $C$, as seen by $A$. The term $p_B^z \,p_C^z$ and the lack of $1/2$ factors in front of the squared momenta correctly take into account the {\it relative} forces originating in the potential in (\ref{H3D}) (e.g., an interaction between $A$ and $C$ will also affect the position of $B$ relative to $A$ even if $B$ and $C$ do not interact, see \cite{vanrietvelde2018change} for further details).
The term $\f{1}{(q_B^z)^2} (R_C^y)^2- \f{p_C^x}{q_B^z}  R_C^y$ constitutes an {\it effective potential} that generally becomes infinitely repulsive on approach of particle $A$ (where our condition (\ref{constraintDiscrete1}) no longer holds) because $B$ is used to relationally define the non-negative $z$-direction from $A$'s perspective. This singularity in the effective potential can be viewed in some analogy to the coordinate singularity in the Schwarzschild spacetime: it is not a physical singularity, but merely signifies that the chosen phase space coordinates become invalid, here due to the global gauge-fixing issues elaborated on above. Indeed, the physical Hamiltonian $H_{\rm tot}$ does not feature any such singular behaviour that is independent of the actual potential $V$. 

This Hamiltonian can also be expressed in terms of the (gauge-fixed) canonically conjugate Dirac observables (\ref{redcoord3d}) as
\be \label{HamiltonRed3DinDobs}
H_{BC|A} = \f{1}{2}  h^{\mu \nu}(\rho_B,\rho_C,u_C) p_\mu p_\nu + V(\rho_B, \rho_C, u_C) \, ,
\ee
where the index $\mu$ is such that $p_\mu$ runs over $p_B^{\rho},p_C^\rho,p_C^u$ in \eqref{redcoord3d} and we have the physical configuration space metric
\ba \label{RedMetric}
&&\!\!\!\!\!h^{\mu \nu} = 2\times \\
&&\!\!\!\!\!\!\!\!\!\!\!\!\!\!\!\small{\begin{pmatrix}
e^{-2 \rho_B} & -\f{1}{2} e^{-\rho_B -\rho_C} \f{u_C}{\sqrt{1+u_C^2}} & -\f{1}{2} e^{-\rho_B -\rho_C} \sqrt{1+u_C^2}\\
-\f{1}{2} e^{-\rho_B -\rho_C} \f{u_C}{\sqrt{1+u_C^2}} & e^{-2 \rho_C} & -\f{1}{2} e^{-\rho_B -\rho_C} \sqrt{1+u_C^2} \\
-\f{1}{2} e^{-\rho_B -\rho_C} \sqrt{1+u_C^2}  & -\f{1}{2} e^{-\rho_B -\rho_C} \sqrt{1+u_C^2} & (e^{-2 \rho_B} + e^{-2 \rho_C}) (1+u_C^2)^2 + e^{-\rho_B -\rho_C} u_C (1+u_C^2)^\f{3}{2} 
\end{pmatrix}}\nn
\ea
with determinant
\be \label{RedVolume} h = 6 e^{-2 \rho_B - 2 \rho_C} (1 + u_C^2)^\f{3}{2} \, \left( (e^{-2 \rho_B} + e^{-2 \rho_C}) \sqrt{1+u_C^2} + e^{- \rho_B - \rho_C} u_C \right) \, . \ee
It can be checked that $h>0$ for all finite values of $\rho_B,\rho_C,u_C$ so that $h=|h|$. However, clearly for $\rho_B,\rho_C\to-\infty$ and $u_C\to\pm\infty$, i.e.\ the configurational pathologies, the metric components diverge. This expression will be useful for directly quantizing $A$'s classical perspective, i.e.\ for reduced quantization, and subsequent comparison with Dirac quantization of the perspective-neutral structure.

It is instructive to look at what (\ref{HamiltonRed3D}) becomes, had we permitted the particles to have different masses $m_i$:
\ba
H_{BC|A} = &\f{1}{2}( \f{1}{m_A} + \f{1}{m_B}) (p_B^z)^2 + \f{1}{2}( \f{1}{m_A} + \f{1}{m_C}) (p_C^z)^2 +\f{1}{2} ( \f{1}{m_A} + \f{1}{m_C}) (p_C^x)^2 + \f{1}{m_A} p_B^z p_C^z \nn
\\ &+ \f{1}{m_B} \f{1}{(q_B^z)^2} (R_C^y)^2 - \f{1}{m_A} \f{1}{q_B^z} p_C^x R_C^y + V(q_B^z,q_C^z,q_C^x) \,.\nn
\ea
Note that in the limit $m_A\rightarrow\infty$ this Hamiltonian becomes of standard form (with effective potential), in agreement with the fact that an infinite mass reference system constitutes an inertial frame.

\subsection{Switching internal perspectives}\label{sec_clswitch3d}

Going from the internal perspective of reference frame $A$ to that of, e.g., particle $C$ amounts to a gauge transformation plus a swap of what one considers the redundant and the relevant Dirac observables (e.g., one would have to exchange the $A$ and $C$ labels in (\ref{conjDirac3d}, \ref{redcoord3d})). This requires one to firstly embed the reduced phase space in $A$ perspective into the perspective-neutral constraint surface, subsequently performing the pertinent gauge transformation and, finally, projecting again to the reduced phase space in $C$ perspective.
We shall only be schematic here as the situation is geometrically transparent. The details of the following discussion can be found in Appendix \ref{app_gaugetr3d} (see also \cite{vanrietvelde2018change}).

There exists a canonical embedding map\footnote{The physical interpretation of $\cp_{BC|A}$ as the physics seen by $A$ singles out this embedding, which otherwise would be ambiguous.}
\ba
\iota_{BC|A}:\cp_{BC|A}\hookrightarrow\cc\,,
\ea
with image $\cc\cap\cg_{BC|A}$, that can be reversed by a `projection'
\ba
\pi_{BC|A}:\cc\cap\cg_{BC|A}\rightarrow\cp_{BC|A}\,,
\ea 
which drops all redundant embedding information, so that $\pi_{BC|A}\circ\iota_{BC|A}=\text{Id}_{\cp_{BC|A}}$. By  exchanging $A$ and $C$ labels, the same construction holds for the reduced phase space $\cp_{AB|C}$  in $C$ perspective.

Switching from $A$ to $C$ perspective requires the gauge transformation $\alpha_{A\to C}$, generated by the constraints $P^a,R^a$, that maps one embedding $\cc\cap\cg_{BC|A}$ to the other $\cc\cap\cg_{AB|C}$. We emphasize that here we take $\cg_{AB|C}$ as being the analogous gauge fixing condition to (\ref{constraintsChi}--\ref{constraintDiscrete12}), except that $A$ and $C$ are everywhere exchanged. That means, we use the analogous gauge fixing procedure as in fig.\ \ref{fig:2}, but seen from the perspective of particle $C$. It is clear that the sought-after transformation amounts to a few translations and rotations of the particles. In Appendix \ref{app_gaugetr3d}, we construct it explicitly and show how this yields a map $
\cs_{A\to C}:\cp_{BC|A}\rightarrow\cp_{AB|C}
$,
that we also spell out explicitly in coordinates, and which satisfies the following commutative diagram:
\begin{center}
\begin{tikzcd}[row sep=huge, column sep = huge]
& \cp_{\rm red}=\mathcal{C}/\!\!\sim \arrow[rd, "\zeta_C"]& \\
\mathcal{C} \cap \mathcal{G}_{BC|A} \arrow[ru, "\zeta^{-1}_A"] \arrow[rr, "\alpha_{A\to C}"]&& \mathcal{C} \cap \mathcal{G}_{AB|C}  \arrow[d, "\pi_{AB|C}"]\\
\mathcal{P}_{BC|A} \arrow[u, "\iota_{BC|A}"] \arrow[rr, "\cs_{A \to C}"] && \mathcal{P}_{AB|C}
\end{tikzcd}
\end{center}
For completeness, we have here also included the perspective-neutral reduced phase space $\cp_{\rm red}=\cc/\!\sim$; $\zeta_A$ is the map that associates with each element of $\cc/\!\!\sim$, i.e.\ gauge orbit in $\cc$, the point on it defined by the intersection with the gauge-fixing surface $\cg_{BC|A}$. Owing to the global gauge fixing issues, this map is not globally defined on $\cp_{\rm red}$ and its inverse is to be understood accordingly. The same applies to $\zeta_C$.

\av{Note that, in the $N=3$ case to which this paper restricts itself, $\alpha$ is an isomorphism, and all perspectives are therefore fully equivalent. This is due to the fact that all perspectives leave out the same set of points (those for which the three particles are aligned). This will however not be the case anymore for $N>3$, where there can be inequivalent perspectives, due to the different possible choices of three-points subsets in this case; see the discussion at the end of Section \ref{quantum3D}.}

Altogether, our change of internal perspective first maps the old perspective back into the perspective-neutral structure, carries out a gauge transformation, and finally projects into a new perspective. The diagram highlights the similarity to coordinate transformations: defining $\varphi_A:=\pi_{BC|A}\circ\zeta_A$, and similarly for $C$, we have $\cs_{A\to C}=\varphi_C\circ\varphi_A^{-1}$.

\section{Quantization and relative states}\label{quantum3D}

We shall now quantize the perspective-neutral structure using the Dirac method. This will endow the Dirac quantized theory with the interpretation of the perspective-neutral quantum theory from which we will then derive the quantum descriptions relative to internal frame perspectives through a quantum symmetry reduction procedure that will constitute the `quantum coordinate maps'.  We shall support this interpretation by exploiting these quantum coordinate maps to derive the transformations that take one from the internal perspective of one quantum reference frame to that of another. On account of the exhibited absence of globally valid internal perspectives, the reduced quantum theories and the transformations between them will also fail to be globally valid, when we take distributional states (classical configurations) into account. For general Hilbert space elements, however, the pathological configurations, being a set of measure zero \cite{littlejohn1995} in the integration, will essentially be irrelevant. Our exposition below will thereby serve to substantiate our discussion in \cite{vanrietvelde2018change} by extending it to a rather general particle model. In order to compare the symmetry reduced quantum theories, obtained upon applying the quantum coordinate maps to the perspective-neutral quantum theory, with the direct quantization of the classical frame perspectives, we shall begin with the reduced quantization of the latter. As we shall see, the two quantizations will not agree in general, exemplifying the generic difference between Dirac and reduced quantization. We will conclude from this that in general we have to give primacy to Dirac quantization because it treats all degrees of freedom on an equal footing and translates the classical redundancy in the description into the quantum theory which ultimately is a prerequisite for defining a notion of quantum frame covariance.

Given that we will encounter a considerable number of Hilbert spaces and transformations between them, we organize the main steps in fig.\ \ref{commutativeDiagram3D} to support the orientation within our construction.

\begin{figure}
\centering
    \begin{tikzcd}
\textrm{\footnotesize original phase space } T^*\cq\simeq\mathbb{R}^{18} \arrow[rrrr, "\vec{P}=\vec{R}=\vec{\chi}=\vec{\phi}=0"] \arrow[d, "\textrm{Dirac quantization}"] & & & & \cp_{BC|A} \arrow[ddd, "\textrm{reduced quantization}"]\\
\mathcal{H}^{\rm kin} \arrow[d, "\delta(\hat{P}^a)"] 
\\
\mathcal{H}^{\rm TI} \arrow[d, "\delta(\hat{R}^a)"] \arrow[r, "{\mathcal{T}}_{A,BC}"]
&  \mathcal{H}^{\rm TI}_{A,BC} \arrow[d, "\delta(\hat{R}^a_B+\hat{R}^a_C)"] \arrow[r, "_A \bra{\vec{\chi}=0}"]
& \mathcal{H}^{\rm TI}_{BC|A} \arrow[d, "\delta(\hat{R}^a_B+\hat{R}^a_C)"] 
\\
\mathcal{H}^{\rm phys} \arrow[r, "{\mathcal{T}}_{A,BC}"]
& \mathcal{H}^{\rm phys}_{A,BC} \arrow[r, "_A \bra{\vec{\chi}=0}"]
& \mathcal{H}^{\rm phys}_{BC|A} \arrow[r, "{\mathcal{R}}_{B,C}"]
& \mathcal{H}^{\rm phys}_{B,C|A} \arrow[r, "\bra{\vec{\phi}=0}"]
& \mathcal{H}_{BC|A} 
\end{tikzcd}
\caption{\small Diagram of the two quantization methods of sec.\ \ref{quantum3D} for the three-body problem and their relation \av{(see also Table \ref{tablenotations} for a summary of the notations used)}. Each column represents a step from the Dirac to the reduced quantum theory, as explained in the main text. Vertical arrows on the left side correspond to constraint imposition. The horizontal arrows between Hilbert spaces are all isometries and correspond to the quantum reduction steps. They are a sequence of: (1) trivialization of the translation generators, (2) conditioning on the classical conditions $\vec{\chi}=0$ that fix the translational gauge freedom, (3) trivialization of the rotation generators, and (4) conditioning on the classical conditions $\vec{\phi}=0$ fixing the rotational gauge freedom. The two squared diagrams among Hilbert spaces on the lower left are commutative. For better visualization, we have summarized the relevant phase and Hilbert spaces appearing in this diagram in a table. The last reduction map $\bra{\vec{\phi}=0}$ is not globally defined owing to the global gauge fixing issues. It does, however, act as an isometry on its image. }\label{commutativeDiagram3D}
\end{figure}

\subsection{Reduced quantization -- quantizing classical internal perspectives} \label{redQ3D}

We begin by quantizing the reduced phase space $\cp_{BC|A}$ in $A$ perspective. For convenience, we choose the canonically conjugate set (\ref{DobsClass}), as the configuration Dirac observables $\rho_B,\rho_C,u$ take value in all of $\mathbb{R}$.\footnote{Otherwise, we would have to deal with global non-trivialities \cite{isham2} in the quantization.} We promote these variables directly to operators satisfying the canonical commutation relations\footnote{Henceforth, we work in units where $\hbar=1$.}
\ba
[\hat{\rho}_B,\hat{\pi}^\rho_B]=[\hat{\rho}_C,\hat{\pi}^\rho_C]=[\hat{u}_C,\hat{\pi}^u_C]=i\,,
\ea
on the Hilbert space $\ch_{BC|A}=L^2(\mathbb{R}^3,\mathrm{d}\rho_B\,\mathrm{d}\rho_C\,\mathrm{d}u_C)$, see fig.\ \ref{commutativeDiagram3D}. We have called our momentum operators $\hat{\pi}$ in order to distinguish them from the momentum operators that we will introduce for the Dirac quantization. In position representation, we will represent them as usual $\hat\pi^\rho_B=-i\partial_{\rho_B}$, etc. An arbitrary reduced quantum state can now be written as
\ba
\ket{\psi}_{BC|A}=\int \,\mathrm{d}\rho_B\,\mathrm{d}\rho_C\,\mathrm{d}u_C\,\psi_{BC|A}(\rho_B,\rho_C,u_C)\,\ket{\rho_B}\ket{\rho_C}\ket{u_C}\,\label{Qredstate}
\ea
and we choose the standard inner product $\int_{\mathbb{R}^3}\mathrm{d}\rho_B\mathrm{d}\rho_C\mathrm{u}_C\,\phi^*_{BC|A}(\rho_B, \rho_C, u_C)\psi_{BC|A}(\rho_B, \rho_C, u_C)$.
To quantize the reduced Hamiltonian (\ref{HamiltonRed3DinDobs}), we choose the standard operator ordering in which the kinetic term in \eqref{HamiltonRed3DinDobs} is promoted to an operator which is equivalent to (minus) the covariant Laplace-Beltrami operator associated with the metric $h^{\mu\nu}$ in \eqref{RedMetric},\footnote{More precisely, the covariant Laplace-Beltrami operator is usually (e.g., see  \cite{kucha1986covariant})  given in the form $-\Delta_h = |h|^{-1/4}\,\hat p_\mu\,|h|^{1/2}h^{\mu\nu}\,\hat p_\nu\,|h|^{-1/4}$, where $\hat p_\mu =-i|h|^{-1/4}\partial_\mu\,|h|^{-1/4}$, and both $\Delta_h$ and $\hat p_\mu$ are self-adjoint  with respect to the inner product $\int\mathrm{d}q^n\,\sqrt{|h|}\,\tilde\phi^*(q^\mu)\tilde\psi(q^\mu)$, i.e.\ one which differs from ours through the density $\sqrt{|h|}$. However, it can be easily checked that $-\Delta_h$ in the measure $\mathrm{d}q^n\sqrt{|h|}$ is equivalent to $|h|^{-\f{1}{4}} \, \hat{\pi}_\mu \, |h|^{\f{1}{2}} \, {h}^{\mu \nu} \, \hat{\pi}_\nu \, {|h|^{-\f{1}{4}}}$ in the measure $\mathrm{d}q^n$ if the wave functions in the two representations are related by $\tilde\psi =|h|^{1/4}\psi$.}
which yields
\be \label{qHred} 
\hat{H}^{\rm red}_{BC|A} = \f{1}{2} |h|^{-\f{1}{4}} \, \hat{\pi}_\mu \, |h|^{\f{1}{2}} \, h^{\mu \nu} \, \hat{\pi}_\nu \, |h|^{-\f{1}{4}} \,  + \, V(\hat{\rho}_B, \hat{\rho}_C, \hat{u}_C) \, .
\ee

\pah{ Being the direct quantization of $A$'s classical frame perspective on the dynamics of particles $B$ and $C$, we can interpret the result for the time being as one {\it a priori possible} quantum theory describing $B$ and $C$ relative to $A$. We will later argue, however, that this is not the appropriate quantum theory describing the perspective of the {\it quantum} reference frame associated with $A$.} Notice that this quantum theory does not include total collisions (which, in any case, we had ruled out dynamically) or total collinearity of the three particles because neither did $\cp_{BC|A}$. Indeed, collisions happen for $\rho_B,\rho_C\rightarrow-\infty$ and total collinearity as $u_C\rightarrow\pm\infty$. Given the normalization conditions in $\ch_{BC|A}$, any $\psi_{BC|A}(\rho_B,\rho_C,u_C)$ will have to have vanishing support there. In this sense, it thus is a quantum theory that is not globally valid. 
Also, the Hamiltonian (\ref{qHred}), while precluding collisions, can still evolve to states with collinear situations in finite time even for initial states that do not have support on it. However, since the pathological configurations comprise a measure zero set, this is not a serious problem and the evolution should be well-defined for a dense set of states. It is only an issue when considering distributional states with support on these pathologies. Given the absence of global relational perspectives, we have to accept this peculiarity for the direct quantization of the classical perspective of $A$.

Later, we will compare this reduced quantized theory to the reduced theory obtained by symmetry reducing the Dirac quantized model. While their structure will look very similar, they will be inequivalent, unless certain somewhat artificial choices are made in Dirac quantization. On the one hand, this will demonstrate that a quantum frame perspective will generally not be equivalent to the direct quantization of a classical frame perspective. On the other, it will nevertheless provide intuition for why we will identify the quantum symmetry reduced theories with quantum frame perspectives.

\subsection{Dirac quantization -- the perspective-neutral quantum theory}

Next, we quantize the perspective-neutral structure the Dirac way. That is, we quantize the original phase space $T^*\cq\simeq\mathbb{R}^{18}$, by promoting all canonical pairs $(q_i^a,p_i^a)$ to operators on a {\it kinematical} Hilbert space $\ch^{\rm kin}:=L^2(\mathbb{R}^9)$, satisfying the canonical commutation relations $[\hat{q}_i^a,\hat{p}_j^b]=i\,\delta^{ab}\,\delta_{ij}$. This permits us to quantize the constraints (\ref{constraints}) and our aim would be to find {\it physical} states $\ket{\psi}^{\rm phys}$ that solve them:
\begin{subequations}\label{Qconstraints}
\begin{equation}\label{QconstraintP}
 \hat{P}^a \ket{\psi}^{\rm phys} = \sum_{i=A,B,C} \hat{p}_i^a \ket{\psi}^{\rm phys}= 0\,,
\end{equation}
\begin{equation}\label{QconstraintR}
\hat{R}^a \ket{\psi}^{\rm phys} = \sum_{i=A,B,C} \epsilon^{abc} \ \hat{q}_i^b \ \hat{p}_i^c \ket{\psi}^{\rm phys} = 0\,.
\end{equation}
\end{subequations}

In addition, we quantize the Hamiltonian (\ref{H3D}) in a standard way, as 

\ba \label{HqDirac}
\hat{H}_{\rm tot} = \frac{1}{2}\,( \hat{\vec{p}}_A{}^2+ \hat{\vec{p}}_B{}^2+ \hat{\vec{p}}_C{}^2) + V(\widehat{|\vec{ q}_{AB} |},\widehat{|\vec{ q}_{AC} |},\widehat{|\vec{ q}_{BC} |}) \, .
\ea

The classical constraint algebra (\ref{classalgebra}) of the Euclidean group directly attains a quantum representation:
\ba
[\hat{P}^a,\hat{P}^b] = 0\,,\q\q\q\q
[\hat{R}^a,\hat{R}^b] = i\epsilon^{abc} \hat{R}^c \,,\q\q\q\q
[\hat{P}^a,\hat{R}^b] = -i \epsilon^{abc} \hat{P}^c \,.
\ea
Notice that the three conditions (\ref{QconstraintR}) are equivalent to a single (quadratic) constraint:
\begin{equation}\label{QconstraintR2}
\vec{\hat R}\,{}^2\,\ket{\psi}^{\rm phys}= 0\,.
\end{equation}
The shape of this algebra permits us to decompose the constraint imposition into convenient steps: 
\begin{enumerate}
\item We construct the {\it translation invariant Hilbert space} $\ch^{\rm TI}$ (see fig.\ \ref{commutativeDiagram3D}) by firstly solving the translation constraints $\hat{P}^a$. Given that the latter have a {\it continuous} spectrum around zero, their solutions will not be normalizable in $\ch^{\rm kin}$ so that $\ch^{\rm TI}$ will not actually be a proper subspace of it. A new translation invariant inner product, normalizing translation invariant states will be required. This will be the topic of sec.\ \ref{sec_Pimpose}.
\item Since $[\hat{P}^a,\hat{R}^b] = -i \epsilon^{abc} \hat{P}^c$, the angular momentum constraints $\hat{R}^a$ will commute with the translation generators $\hat{P}^a$ on $\ch^{\rm TI}$. Hence, the $\hat{R}^a$ will leave $\ch^{\rm TI}$ invariant and we can consistently treat them as translation invariant observables on it. We could therefore then simply impose (\ref{QconstraintR}) on $\ch^{\rm TI}$. Given that the spectrum of the $\hat{R}^a$ is {\it discrete}, solutions to them will be normalizable in $\ch^{\rm TI}$ and so the translation invariant inner product will also constitute the physical inner product. That is, the physical Hilbert space $\ch^{\rm phys}$ (see fig.\ \ref{commutativeDiagram3D}) of translation and rotation invariant states (\ref{Qconstraints}) will be a proper subspace of $\ch^{\rm TI}$.

\end{enumerate}

Notice that the reverse ordering of these steps would be more cumbersome as the translation generators would a priori not leave a rotation invariant Hilbert space invariant.

Nevertheless, for our purposes it will be even more convenient to proceed slightly differently. Our ultimate goal is to construct the transformation that switches the internal perspective from $\ch_{BC|A}$ to $\ch_{AB|C}$, i.e.\ from $A$ to $C$ perspective. To that end, it will not actually be necessary to explicitly construct $\ch^{\rm phys}$, although the transformation will switch via $\ch^{\rm phys}$ from one perspective to another. While constructing $\ch^{\rm phys}$ is certainly feasible, rotation invariant states of three particles will be unnecessarily convoluted and we shall therefore abstain from explicitly exhibiting them. 

Instead, we will sidestep the construction of $\ch^{\rm phys}$ by including an additional step in-between steps 1 and 2 above. Indeed, we will insert a partial reduction step through a map ${\mathcal{T}}_{A,BC}$ that `trivializes' the translation generator constraints, pushing all redundancy onto particle $A$ and yielding the Hilbert space $\ch^{\rm TI}_{A,BC}$ in fig.\ \ref{commutativeDiagram3D}. This step, followed by a subsequent conditioning ${}_A\bra{\vec{q}=0}$, is the quantum analog of imposing the classical gauge condition $\vec{q}_A=0$ and will yield the translation invariant two-body Hilbert space $\ch^{\rm TI}_{BC|A}$, see fig.\ \ref{commutativeDiagram3D}. At this stage, we have a two-body problem of $B$ and $C$, and imposing rotational invariance will become substantially simpler. In this manner, we explicitly construct $\ch^{\rm phys}_{BC|A}$ in fig.\  \ref{commutativeDiagram3D}, replacing step 2 above that would yield $\ch^{\rm phys}$. We show, however, that the imposition of the rotation constraints $\delta(\hat{R}^a)$ commutes with the trivialization map ${\mathcal{T}}_{A,BC}$ and that $\ch^{\rm phys}_{BC|A}$ will thereby be equivalent to $\ch^{\rm phys}$. This procedure will be detailed in secs.\ \ref{translationalRed} and \ref{rotationalRed}. 

The remaining reduction steps  to $\ch_{BC|A}$ in fig.\ \ref{commutativeDiagram3D} will be discussed in sec.\ \ref{sec_recovery}.

\begin{table}
\centering
\renewcommand{\arraystretch}{1.5}

    \begin{tabular}{ | c | c |}
\hline
 $\mathcal{P}_{BC|A}$ & Classical phase space after reduction to the perspective of A \\
 $\mathcal{H}^\textrm{kin}$ & Kinematical (`total') Hilbert space  \\
 $\mathcal{H}^\textrm{TI}$ & Translation-invariant Hilbert space \\
 $\mathcal{H}^\textrm{TI}_{A,BC}$ & Image of the prior under translational trivialization $\mathcal{T}_{A,BC}$ \\
 $\mathcal{H}^\textrm{TI}_{BC|A}$ & Image of the prior under reduction by $_A\bra{\chi=0}$ \\
 $\mathcal{H}^\textrm{phys}$ & Physical (i.e.\ translation and rotation-invariant) Hilbert space \\
 $\mathcal{H}^\textrm{phys}_{A,BC}$ & Image of the prior under transformation $\mathcal{T}_{A,BC}$  \\
 $\mathcal{H}^\textrm{phys}_{BC|A}$ & Image of the prior under reduction by $_A\bra{\chi=0}$  \\
 $\mathcal{H}^\textrm{phys}_{B,C|A}$ & Image of the prior under the rotational trivialization $\mathcal{R}_{B,C}$ \\
 $\mathcal{H}_{BC|A}$ & Can be defined as both:\\
 &(i) Image of the prior under conditioning by $\bra{\theta_B, \varphi_B, \varphi_C = 0}$\\ &(ii) Result of the quantization of $\mathcal{P}_{BC|A}$ \\
 \hline
\end{tabular}

\caption{\av{A summary of the notations used.}}
 \label{tablenotations}
\end{table}

\subsubsection{Constructing the translation invariant Hilbert space}\label{sec_Pimpose}

Given the linearity of the translation generators $\hat{P}^a$, it is not difficult to construct the translation invariant Hilbert space $\ch^{\rm TI}$. As this step is already carried out in \cite{vanrietvelde2018change},\footnote{In \cite{vanrietvelde2018change}, this step is carried out for the one-dimensional three-body problem, however, extension to three dimensions is trivial and amounts to simply `attaching a vector symbol $\,\,\, \vec{}\,\,\, $ to the basic variables'.} we shall be brief here and refer to this reference for further details.

We invoke {\it group averaging} \cite{Marolf:1995cn,Marolf:2000iq,Thiemann:2007zz} to define an (improper) projector onto solutions of (\ref{QconstraintP}):\footnote{See \cite{Kempf:2000qz} for an alternative method that adapts the underlying Hilbert space topology.}
\begin{equation}\label{projp}
\begin{split}
\delta(\vec{\hat P}) : \quad  & \mathcal{H}^{\rm kin}  \to  \mathcal{H}^{\rm TI} \\
& \ket{\psi}^{\rm kin}  \mapsto  \ket{\psi}^{\rm TI}:=\Big(\f{1}{(2\pi)^3} \int_{-\infty}^{+\infty}\, \mathrm{d}s^x\,\mathrm{d}s^y\,\mathrm{d}s^z \ e^{i s^a \hat{P}^a}  \Big) \ket{\psi}^{\rm kin}\,. 
\end{split}
\end{equation}
A general translation invariant state then takes any of the following forms 
\ba
\ket{\psi}^{\rm TI}&=& \int \mathrm{d}^3 \vec{p}_B \ \mathrm{d}^3\vec{p}_C \ \psi^{\rm TI}_{BC|A}(\vec{p}_B,\vec{p}_C) \ket{-\vec{p}_B-\vec{p}_C}_A \ket{\vec{p}_B}_B \ket{\vec{p}_C}_C\nn\\
&=& \int \mathrm{d}^3 \vec{p}_A \ \mathrm{d}^3\vec{p}_C \ \psi^{\rm TI}_{AC|B}(\vec{p}_A,\vec{p}_C) \ket{\vec{p}_A}_A \ket{-\vec{p}_A-\vec{p}_C}_B \ket{\vec{p}_C}_C\label{physstate1d}\\
&=& \int \mathrm{d}^3 \vec{p}_A \ \mathrm{d}^3\vec{p}_B \ \psi^{\rm TI}_{AB|C}(\vec{p}_A,\vec{p}_B) \ket{\vec{p}_A}_A \ket{\vec{p}_B}_B \ket{-\vec{p}_A-\vec{p}_B}_C\,,\nn
\ea
depending on which particle's momentum is solved for, where for later use we have defined
\ba
\psi^{\rm TI}_{BC|A}(\vec{p}_B,\vec{p}_C)&:=&\psi^{\rm kin}(-\vec{p}_B-\vec{p}_C,\vec{p}_B,\vec{p}_C)\,,\nn\\
\psi^{\rm TI}_{AC|B}(\vec{p}_A,\vec{p}_C)&:=&\psi^{\rm kin}(\vec{p}_A,-\vec{p}_A-\vec{p}_C,\vec{p}_C)\,,\label{redphysstate}\\
\psi^{\rm TI}_{AB|C}(\vec{p}_A,\vec{p}_B)&:=&\psi^{\rm kin}(\vec{p}_A,\vec{p}_B,-\vec{p}_A-\vec{p}_B)\,.\nn
\ea
This yields three different descriptions of the same translation invariant state $\ket{\psi}^{\rm TI}$ and will be of use later, corresponding to different internal frame perspectives.
The translation invariant inner product, normalizing these states, reads
\begin{equation}
(\psi^{\rm TI},\phi^{\rm TI})_{\rm TI} := {}^{\rm kin}\bra{\psi} \delta(\vec{\hat P})\ket{\phi}^{\rm kin}\,,
\end{equation}
where $\langle\cdot|\cdot\rangle$ denotes the inner product of $\ch^{\rm kin}$. Based on these structures, one can Cauchy complete the space of solutions of (\ref{QconstraintP})  to a Hilbert space $\ch^{\rm TI}$.

The conjugate Dirac observables (in perspective-neutral form) (\ref{conjDirac3d}) are, in particular, translation invariant. We can thus represent them already on $\ch^{\rm TI}$. We choose a symmetric factor ordering:
\ba\label{Dobserv}
\Hat{\rho}_{BA} &=& \reallywidehat{\ln{\sqrt{\vec{q}_{BA}^{\ 2}}}}, \q\q\q\q\q\q\q\q\q\q\q\q\q\q\q\q \Hat{p}_{\rho_{BA}} =  \f{1}{2} ( {\vec{\hat p}}_{B} \cdot {\vec{\hat q}}_{BA} + {\vec{\hat q}}_{BA}   \cdot {\vec{\hat p}}_{B})\,,\nn\\
\Hat{\rho}_{CA}& =&\reallywidehat{\ln{\sqrt{\vec{q}_{CA}^{\ 2}}}}, \q\q\q\q\q\q\q\q\q\q\q\q\q\q\q\q \Hat{p}_{\rho_{CA}} =  \f{1}{2} ( {\vec{\hat p}}_{C} \cdot {\vec{\hat q}}_{CA} + {\vec{\hat q}}_{CA}   \cdot {\vec{\hat p}}_{C}) \,,\nn\\
\Hat{u} &=& - \reallywidehat{|\vec{q}_{BA} \times \vec{q}_{CA}|^{-1}}\,\,\cdot {\vec{\hat q}}_{BA} \cdot {\vec{\hat q}}_{CA} = - \widehat{\cot \gamma} \,,\\
    \Hat{p}_u &= &\f{1}{2} \Big( ( -\hat{u}\,\reallywidehat{(1+u^2)^{-1}}\, {\vec{\hat q}}_{CA} - \reallywidehat{(1+u^2)^{-1/2}}  \, e^{\hat{\rho}_{CA}-\hat{\rho}_{BA}} \,\hat{\vec{q}}_{BA} ) \cdot {\vec{\hat p}}_{C} \nn\\
    &&\q\q\q\q\q\q\q\q\q+ {\vec{\hat p}}_{C} \cdot ( -\hat{u}\,\reallywidehat{(1+u^2)^{-1}}\, {\vec{\hat q}}_{CA} - \reallywidehat{(1+u^2)^{-1/2}} \, e^{\hat{\rho}_{CA}-\hat{\rho}_{BA}} \,{\vec{\hat q}}_{BA} ) \Big) \nn\\
    &= & \f{1}{2} \left( ( \widehat{\sin{\gamma}}\, \widehat{\cos{\gamma}}\, {\vec{\hat q}}_{CA} - \widehat{\sin{\gamma}} \, \, e^{\hat{\rho}_{CA}-\hat{\rho}_{BA}} \,{\vec{\hat q}}_{BA} ) \cdot {\vec{\hat p}}_{C} \right.\nn\\
    &&\q\q\q\q\q\q\q\q\q\q\left.+ {\vec{\hat p}}_{C} \cdot ( \widehat{\sin{\gamma}}\, \widehat{\cos{\gamma}} \,{\vec{\hat q}}_{CA} - \widehat{\sin{\gamma}} \,\, e^{\hat{\rho}_{CA}-\hat{\rho}_{BA}} \,{\vec{\hat q}}_{BA} ) \right)\,.\nn
\ea
Some of these operators have to be understood in terms of spectral decomposition. As noted above, $\ch^{\rm phys}$ is a proper subspace of $\ch^{\rm TI}$ so that these operators here already define a representation of the Dirac observables (\ref{conjDirac3d}) on $\ch^{\rm phys}$ too. It is clear that some of these Dirac observable operators will be unbounded on states with support on total collisions. Since these pathological configurations are a set of measure zero \cite{littlejohn1995}, there is still a chance that these operators may be densely defined and ultimately essentially self-adjoint. However, we shall not enter into such subtleties here, as our focus is on symmetry reduction to a quantum frame perspective; our construction will thus be somewhat informal.  We also recall that we have dynamically ruled out total collisions through a potential that becomes infinitely repulsive there. The quantized translation and rotation invariant Hamiltonian (\ref{HqDirac}) is an observable on both $\ch^{\rm TI}$ and $\ch^{\rm phys}$.
In particular, on $\ch^{\rm phys}$ (which we do not construct explicitly), this will mean that, starting with finite-energy states that do {\it not} have support on total collisions, will not dynamically produce states that do if evolved with $\hat{H}_{\rm tot}$. On these dynamically accessible states, $\hat{\rho}_{BA},\hat{p}_{\rho_{BA}},\hat{\rho}_{CA},\hat{p}_{\rho_{CA}}$ and $\hat{p}_u$ will remain bounded and only $\hat{u}$ will be unbounded for states having support on collinear configurations.

It must be emphasized that, clearly, one can promote other Dirac observables to self-adjoint operators on both $\ch^{\rm TI}$ and $\ch^{\rm phys}$. For example. the `basis set' of Dirac observables (\ref{indirac3d}) does not feature any pathologies and can  be turned into self-adjoint operators, e.g.\ through a symmetric factor ordering. Hence, the perspective-neutral quantum theory will not have any issues per se. The pathologies are purely related to choosing observables that (i) have a direct interpretation in the specific perspective of $A$ and (ii) are canonically conjugate (where defined) so that later we can relate them to the reduced quantum theory on $\ch_{BC|A}$. As argued in sec.\ \ref{sec_mach3d}, better conjugate observables do not exist in this model due to the Gribov problem.

\subsubsection{Translational reduction}\label{translationalRed}

As explained above, instead of directly continuing with the imposition of the rotation constraints, we firstly include the step of translational reduction to effectively produce a two-body problem that simplifies solving the (then transformed) constraints. This step is the quantum analog of gauge fixing to $\vec{q}_A=0$ in (\ref{constraintsChi}). Clearly, in the quantum theory we can {\it not} gauge fix because imposing the translation constraints (\ref{projp}) directly produces translation invariant states.\footnote{As a consequence of the uncertainty relations, projecting onto the zero-eigenvectors of the total momentum leads to a maximal spread in gauge variables conjugate to it and thereby to translation invariance.} In order to remove redundant degrees of freedom and reduce the quantum theory, we thus have to proceed differently. As shown in \cite{vanrietvelde2018change}, this works by firstly trivializing the constraints and subsequently projecting onto the classical gauge fixing conditions. Some of the below is a repetition of steps carried out in \cite{vanrietvelde2018change} and so we shall be brief.  

We define the {\it constraint trivialization} map, $\ct_{A,BC}:\ch^{\rm TI}\rightarrow\ch^{\rm TI}_{A,BC}$, see fig.\ \ref{commutativeDiagram3D},
\begin{equation}\label{}
{\mathcal{T}}_{A,BC} = \prod_a \exp{ \Big( i\, \Hat{q}_A^a (\Hat{p}_B^a + \Hat{p}_C^a) \Big) }\,,
\end{equation}
which is unitary on $\ch^{\rm kin}$, trivializes the translation generators into the $A$ tensor factor
\ba\label{Ptrivial}
\ct_{A,BC}\,\hat{P}^a\,\ct_{A,BC}^\dag = \hat{p}_A^a\,,
\ea
where $\dag$ is defined with respect to $\ch^{\rm kin}$, and maps translation invariant states to
\begin{equation}\label{TIABC}
\ket{\psi}^{\rm TI}_{A,BC} =  {\mathcal{T}}_{A,BC} \ket{\psi}^{\rm TI} = \ket{\vec{p}=0}_A \otimes \Big( \int \mathrm{d}^3 \vec{p}_B \ \mathrm{d}^3 \vec{p}_C \ \psi^{\rm TI}_{BC|A}(\vec{p}_B,\vec{p}_C) \ket{\vec{p}_B}_B \ket{\vec{p}_C}_C \Big)\,.
\end{equation}
Hence, the $A$ tensor factor of the state carries no relevant information about the original state and has become redundant. We can remove it through a Page-Wootters-like \cite{Page:1983uc} {\it conditioning on the classical gauge-fixing conditions} (\ref{constraintsChi}),\footnote{Equivalence of this reduction procedure with the Page-Wootters formalism has been demonstrated in later work \cite{Hoehn:2019owq,Hoehn:2020epv}.} producing states in a novel Hilbert space $\ch^{\rm TI}_{BC|A}$ in fig.\ \ref{commutativeDiagram3D}
\ba
\ket{\psi}^{\rm TI}_{BC|A}:=(2\pi)^{3/2}{}_A\braket{\vec{q}=0\,|\,\psi}^{\rm TI}_{A,BC}=  \int \mathrm{d}^3 \vec{p}_B \ \mathrm{d}^3 \vec{p}_C \ \psi^{\rm TI}_{BC|A}(\vec{p}_B,\vec{p}_C) \ket{\vec{p}_B}_B \ket{\vec{p}_C}_C\,.
\ea
As shown in \cite{vanrietvelde2018change}, this defines an isometry from $\ch^{\rm TI}$ to $\ch^{\rm TI}_{BC|A}$, where the inner product on the latter is just the standard one with integration over $B$ and $C$ variables only.

Crucial for us is also how it transforms basic translation invariant operators
\ba\label{Tobs}
\ct_{A,BC}\,(\vec{\hat{q}}_i-\vec{\hat{q}}_A)\,\ct^\dag_{A,BC}=\vec{\hat{q}}_i\,,\q\q\q\q\q\q \ct_{A,BC}\,\vec{\hat{p}}_i\,\ct^\dag_{A,BC}=\vec{\hat{p}}_i\,,\q\q\q i=B,C\, \, ,
\ea
and the Hamiltonian (\ref{HqDirac}), which becomes\footnote{We have not written the terms which are null on $\ch^{\rm TI}_{A,BC}$.}
\be \label{HqDiracTI}
\hat{H}^\textrm{tot}_{A,BC} := \ct_{A,BC}\,\hat{H}^\textrm{tot}\,\ct^\dag_{A,BC} = \vec{\hat p}_B{}^2 + \vec{\hat p}_C{}^2 + \vec{\hat p}_B \cdot \vec{\hat p}_C + V(\widehat{|\vec{ q}_{B} |},\widehat{|\vec{ q}_{C} |},\widehat{|\vec{ q}_{BC} |}) \, .
\ee

The Dirac observables (\ref{Dobserv}) simplify under  transformation (\ref{Tobs}); we shall not spell out their transformed form as it is clear that it amounts to simply dropping the $A$-label from all expressions in (\ref{Dobserv}).\footnote{In some cases, this has to be justified by considering the operator functions as Taylor series away from pathological configurations.} 
As one can easily check using (\ref{Ptrivial}, \ref{Tobs}), this translation generator trivialization map leaves the rotation generators (\ref{QconstraintR}) invariant:
\ba
\ct_{A,BC}\,\hat{R}^a\,\ct^\dag_{A,BC}=\hat{R}^a\,.
\ea
This implies that imposing (\ref{QconstraintR2}) and applying the trivialization map $\ct_{A,BC}$ {\it commutes} on $\ch^{\rm TI}$ such that the square on the lower left of fig.\ \ref{commutativeDiagram3D} is indeed commutative. Noting the shape of (\ref{TIABC}), this leads to the following equivalences
\ba
\hat{R}^a\,\ket{\psi}^{\rm TI}=0\q\q\Leftrightarrow\q\q(\hat{R}^a_B+\hat{R}^a_C)\,\ket{\psi}^{\rm TI}_{A,BC}=0\q\q\Leftrightarrow\q\q (\hat{R}^a_B+\hat{R}^a_C)\,\ket{\psi}^{\rm TI}_{BC|A}=0\,.\nn
\ea
Hence, we lose no information by imposing two-body rotational invariance on states in $\ch^{\rm TI}_{BC|A}$, rather than three-body rotational invariance in $\ch^{\rm TI}$, thereby simplifying the task.

\av{Note how this means that we effectively turned the situation into a two-body problem. Indeed, the use of the constraint trivialization map $\ct_{A,BC}$ mapped us to a description in which $A$'s degrees of freedom have become pure gauge, so that only $B$ and $C$'s degrees of freedom are potentially meaningful (although they still include some rotational gauge). However, the ``two bodies'' here are not strictly speaking $B$ and $C$; rather, they correspond to $B$ and $C$'s positions relative to $A$. The intuition is that, as in \cite{vanrietvelde2018change}, trivialising the translational gauge effectively amounts to eliminating one of the particles by moving to a description in which positions are defined relative to it.}

\subsubsection{Rotational invariance}\label{rotationalRed}

We continue with the two-body problem on $\ch^{\rm TI}_{BC|A}\simeq L^2(\mathbb{R}^3)\otimes L^2(\mathbb{R}^3)$. In order to impose rotational invariance, it will be more convenient to Fourier transform to configuration space and to switch to polar coordinates $(r,\theta,\phi)$, so we can use spherical harmonics $Y^{j,m}(\theta,\varphi)$ to define a basis for $L^2(\mathbb{R}^3)$
\begin{equation}\label{}
\ket{r; j,m} =   \int \mathrm{d}\Omega \ Y^{j,m}(\theta,\varphi) \ket{r, \theta, \varphi}\,,
\end{equation}
where $r=|\vec{q}\,|$. Here, $\ket{j,m}$ are the usual simultaneous eigenstates of the angular momentum operators $\hat{R}^2$ and $\hat{R}^z$. This construction is done more carefully in Appendix \ref{appendixSpherical}.

According to the Clebsch-Gordan coefficients for the decomposition of angular momentum eigenstates, we can then write an arbitrary zero-total-angular-momentum state in $\ch^{\rm TI}_{BC|A}$ as:
\begin{equation}\label{psiDecomposed}
\ket{\psi}^{\rm phys}_{BC|A} = \int \mathrm{d} r_B \ \mathrm{d} r_C \ r_B^2 \,r_C^2 \sum_{j=0}^\infty \psi_{BC|A}^{\rm phys}(r_B, r_C; j) \ket{\Phi(r_B,r_C;j)}\,,
\end{equation}
where
\begin{equation}\label{Phi}
\ket{\Phi(r_B,r_C;j)} = \sum_{|m| \leq j} \frac{(-1)^{j-m}}{\sqrt{2j+1}} \ket{r_B;j,-m}_B \ket{r_C;j,+m}_C \,.
\end{equation}
Notice that these states lie in a proper subspace of $\ch^{\rm TI}_{BC|A}$, which we will label by $\ch^{\rm phys}_{BC|A}$ (see fig.\ \ref{commutativeDiagram3D}), given that it now also includes rotational invariance. The transformations of the Dirac observables (\ref{Dobserv}) alluded to in sec.\ \ref{translationalRed} (i.e., those in (\ref{Dobserv}) with all $A$-labels dropped), are also observables on $\ch^{\rm phys}_{BC|A}$. From the discussion above it follows that this Hilbert space is equivalent to $\ch^{\rm phys}$, including the evaluation of the (transformed) Dirac observables.

\subsection{Rotational reduction to $A$'s perspective}\label{sec_recovery}

We have implemented translational invariance, a subsequent translational reduction and rotational invariance. There is therefore still redundancy in the description which we have to remove in order to fully construct $A$'s quantum frame perspective from the perspective-neutral Dirac quantized theory and to compare the result to the reduced quantum theory on $\ch_{BC|A}$ of sec.\ \ref{redQ3D}, In particular, we now also have to carry out a rotational reduction, which will be the quantum analog of the classical gauge fixing $\vec{\phi}=0$ in (\ref{constraintPhi12}--\ref{constraintDiscrete12}) that fixes the rotational gauge freedom for generic configurations. Since we cannot directly gauge fix in the Dirac quantized theory (the system is already translation and rotation invariant after imposition of the quantum constraints), we have to proceed differently. As explained through the translational reduction in sec.\ \ref{translationalRed} (and in \cite{vanrietvelde2018change}), the quantum analog is (1) trivialize the relevant constraints, and (2) condition on the classical gauge fixing conditions. We will now show how this works for the rotation generators.

\subsubsection{Rotational reduction}\label{sssec_rotred}

We define a {\it rotation generator trivialization} map
\begin{equation}\label{RBC}
{\mathcal{R}}_{B,C} = \exp{\Big( i\, \Hat{\varphi}_B \hat{R}_C^z \Big) } \exp{ \Big( i\, \Hat{\theta}_B \hat{R}_C^y \Big)} \exp{ \Big( i\, \Hat{\varphi}_B \hat{R}_C^z \Big) }\,,
\end{equation} 
where the angular operators $\Hat{\theta}$ and $\Hat{\varphi}$ can be defined in terms of the states $ \ket{r, \theta, \varphi}$, see Appendix \ref{appendixSphericalObservables}. As shown in Appendix \ref{proofRotational}, geometrically, this sequence of rotations of particle $C$ effectively rotates its polar coordinate system until its polar angle $\theta_C$ is measured relative to the direction of $B$. Hence, $\theta_C$ will coincide with the relative angle $\gamma$ between $B$ and $C$ (essentially the Dirac observable $u$ in (\ref{conjDirac3d})), see also figs.\ \ref{fig:1a} and \ref{fig:2}. This procedure is thus reminiscent of the one followed classically. 
Appendix \ref{proofRotational} also proves that ${\mathcal{R}}_{B,C}$ maps $\ket{\psi}^{\rm phys}_{BC|A}$ as defined in (\ref{psiDecomposed}) to
\begin{equation}\label{PsiB,C|A}
\ket{\psi}^{\rm phys}_{B,C|A} = {\mathcal{R}}_{B,C} \ket{\psi}^{\rm phys}_{BC|A} = \int \mathrm{d} r_B \ \mathrm{d} r_C \ r_B^2\, r_C^2 \sum_{j=0}^\infty (-1)^j \psi^{\rm phys}_{BC|A}(r_B, r_C; j) \ket{r_B;0,0}_B \ket{r_C;j,0}_C \,,
\end{equation}
i.e.\ to a state of zero total angular momentum in $B$ and zero angular momentum around $z$ in $C$, and that this is an isometry from $\ch^{\rm phys}_{BC|A}$ to a new Hilbert subspace $\ch^{\rm phys}_{B,C|A}:=\calr_{B,C}(\ch^{\rm phys}_{BC|A})\subset \ch^{\rm TI}_{BC|A}$ (see fig.\ \ref{commutativeDiagram3D}).
In other words, $\mathcal{H}^{\rm phys}_{B,C|A}$ is spanned by states of the form $\ket{r_B;0,0}_B \ket{r_C;j,0}_C$.

As can be expected from the shape of (\ref{PsiB,C|A}) and shown in Appendix \ref{app_RBCtrivial}, $\calr_{B,C}$ indeed trivializes the rotation constraints with $\calr_{B,C}\,(\vec{\hat{R}}_B+\vec{\hat{R}}_C)^2\,\calr_{B,C}^{-1}$ being equivalent to
\ba
(\vec{\hat{R}}_B)^2 \ket{\psi}^{\rm phys}_{B,C|A}=
 \hat{R}_C^z \ket{\psi}^{\rm phys}_{B,C|A}=0\,.
\ea

It is also necessary to check how the Dirac observables (\ref{Dobserv}) transform under $\calr_{B,C}$. To this end, recall that we have carried out a translational reduction in sec.\ \ref{translationalRed}, which corresponded to simply dropping the $A$ labels from all expressions in (\ref{Dobserv}). In Appendix \ref{app_RBCobs}, we show that (these transformations are only valid on states without support on pathological configurations):\footnote{Notice that these observables are, in fact, somewhat better behaved as those in (\ref{Dobserv}) because the operator $\theta_C$ appearing here is fully defined on all states, thanks to the conditions in Appendix \ref{appendixSphericalObservables} which we had not applied similarly to $\gamma$ appearing in (\ref{Dobserv}).}
\ba
\calr_{B,C}\,\Hat{\rho}_B\,\calr_{B,C}^{-1}&=& \Hat{\rho}_B\,,\q\q\q\q\q\q\q\q\q\q\q\q\q
\calr_{B,C}\,\Hat{p}^{\rho}_B\,\calr_{B,C}^{-1}= \Hat{p}^{\rho}_B\,,\nn\\
\calr_{B,C}\,\Hat{\rho}_C\,\calr_{B,C}^{-1}&=& \Hat{\rho}_C\,,\q\q\q\q\q\q\q\q\q\q\q\q\q
\calr_{B,C}\,\Hat{p}^{\rho}_C\,\calr_{B,C}^{-1}= \Hat{p}^{\rho}_C\,,\label{transbcobs}\\
{\mathcal{R}}_{B,C} \,\Hat{u}_C\, {\mathcal{R}}_{B,C}^{-1} &=& - \widehat{\cot \, \theta_C} \,,\nn\\
    {\mathcal{R}}_{B,C} \,\Hat{p}_C^u\,{\mathcal{R}}_{B,C}^{-1} 
    &=& - \f{1}{2} \left( - \widehat{\cos{\theta_C}}\, \widehat{\sin{\theta_C}} \,{\vec{\hat q}}_{C} \cdot {\vec{\hat p}}_{C} + \Hat{r}_C\, \widehat{\sin{\theta_C}} \,\Hat{p}_C^z
    - {\vec{\hat p}}_{C} \cdot \,{\vec{\hat q}}_{C} \,\widehat{\cos{\theta_C}}\, \widehat{\sin{\theta_C}} \right.\nn\\
    &&\q\q\q\q\q\q\q\left.+ \Hat{p}_C^z\, \Hat{r}_C\, \widehat{\sin{\theta_C}} \right)\,.\nn
\ea
The key point of those transformations is that $\widehat{\gamma}$ gets mapped to $\Hat{\theta}_C$:
\be {\mathcal{R}}_{B,C} \,\widehat{\gamma} \,{\mathcal{R}}_{B,C}^{-1} = \Hat{\theta}_C\,. \ee
This is in harmony with the above observation that the transformation ${\mathcal{R}}_{B,C}$ has the property of transforming the description of particle $C$ so that its polar angle now physically corresponds to the relative angle that it forms with the direction of $B$.
Under these transformations, the Hamiltonian (\ref{HqDiracTI}) is mapped to\footnote{We have not written the terms which are null on $\mathcal{H}^{\rm phys}_{B,C|A}$. Furthermore, for brevity we have chosen a factor ordering where the momenta always stand on the right. Despite its appearance, this Hamiltonian is actually a symmetric operator.}
\be \label{HqDiracTrans} \begin{split}
&\hat{H}^\textrm{tot}_{B,C|A} :=  {\mathcal{R}}_{B,C} \,\hat{H}^\textrm{tot}_{A,BC}\,{\mathcal{R}}_{B,C}^{-1} \\
&= e^{-2 \hat{\rho}_B} \hat{p}_B^\rho{}^2 + e^{-2 \hat{\rho}_C} \hat{p}_C^\rho{}^2 + \left( (e^{-2 \hat{\rho}_B} + e^{-2 \hat{\rho}_C}) (1+ \hat{u}_C^2)^2 + e^{- \hat{\rho}_B - \hat{\rho}_C} \hat{u}_C \reallywidehat{(1+u_C^2)^\f{3}{2}} \right) \hat{p}_C^u{}^2 \\
&- e^{- \hat{\rho}_B - \hat{\rho}_C} \left( \reallywidehat{\sqrt{1+u_C^2}} (\hat{p}_C^u \hat{p}_B^\rho + \hat{p}_C^u \hat{p}_C^\rho ) + \reallywidehat{\f{u_c}{\sqrt{1+u_C^2}}} \hat p_B^\rho \hat p_C^\rho \right) + 2 i e^{-2 \hat{\rho}_B} \hat{p}_B^\rho + 2 i e^{-2 \hat{\rho}_C} \hat{p}_C^\rho\\
& - i \left( 4 (\hat{u}_C + \hat{u}_C^3)(e^{-2 \hat{\rho}_B} + e^{-2 \hat{\rho}_C}) + 2 e^{- \hat{\rho}_B - \hat{\rho}_C} (1+ 2 \hat{u}_C^2) \reallywidehat{\sqrt{1+u_C^2}} \right) \hat p_C^u \\
&-\f{9}{4} (e^{-2 \hat{\rho}_B} + e^{-2 \hat{\rho}_C}) (1+ \hat{u}_C^2) -\f{9}{4} e^{- \hat{\rho}_B - \hat{\rho}_C} \hat{u}_C \reallywidehat{\sqrt{1+u_C^2}} + V(\hat{\rho}_B,\hat{\rho}_C,\hat{u}_C) \, .
\end{split}
\ee

We now wish to complete the rotational reduction, removing -- in analogy to the translational reduction in sec.\ \ref{translationalRed} (and in \cite{vanrietvelde2018change}) -- any redundant information from our states (\ref{PsiB,C|A}) and recovering the reduced quantum theory on $\ch_{BC|A}$ in $A$ perspective from sec.\ \ref{redQ3D}. This second step will, again, be achieved by a conditioning on the classical gauge fixing conditions and a variable change. 

To this end, it is convenient to rewrite the states (\ref{PsiB,C|A}) in polar coordinates as follows
\ba\label{rotredstate1}
\ket{\Psi}^{\rm phys}_{B,C|A}\!= \!\! \int \mathrm{d}r_B \ r_B^2 \ \mathrm{d}\Omega_B \ \mathrm{d}r_C \ r_C^2  \mathrm{d}\Omega_C  \psi^{\rm phys}_{BC|A}(r_B,r_C,\theta_C) \ket{r_B,\theta_B,\varphi_B}_B \ket{r_C,\theta_C,\varphi_C}_C\,,\nn\\
\ea
where, using the ingredients of Appendix \ref{proofRotational}, we have defined 
\ba\label{rotredstate2}
\psi^{\rm phys}_{BC|A}(r_B,r_C,\theta_C):=\sum_{j=0}^\infty (-1)^j \psi^{\rm phys}_{BC|A}(r_B,r_C;j) \frac{\sqrt{2j+1}}{4 \pi} P_j(\cos \theta_C)\,,
\ea
and $P_j$ is the Legendre polynomial of degree $j$. As a result, we now have a wave function of the polar coordinates and we see that it depends only on the physically meaningful $r_B,r_C,\theta_C$ that also survived the classical gauge fixing (see fig.\ \ref{fig:2}). In contrast, the three other configuration degrees of freedom $\theta_B,\varphi_B,\varphi_C$ are pure gauge, and thus the distribution of any $\ket{\psi}^{\textrm{phys}}_{B,C|A}$ is uniform with respect to each of them; these degrees of freedom are hence now redundant and we can remove them without loss of information. This is completely analogous to the translational case in sec.\ \ref{translationalRed} and we shall now condition on the classical gauge fixing conditions $\vec{\phi}=0$ (in polar coordinates) in (\ref{constraintsChi}--\ref{constraintPhi3})
\ba\label{gaugeproj}
\ket{\psi}_{BC|A}&:=&\braket{\theta_B=0,\varphi_B=0,\varphi_C=0\,|\,\psi}^{\rm phys}_{B,C|A}\nn\\
&=&\int \mathrm{d}r_B \ r_B^2  \ \mathrm{d}r_C \ r_C^2 \ \mathrm{d}\theta_C\,\sin\theta_C \ \psi^{\rm phys}_{BC|A}(r_B,r_C,\theta_C) \ket{r_B}\ket{r_C}\ket{\theta_C}\,,
\ea
(where we have made use of the normalization (\ref{polnorm}) in Appendix \ref{appendixSpherical}). This is an isometry because, as one can easily check, ${}^{\rm phys}_{BC|A}\braket{\phi|\psi}^{\rm phys}_{BC|A}\equiv{}_{BC|A}\braket{\phi|\psi}_{BC|A}$. However, we recall that the gauge fixing conditions $\theta_B=0,\varphi_B=0,\varphi_C=0$ are not globally valid. While this is a problem for distributional states with support on pathological configurations, this is not one for generic Hilbert space states since the troublesome configurations comprise a set of measure zero.

We shall now show that the resulting symmetry reduced quantum state lies in $\ch_{\rm BC|A}$, i.e.\ in the Hilbert space corresponding to directly quantizing $A$'s classical perspective. To this end, we switch variables $(r_B,r_C,\theta_C)\mapsto(\rho_B,\rho_C,u_C)$ in line with sec.\ \ref{redQ3D}, producing 
\ba
\ket{\psi}_{BC|A}&=&\int_{-\infty}^{\infty} \mathrm{d}\rho_B   \ \mathrm{d}\rho_C \  \mathrm{d}u_C\,  e^{3 \rho_B} e^{3 \rho_C} \Big( \f{1}{1+u_C^2} \Big)^\f{3}{2} \nn\\
&&\q\q\q\q\times\psi^{\rm phys}_{BC|A}(\rho_B,\rho_C,u_C) \ket{r_B(\rho_B)}\ket{r_C(\rho_C)}\ket{\theta_C(u_C)}\,.\label{variablechange}
\ea
Due to (\ref{polnorm}) in Appendix \ref{appendixSpherical}, we have the normalization
\ba
\braket{r_B|r_B'}&=&\f{\delta(r_B-r_B')}{r_B^2}\,,\q\q\q\q\q \braket{r_C|r_C'}=\f{\delta(r_C-r_C')}{r_C^2}\,,\nn\\
\braket{\theta_C|\theta'_C}&=&\f{\delta(\theta_C-\theta'_C)}{\sin\theta_C}\,,
\ea
which differs from that in (\ref{Qredstate}). Hence, redefining
\be \label{wfredefine}
\psi_{BC|A}(\rho_B,\rho_C,u_C) := \Bigg(e^{3 \rho_B} e^{3 \rho_C} \Big( \f{1}{1+u_C^2} \Big)^\f{3}{2} \Bigg)^\f{1}{2} \psi^{\rm phys}_{BC|A} (\rho_B,\rho_C,u_C)\,, 
\ee
and
\ba
\ket{\rho_B}\ket{\rho_C}\ket{u_C}:=\Bigg(e^{3 \rho_B} e^{3 \rho_C} \Big( \f{1}{1+u_C^2} \Big)^\f{3}{2} \Bigg)^\f{1}{2}  \ket{r_B(\rho_B)}\ket{r_C(\rho_C)}\ket{\theta_C(u_C)}\,,
\ea
we recover exactly the shape of the reduced states (\ref{Qredstate}) with correct normalization
\ba\label{Qredstate2}
\ket{\psi}_{BC|A}=\int_{-\infty}^{\infty} \mathrm{d}\rho_B   \ \mathrm{d}\rho_C \  \mathrm{d}u_C\, \psi_{BC|A}(\rho_B,\rho_C,u_C) \, \ket{\rho_B}\ket{\rho_C}\ket{u_C}\,.
\ea
This redefinition is not a surprise because the Dirac quantization brought us to a kinematical Hilbert space with measure $\mathrm{d} \mu = \prod_i \mathrm{d}^3\vec{q}_i$, from which the measure in polar coordinates on $\mathcal{H}^{\textrm{phys}}_{B,C|A}$ is directly inherited, whereas the reduced quantization started with a measure $\mathrm{d} \mu =  \mathrm{d} \rho_B \, \mathrm{d} \rho_C \, \mathrm{d} u_C $ on $\mathcal{H}_{BC|A}$. 

We need to check whether also the basic observables generating the commutator algebra behave correctly. It is clear that the translationally and rotationally reduced Dirac observables (\ref{transbcobs}) are unaffected by the projection (\ref{gaugeproj}). The variable shift $(r_B,r_C,\theta_C)\mapsto(\rho_B,\rho_C,u_C)$ could have been carried out prior to the projection. However, this shift will affect the precise representation of these observables. Indeed, the configuration Dirac observables on $\mathcal{H}^{\textrm{phys}}_{B,C|A}$, $\Hat{\rho}_B$, $\Hat{\rho}_C$ and $\mathcal{R}_{B,C} \Hat{u}_C \hat{\mathcal{R}}_{B,C}^{-1}=-\widehat{\cot\theta_C}$, will act simply as multiplication operators both before and after the projection and regardless of the wave function redefinition (\ref{wfredefine}). They just directly become the observables of the reduced theory. On the other hand, as can be checked, the momentum Dirac observables, $\Hat{p}_B^\rho$, $\Hat{p}_C^\rho$ and $\mathcal{R}_{B,C} \,\Hat{p}_C^u \,{\mathcal{R}}_{B,C}^{-1}$ are represented in terms of derivatives on wave function $\psi^{\rm phys}_{BC|A}$ in (\ref{variablechange}) as:\footnote{This ultimately follows from their representation in terms of Cartesian coordinates $\vec{q}_i$ on $\ch^{\rm kin}$ in the standard measure of sec.\ \ref{sec_Pimpose} and the subsequent transformations.}
\begin{subequations}
\be \Hat{p}_B^\rho = - i \partial_{\rho_B} - \f{3}{2} i\,, \ee
\be \Hat{p}_C^\rho = - i \partial_{\rho_C} - \f{3}{2} i \,,\ee
\be \mathcal{R}_{B,C} \,\Hat{p}_C^u \,{\mathcal{R}}_{B,C}^{-1} = - i \partial_{u_C} + \f{3}{2} i \f{u_C}{1+u_C^2}\,. \ee
\end{subequations}
It is straightforward to prove that, due to the redefinition of the wave function normalization, this is equivalent to the following action on $\psi_{BC|A}$ in (\ref{Qredstate2}) 
\ba
\Bigg(e^{3 \rho_B} e^{3 \rho_C} \Big( \f{1}{1+u_C^2} \Big)^\f{3}{2} \Bigg)^\f{1}{2}\, \Hat{p}_B^\rho\, \Bigg(e^{3 \rho_B} e^{3 \rho_C} \Big( \f{1}{1+u_C^2} \Big)^\f{3}{2} \Bigg)^{-\f{1}{2}} &=& - i \partial_{\rho_B}\,, \nn\\
 \Bigg(e^{3 \rho_B} e^{3 \rho_C} \Big( \f{1}{1+u_C^2} \Big)^\f{3}{2} \Bigg)^\f{1}{2}\, \Hat{p}_C^\rho \, \Bigg(e^{3 \rho_B} e^{3 \rho_C} \Big( \f{1}{1+u_C^2} \Big)^\f{3}{2} \Bigg)^{-\f{1}{2}}  &=& - i \partial_{\rho_C} \,,\nn\\
\Bigg(e^{3 \rho_B} e^{3 \rho_C} \Big( \f{1}{1+u_C^2} \Big)^\f{3}{2} \Bigg)^\f{1}{2}\,\calr_{B,C}\, \Hat{p}_C^u \,\hat{\mathcal{R}}_{B,C}^{-1} \,\Bigg(e^{3 \rho_B} e^{3 \rho_C} \Big( \f{1}{1+u_C^2} \Big)^\f{3}{2} \Bigg)^{-\f{1}{2}} &=& - i \partial_{u_C}\,.\nn
\ea
We have thereby proved that our basic Dirac observables (\ref{Dobserv}) ultimately transform correctly under the various reduction maps to the observables $\hat{\pi}_\mu$ of the reduced quantized theory in $A$ perspective. In particular, given that we have carried out a sequence of isometries, the expectation values of these observables will be  equivalent to those of the reduced theory in sec.\ \ref{redQ3D}.

Crucially, notice that the wave function redefinition in (\ref{wfredefine}) takes care of the fact that, since total collisions and collinearity happen as $\rho_B,\rho_C\rightarrow-\infty$ and $u_C\rightarrow\pm\infty$, states in the reduced theory of sec.\ \ref{redQ3D} have a vanishing support there, while the Dirac theory {\it does} admit states with support on such configurations. Indeed, the rescaling factor in (\ref{wfredefine}) vanishes on approach of such pathological configurations and thus extinguishes any support physical states might have there.

This completes our reduction procedure in the quantum theory (see fig.\ \ref{commutativeDiagram3D} for a summary), the result of which we take as $A$'s quantum frame perspective.

\subsubsection{Relation with reduced quantization of $A$'s classical perspective}\label{sssec_relation}

Our complete quantum reduction map, $\Phi'_A:\ch^{\rm TI}\rightarrow\ch_{BC|A}$, given by
\ba
\Phi_A':=\bra{\theta_B=0,\varphi_B=0,\varphi_C=0}\calr_{B,C}\delta(\hat R^a_B+\hat R^a_C)\bra{\vec{\chi}_A=0}\ct_{A,BC}\label{PHI}
\ea
 from the translation invariant Hilbert space $\ch^{\rm TI}$ to what we identify as $A$'s quantum frame perspective thus yields the Hilbert space $\ch_{BC|A}$ and the basic operators of the reduced quantized theory of sec.~\ref{redQ3D}. We recall that this procedure is equivalent to the quantum symmetry reduction, i.e.\ the `quantum coordinate map' $\Phi_A:\ch^{\rm phys}\rightarrow\ch_{BC|A}$, 
\ba\label{compredmapphys}
\Phi_A:=\bra{\theta_B=0,\varphi_B=0,\varphi_C=0}\calr_{B,C}\bra{\vec{\chi}_A=0}\ct_{A,BC}
\ea
 from the perspective-neutral Hilbert space to $\ch_{BC|A}$ (see fig.\ \ref{commutativeDiagram3D}). We also emphasize that $\Phi_A$ will in fact be invertible on its domain in $\ch^{\rm phys}$ despite the appearance of conditionings in its construction. A conditioning such as $\bra{\vec{\chi}_A=0}$ would indeed not be invertible on the kinematical Hilbert space $\ch^{\rm kin}$ as this operation would project  away non-trivial and in general independent information about $A$. Thanks to the redundancy in the description of $\ch^{\rm phys}$ \pah{(in terms of kinematical variables)} though, a conditioning such as $\bra{\vec{\chi}_A=0}$ only removes redundant information from an invariant state. For example, the factor $\ket{\vec{p}_A=0}$ in \eqref{TIABC} carries no more independent information. This operation will thus be invertible for \emph{physical} states. The analogous argument holds for the other conditioning in $\Phi_A$. Hence, $\Phi_A$ will be invertible on its domain and this is crucial for transforming observables from $\ch^{\rm phys}$ to $\ch_{BC|A}$ which requires a conjugation with this map. We emphasize that $\Phi_A$ provides a general method for comparing Dirac with reduced quantization.
  
At first sight it thus seems as though $\Phi_A$ just recovered the direct quantization of $A$'s classical frame perspective from the perspective-neutral quantum theory on $\ch^{\rm phys}$. However, this is not the case because \emph{composite} operators constructed from the elementary ones in Dirac quantization do \emph{not} in general map under $\Phi_A$ to the corresponding composite operators constructed on $\ch_{BC|A}$ in reduced quantization. In particular, as we explain in appendix~\ref{HamiltonianDiscrepancy}, the Hamiltonians of the two quantizations are inequivalent; the total Hamiltonian $\hat H_{\rm tot}$ in \eqref{HqDirac}, which recall is an observable on $\ch^{\rm phys}$, does not reduce under $\Phi_A$ to the Hamiltonian \eqref{qHred} of reduced quantization, i.e.\ $\Phi_A\,\hat H_{\rm tot}\,\Phi_A^{-1}\neq\hat H^{\rm red}_{BC|A}$, where $\Phi_A\,\hat H_{\rm tot}\,\Phi_A^{-1}$ is simply equal to expression (\ref{HqDiracTrans}) with the $\hat{p}$'s replaced by $\hat{\pi}$'s. The difference between the two only involves configuration variables. Both the dynamics and spectra will thus differ. 

This is in line with previous reports in the literature that the Hamiltonians in Dirac and reduced quantization are typically inequivalent, e.g.\ see \cite{Ashtekar:1982wv, Kuchar1986, Romano:1989zb, Schleich:1990gd, Loll:1990rx, McMullan:1988tw, Dolan_1990}. This discrepancy can ultimately be traced back to an ambiguity in the ordering of gauge-invariant operators in Dirac quantization, arising due to the contribution of the volume of gauge orbits to the Hilbert space measure \cite{Kunstatter:1991ds}. As demonstrated in  \cite{Kunstatter:1991ds,Kuchar1986}, for systems with constraints linear in momenta as in our model, one can always modify Dirac quantization through factor ordering in such a way that the Hamiltonians will agree.  While this yields a well-defined procedure, it is somewhat artificial from the point of view of Dirac quantization. We will briefly illustrate this procedure in appendix~\ref{HamiltonianDiscrepancy} and show that it leads to a Hamiltonian on $\ch^{\rm phys}$ that in fact no longer treats all degrees of freedom on an equal footing, but is special to a frame choice. Hence, in order to ensure equivalence between Dirac and reduced quantization, one will have to modify Dirac quantization according to the reduced theory one is interested in and different frame choices will therefore lead to inequivalent modifications; the reduced quantizations of different classical frame perspectives are unitarily inequivalent. This is untenable in light of intending to construct a notion of quantum frame covariance. Instead, we propose to embrace the general inequivalence of Dirac and reduced quantization and accept its consequences.

The conclusion we draw from this is that it is generally \emph{not} equivalent to construct the quantum frame perspective of $A$ from the perspective-neutral quantum theory, or to directly quantize $A$'s classical frame perspective. So which version should one identify as $A$'s quantum frame perspective? Our proposal is to give primacy to the more fundamental perspective-neutral structure in both the classical and quantum theory from which the internal perspectives are derived. In the context of quantum reference frames, this means giving Dirac quantization primacy over the reduced quantization method. \pah{We} therefore identify the quantum symmetry reduced theory obtained from the perspective-neutral structure via $\Phi_A$ as $A$'s \pah{internal} quantum frame perspective \pah{and \emph{not} the direct quantization of $A$'s classical frame perspective}. 

Our reasoning for this is as follows: Dirac quantization treats all degrees of freedom, including whatever we may choose as a reference frame, on an equal footing and in particular translates the classical redundancy in the description of physics into the quantum theory. This redundancy is the prerequisite for being able to describe the same physical situation in many different ways and therefore for establishing a notion of quantum frame covariance. Such a covariance is a necessity in the presence of many physically equally good choices of quantum reference frames within a given problem, and in reduced quantization one generally lacks the linking structure between the quantizations of the different classical frame perspectives which, as mentioned, are unitarily inequivalent. By contrast, in Dirac quantization one obtains a perspective-neutral structure by construction which links the different possible internal frame perspectives in the form of quantum coordinate transformations, as we shall illustrate in the next section, extending \cite{vanrietvelde2018change}. This is the natural quantum analog of classical frame changes through coordinate transformations \pah{and in line with quantum reference frame transformations  being merely changes of description}. In short, quantum covariance is in our view the reason one should give Dirac quantization precedence.

This upgrades the arguments for preferring Dirac over reduced quantization usually found in the literature according to which the former should be considered more general because in the latter the reference system has already been removed prior to quantization such that it can never feature any quantum fluctuations. While it is correct that the reference frame is absent from reduced quantization, this reasoning is somewhat misleading as in Dirac quantization too the reference frame no longer features independent quantum fluctuations after its degrees of freedom have been selected as the redundant ones. This is ultimately the reason why in the quantum symmetry reduced theory the frame degrees of freedom too are absent just like in reduced quantization. However, what our framework underlines is that, while the reference system does not feature \fla{them} in its own perspective, it \emph{can} feature independent quantum fluctuations in the perspective of a \emph{different} quantum reference frame, while maintaining a link between the different perspectives through the perspective-neutral Hilbert space. We will illustrate this next.

\pah{In summary, although experiments will ultimately have to decide which quantization method is the appropriate one,\footnote{\pah{This requires better understanding measurements in the context of quantum frames, e.g.\ by building up on \cite{yang2020switching}.}} our conceptual proposal in the context of quantum reference frames is to retain only the Dirac quantized theory (and the various symmetry reduced theories obtained from it), while discarding the reduced quantizations as unitarily inequivalent theories altogether.\footnote{\pah{
In particular, it is not a problem that the acts of (i) reducing perspective-neutral states via the instantaneous map \eqref{compredmapphys} (cf.\ sec.\ \ref{sssec_rotred}) to ones on $\ch_{BC|A}$ (the Hilbert space of both reduced quantization and symmetry reducing the Dirac quantized theory), and (ii) time evolving with either $\hat H_{\rm tot}$ on $\ch^{\rm phys}$ or $H^{\rm red}_{BC|A}$ on $\ch_{BC|A}$ do not commute, as a result of the two Hamiltonians being unitarily inequivalent. This is just a manifestation of the dynamical inequivalence of Dirac and reduced quantization and we henceforth only work with the former. Indeed, in our proposal, the correct Hamiltonian on $\ch_{BC|A}$ is the quantum symmetry reduced one $\Phi_A\hat H_{\rm tot}\Phi_A^{-1}$, and (i) and (ii) \emph{do} commute when replacing $H^{\rm red}_{BC|A}$ with it. The quantum frame perspectives are thus dynamically consistent with the perspective-neutral descriptions.}}}

\subsection{Transformations between relative states}

In this section, we give the transformation of the quantum state when changing from the perspective of A to the perspective of C, expanding the framework of \cite{vanrietvelde2018change}. It is clear that, starting from the perspective-neutral theory, we could equally well have chosen to go into the perspective of system C, and that this can be achieved by following the same steps explained previously, but swapping all A and C labels. Once in a specific perspective, we can transform all the way back to the physical Hilbert space $\mathcal{H}^{\rm phys}$, exploiting the invertibility of $\Phi_A$ (cf.\ sec.~\ref{sssec_relation}), as shown in the following diagram:

\begin{tikzcd}
& \mathcal{H}^{\textrm{phys}}  \arrow[rd, "\mathcal{T}_{C,AB}"]& \\
\mathcal{H}^{\textrm{phys}}_{A,BC} \arrow[ru, "\mathcal{T}_{A,BC}^{-1}"] & & \mathcal{H}^{\textrm{phys}}_{C,AB} \arrow[d, "_C\bra{\chi=0}"]\\
\mathcal{H}^{\textrm{phys}}_{BC|A} \arrow[u, "\ket{\vec{p}=0}_A\otimes(\cdot)_{BC|A}^{\rm phys}"] \arrow[rr, "\mathcal{S}^T_{A \to C}"]  & & \mathcal{H}^{\textrm{phys}}_{AB|C} \arrow[d, "\mathcal{R}_{B,A}"] \\
\mathcal{H}^{\textrm{phys}}_{B,C|A} \arrow[u, "\mathcal{R}_{B,C}^{-1}"]  & & \mathcal{H}^{\textrm{phys}}_{B,A|C} \arrow[d, "\bra{\theta_B, \varphi_B, \varphi_A = 0}"] \\
\mathcal{H}_{BC|A} \arrow[u, "\ket{j_A =0; m_B=0, m_C =0} \otimes(\cdot)_{BC|A} "] \arrow[rr, "\mathcal{S}_{A \to C}"] & & \mathcal{H}_{BA|C} \\
\end{tikzcd}

\noindent where $(\cdot)_{BC|A}$ means inserting the input state from $\ch_{BC|A}$ into the brackets, upon which the new tensor factor equipping the arrow is appended to the input state. The quantum reference frame transformation from $A$'s to $C$'s perspective thus takes the form of a quantum coordinate transformation, $\cs_{A\to C}:\ch_{BC|A}\rightarrow\ch_{AB|C}$, 
\ba
\cs_{A\to C}:=\Phi_C\circ\Phi_A^{-1}
\ea
which passes through the perspective-neutral Hilbert space $\ch^{\rm phys}$.
Note, however, that in practice we never need to explicitly go back to $\ch^{\rm phys}$, which we also never fully constructed. Instead, we can sidestep it, as we now explain.

It is worth stressing that, in order to transform from $\mathcal{H}_{BC|A}$ to $\mathcal{H}^{\rm phys}_{B,C|A}$, one needs to tensor by $\ket{j_A =0; m_B=0, m_C =0} \otimes (\cdot)_{BC|A} = \frac{1}{2\sqrt{2}\pi}\int d\Omega_B d \phi_C \ket{\theta_B; \phi_B; \phi_C} \otimes(\cdot)_{BC|A}$, which corresponds to averaging over the classical gauge conditions. This operation inverts the conditioning on the classical gauge fixing conditions, and restores the gauge invariance of the model.

Informally, the steps from $\mathcal{H}_{BC|A}$ to $\mathcal{H}_{AB|C}$ can be written as
\begin{equation}\label{QRFtransformation}
	\mathcal{S}_{A \rightarrow C} = \bra{\theta_B=0; \phi_B =0; \phi_C =0}\mathcal{R}_{B,C}\mathcal{S}^T_{A\rightarrow C}\mathcal{R}_{B,A}^{-1}\ket{j_A =0; m_B=0, m_C =0} \otimes(\cdot)_{BC|A}\,.
\end{equation}
Here, $\cs^T_{A\to C}:\ch_{BC|A}^{\rm phys}\rightarrow\ch_{AB|C}^{\rm phys}$ is the three-dimensional analog of the relative state transformation in the translational case of \cite{vanrietvelde2018change, Giacomini:2017zju}, which provides a shortcut, sidestepping $\ch^{\rm phys}$, in the above diagram and mapping directly between the Hilbert spaces that have been reduced with respect to translation invariance only:
\ba\label{transtrafo}
\cs^T_{A\to C}:=\cp_{CA}\,\exp(i\,\vec{\hat q}_C\cdot\vec{\hat p}_B)\,,
\ea
where $\cp_{CA}$ is now a three-dimensional parity swap:
\ba
\cp_{CA}\ket{\vec{p}}_A=\ket{-\vec{p}}_C\,.
\ea

Explicitly evaluated on some arbitrary initial state in $\ch_{BC|A}$
\begin{equation}
	\ket{\psi}_{BC|A} = \int dr_B dr_C d\theta_C \sin (\theta_C) r_B^2 r_C^2 \psi_{BC|A}(r_B, r_C, \theta_C) \ket{r_B}_B \ket{r_C}_C \ket{\theta_C}_C
\end{equation}
this yields $\ket{\psi}_{BA|C} = \mathcal{S}_{A \rightarrow C} \ket{\psi}_{BC|A}$, where
\begin{equation} \label{eq:statefromC}
	\ket{\psi}_{BA|C} = \int dr_A \,dr_B \,d\theta_A \sin (\theta_A) r_B^2 \,r_A^2\, \psi_{BC|A}(d_{AB}, r_A, \gamma_{B|C}) \ket{r_A}_A \ket{r_B}_B \ket{\theta_A}_A,
\end{equation}
and where $d_{AB}$ is the distance between A and B and $\gamma_{B|C}$ is the old angle $\theta_C$, both expressed in terms of the variables relative to C as
\begin{align}
	d_{AB} &= \sqrt{r_A^2 + r_B^2 -2 r_A r_B \cos \theta_A},\\
	\cos\gamma_{B|C} &= \frac{r_A - r_B \cos \theta_A}{\sqrt{r_A^2 + r_B^2 -2 r_A r_B \cos \theta_A}}.
\end{align}
This is a result which has to be expected geometrically. It corresponds to replacing the relative distances between $A$ and $B$ by that between $C$ and $B$ and having now the angle $\theta_A$ as the one between $A$ and $B$ as seen from $C$.

It is easy to show that, similarly to what happens in the one-dimensional case explored in \cite{Giacomini:2017zju} and \cite{vanrietvelde2018change}, entanglement and superposition are QRF-dependent features. Here, we illustrate this operational consequence of the framework on the angular degrees of freedom, whose treatment is a novel aspect of this work. For the purpose of illustrating this effect, it is enough to consider, in the initial QRF of $A$, a (non-normalised) separable state consisting in a superposition of two angles, and sharply localised in the two radial components of $B$ and $C$. For example, assuming $\theta_i\neq0$ and $r_i\neq0$, $i=1,2$, so that we are not dealing with total collisions or collinearity,
\begin{equation}
    \ket{\psi}_{BC|A} = r_1^2 r_2^2 \ket{r_1}_B \ket{r_2}_C \frac{1}{\sqrt{2}}\left( \sin \theta_1 \ket{\theta_1}_C + \sin \theta_2 \ket{\theta_2}_C \right).
\end{equation}
From the perspective of $C$, it is straightforward to show that, via Eq.~\eqref{eq:statefromC}
\begin{equation}
    \ket{\psi}_{AB|C} = r_1 r_2^2  \ket{r_2}_A \frac{1}{\sqrt{2}}\left( q_1 \cos \theta_1 \ket{q_1}_B \ket{\phi_1}_A +  q_2 \cos \theta_2 \ket{q_2}_B \ket{\phi_2}_A \right),
\end{equation}
where, for $i=1,2$, we have
\begin{equation}
    \begin{split}
        & q_i = \sqrt{r_1^2 + r_2^2 -2 r_1 r_2 \cos \theta_i}\\
        & \cos \phi_i = \frac{r_2 - r_1 \cos \theta_i}{\sqrt{r_1^2 + r_2^2 -2 r_1 r_2 \cos \theta_i}}.
    \end{split}
\end{equation}
From the previous expression, it is clear that entanglement is QRF-dependent, thus generalising the result of \cite{Giacomini:2017zju} and \cite{vanrietvelde2018change} to the three-dimensional case.

\subsection{Remarks on the unitarity of the quantum frame transformations}

Let us now come to the important question of whether our quantum frame transformations in \eqref{QRFtransformation} are unitary. We will be somewhat informal in our answer, in line with the previous discussion, as a rigorous statement would require substantially more efforts that go beyond the scope of this manuscript. Our answer has two parts, one pertaining to proper Hilbert space states, the other to distributional states.

For proper Hilbert space elements, the transformation will, informally at least, be unitary in the sense of being both invertible and an isometry. Firstly, the operator $\cs^T_{A\to C}$ in \eqref{transtrafo} is clearly unitary (see also \cite{Giacomini:2017zju,vanrietvelde2018change}). The more subtle part are the transformations $\calr_{B,C}$ and $\calr_{B,A}$ and the conditioning on the classical gauge fixing condition, such as $\bra{\theta_B=0;\phi_B=0;\phi_C=0}$, which is not globally defined on the classical configuration space, owing to the mechanical analog of the Gribov problem. However, as noted before, the pathological configurations -- total collisions and collinearity -- comprise a set of measure zero in the gauge invariant configuration space \cite{littlejohn1995}. It follows from appendix~\ref{proofRotational} that $\calr_{B,C}$ is invertible and isometric on its domain and, given that the pathological configurations are a measure zero set, this should be a dense subspace of $\ch_{BC|A}^{\rm phys}$ such that it is unitary (similarly with $A,C$ interchanged).
The above conditioning, on the other hand, is well-defined on states of the form ${\ket{j_A=0;m_B=0,m_C=0}\otimes\ket{\psi}_{BC|A}}$, which comprise the Hilbert space $\ch_{B,C|A}^{\rm phys}$ (cf.\ the expressions in equations (\ref{rotredstate1},\ref{rotredstate2})), where ${\ket{j_A=0;m_B=0,m_C=0}=\f{1}{2\sqrt{2}\pi}\int d\Omega_B d\phi_C\ket{\theta_B;\phi_B;\phi_C}}$. As observed in secs.~\ref{sssec_rotred} and \ref{sssec_relation}, the conditioning is an isometry and invertible, as it only removes redundant information; indeed, as discussed above, the inverse operation is given by appending the tensor factor ${\ket{j_A=0;m_B=0,m_C=0} \otimes(\cdot)_{BC|A}}$. We therefore conclude that, for Hilbert space elements, the quantum frame transformation $\cs_{A\to C}$ in \eqref{QRFtransformation} is unitary. While we have argued here for three particles, the same will hold for arbitrary $N$, as the pathological configurations always constitute a measure zero set in the gauge-invariant configuration space.

What about distributional states with support on the classically pathological configurations? For those states, the quantum frame transformation $\cs_{A\to C}$ in \eqref{QRFtransformation} is \emph{not} defined, in particular, because the rotational trivializations $\calr_{B,C}$ and $\calr_{B,A}$ are not defined (cf.\ appendix~\ref{proofRotational}). Note, however, that the classical gauge fixing conditions associated with $A$'s perspective will fail if and only if the gauge fixing conditions of $C$'s (and $B$'s) perspective fail; for three particles  collinearity and total collision mean the failure of either internal perspective, see fig.~\ref{fig:2}. All the valid configurations of $A$'s perspective can therefore be mapped into valid configurations in $C$'s perspective and vice versa; in this sense the quantum frame transformations $\cs_{A\to C}$ is invertible for distributional states for three particles too.

The situation for distributional states will, however, be different in general with $N>3$. In that case, it will generally no longer be true that the gauge associated with one frame's perspective fails if and only if the gauge associated with the second frame's perspective fails. The two frames may use non-overlapping subsets of particles to set up there local orientation relationally. When the particles used by the first frame are collinear or collide, the particles employed by the second frame may be in a generic configuration. In such a case, the transformation between the first and second frame will only be defined on those distributional states which correspond to global particle configurations that are not pathological for both frame perspectives. This means that, in those cases, the quantum reference frame transformation will only be invertible on a subset of distributional states that are valid in either perspective. The absence of globally valid classical frame perspectives thus only plays a significant role for distributional states.

\section{Conclusions}\label{sec_conc}

In this article, we have expanded on the systematic approach to quantum reference frame transformations started in \cite{vanrietvelde2018change}, by applying it to the relational $N$-body problem in three-dimensional space with translational and rotational invariance. As discussed in detail, this model is particularly interesting because it is subject to a mechanical analog of the Gribov problem in gauge field theories and thereby does not permit globally valid reference frame perspectives.  Since this is a property of generic systems with gauge symmetry, this example serves as an illustration how our approach can also handle challenges that arise in generic situations. 

In particular, the $N$-body problem features a number of technical challenges -- all of which are a consequence of the compactness of the rotational gauge orbits --, such as a constraint surface that is foliated by gauge orbits of different dimension and an absence of globally valid pairs of canonically conjugate Dirac observables. As a result, our descriptions relative to specific frame perspectives are not valid on all states; in particular, the invariant observables which we choose to describe the physics relative to specific choices of frames become unbounded on collisional or collinear configurations (a set of measure zero). But given the absence of global perspectives, this is inevitable and will appear, in one form or another, in most interesting systems. These challenges notwithstanding, we demonstrate how one can use the quantum reduction method, consisting of a constraint trivialization and a conditioning on the classical gauge fixing conditions, to systematically construct transformations between different quantum reference frame perspectives in the three-body problem. In particular, for Hilbert space states, these transformations are unitary, despite the absence of classically global internal perspective. This is because the pathological configurations comprise a set of measure zero in the gauge-invariant configuration space. For distributional states, on the other hand, this transformation will not be invertible in general situations with $N>3$. We have also shown that the transformations lead to a quantum frame dependence of entanglement.
This work thereby generalizes the approach to constructing quantum reference frame transformations of both \cite{vanrietvelde2018change, Giacomini:2017zju}, which only considered one-dimensional particle models, to a three-dimensional setting including rotations.

This article also substantiates our conceptual argumentation in \cite{vanrietvelde2018change}. Specifically, it shows that also in this more general setting, and despite the absence of global relational perspectives, our interpretation of the classical constraint surface and the Dirac quantized physical Hilbert space as perspective-neutral structures remains valid. Similarly, our interpretation of (specifically) gauge-fixed reduced phase spaces and quantum symmetry reduced Hilbert spaces as the physics seen from a particular frame perspective is supported by the $N$-body problem. In particular, our work also sheds novel light on the relation between Dirac and reduced quantization from the perspective of quantum reference frames. As we have seen, quantizing a classical frame perspective is generally inequivalent to constructing the quantum frame perspective by symmetry reducing the perspective-neutral structure obtained through Dirac quantization. The general inequivalence of Dirac and reduced quantization \cite{Ashtekar:1982wv,guillemin1982geometric, tian1998analytic,hochs2008guillemin,gotay1986constraints,kucha1986covariant,Ashtekar:1991hf,Schleich:1990gd,Kunstatter:1991ds,Hajicek:1990eu,Romano:1989zb,Dittrich:2016hvj,Dittrich:2015vfa,Loll:1990rx} can be phrased more physically  as `frame associated symmetry reduction and quantization don't commute'. Based on quantum frame covariance, we have argued that precedence should be given to Dirac quantization and hence the perspective-neutral structure from which internal frame perspectives are derived.

As an outlook, our method is further developed by applying it to quantum cosmological models in \cite{Hoehn:2018aqt,Hoehn:2018whn} to demonstrate how one can also witch temporal reference systems, i.e.\ relational clocks, in a quantum theory. This, in fact,  also inspires a new perspective on the `wave function of the universe'. As will be exhibited in \cite{pqps}, our frame transformation method is actually equivalent to that originally developed in \cite{Bojowald:2010xp,Bojowald:2010qw,Hohn:2011us}, provided one restricts to a semiclassical regime within which the latter was constructed. Finally, an extension of the original quantum reference frame approach of \cite{Giacomini:2017zju} to relativistic systems with {\it internal} spin degrees of freedom is presented in \cite{FEC} together with an analysis of the operational consequences of such extended quantum frame transformations.

\section*{Acknowledgments}
It is a pleasure to thank \v{C}aslav Brukner and Esteban Castro-Ruiz for many discussions on quantum reference frames. PH is grateful to Tim Koslowski for originally proposing the model employed in this article and many discussions on the $N$-body problem. PH also thanks Flavio Mercati for correspondence on Dirac observables in the three-body problem. The project leading to this publication has initially received funding from the European Union's Horizon 2020 research
and innovation programme under the Marie Sklodowska-Curie grant agreement No 657661 (awarded to PH). PH also acknowledges support through a Vienna Center for Quantum Science and Technology Fellowship. F.G. acknowledges support from the John Templeton Foundation, Project 60609, ``Quantum Causal Structures'', from the research platform ``Testing Quantum and Gravity Interface with Single Photons'' (TURIS), the Austrian Science Fund (FWF) through the project I-2526-N27, the doctoral program ``Complex Quantum Systems'' (CoQuS) under Project W1210-N25, and from Perimeter Institute for Theoretical Physics. Research at Perime- ter Institute is supported in part by the Government of Canada through the Department of Innovation, Science and Economic Development and by the Province of Ontario through the Ministry of Colleges and Universities. This publication was made possible through the support of a grant from the John Templeton Foundation. The opinions expressed in this publication are those of the authors and do not necessarily reflect the views of the John Templeton Foundation.

\appendix
\section{The rotation and translation invariant $N$-body problem }

\subsection{The rotation and translation invariant Lagrangian}\label{app_lagr}
We shall take the $N$ particles to be of unit mass and the configuration manifold as $\cq=\mathbb{R}^{3N}$. The tangent bundle is therefore $T\cq\simeq\mathbb{R}^{6N}$. We define the positions relative to the center of mass (which are translation invariant),
\ba
q^a_{0i}\, = \,q_a^i - q_{CM}^i \,=\, q_a^i - \frac{1}{N} \left( \sum_{j=1}^N q_a^i \right) \, ,
\ea
as well as their derivatives $\dot{q}^a_{0i}$. We also define the total angular momentum,
\ba
\vec{L} \, = \, \sum_{i=1}^N \vec{q}_{0i} \times \dot{\vec{q}}_{0i} \, ,
\ea
and the moment of inertia $3 \times 3$ matrix,
\ba
M^{ab} \, = \, \sum_{i=1}^N \, |\vec{q}_{0i}|^2 \delta^{ab} \, - \, q_{0i}^a q_{0i}^b.
\ea

We then define the Lagrangian as\footnote{Of course the definition of this Lagrangian crucially depends on the invertibility of the inertia matrix $M$. Interestingly, the subset of the tangent bundle on which $M$ is not invertible is not exactly equivalent to the subset of the cotangent bundle on which the canonical constraints become dependent. For example, when all the positions of the particles are aligned but not their momenta, the canonical constraints are still independent but $M$ is not invertible.}
\ba
\mathcal{L}&=& \f{1}{2}\,\sum_{i=1}^N{\vec{\dot{q}}_{0i}}^2-\,\,\,\underbrace{\f{1}{2}\,\vec{L} \cdot M^{-1} \cdot \vec{L}}_{E_{\rm kin}^{\rm rot}}\,\,\, - \,\,\,\,\,V\left(\{|\vec{q}_i-\vec{q}_j|\}_{i,j=1}^N\right)\nn\\
&=&\f{1}{2}\,\sum_{i=1}^N\,{\vec{\dot{q}}_i}^2\,\,\,-\,\,\,\underbrace{\f{1}{2N}\,\left(\sum_{i=1}^N\,\vec{\dot{q}}_i\right)^2}_{E_{\rm kin}^{\rm cm}}\,\,\,-\,\,\,\underbrace{\f{1}{2}\,\vec{L} \cdot M^{-1} \cdot \vec{L}}_{E_{\rm kin}^{\rm rot}}\,\,\, - \,\,\,\,\,V\left(\{|\vec q_i-\vec q_j|\}_{i,j=1}^N\right)\,,\label{Lag3d}
\ea
where $V$ is invariant under translations and rotations, and thus depends only on the absolute distances between the particles. We have subtracted the kinetic energy $E_{\rm kin}^{\rm cm}$ of the center of mass, so that only the motion relative to the latter contributes to the energy; and the rotational kinetic energy $E_{\rm kin}^{\rm rot}$, so that the global state of rotation of the system does not contribute to the energy. In consequence, this Lagrangian is singular and features gauge symmetries. It is easy to see that $\mathcal{L}$ is invariant under global translations since $\vec{q}_{0i}$ is translation invariant,
\ba
(q^a_i,\dot{q}^a_i)\,\,\,\mapsto\,\,\,(q^a_i+f^a(t),\dot{q}^a_i+\dot{f}^a(t)) ,\label{gauge1d}
\ea
where the $f^a(t)$ are arbitrary functions of time that do not depend on particle $i$. In addition, $\mathcal{L}$ is invariant under rotations, that is 
\ba
(\vec{q}_i,\dot{\vec{q}}_i)\,\,\,\mapsto\,\,\,(R(t) \cdot \vec{q}_i, \, R(t) \cdot \dot{\vec{q}}_i+\dot{R}(t) \cdot  \vec{q}_i) ,
\ea
where $R(t)$ is a rotation matrix dependent on time. Indeed, if we look at infinitesimal rotations,
\ba
(\vec{q}_i,\dot{\vec{q}}_i)\,\,\,\mapsto\,\,\,(\vec{q}_i + \epsilon \, \vec{\omega}(t) \times \vec{q}_i, \, \dot{\vec{q}}_i + \epsilon\, \vec{\omega}(t) \times \dot{\vec{q}}_i+ \epsilon\, \dot{\vec{\omega}}(t) \times  \vec{q}_i) ,
\ea
then we have the transformations
\begin{subequations}
\ba \f{1}{2}\,\sum_{i=1}^N\,\dot{q}_i^2 - E_{\rm kin}^{\rm cm} \, \to \, \f{1}{2}\,\sum_{i=1}^N\,\dot{q}_i^2 - E_{\rm kin}^{\rm cm} + \epsilon \, \dot{\vec{\omega}} \cdot \vec{L} \, ,\ea
\ba \vec{L} \, \to \, \vec{L} + \epsilon\, M \cdot \dot{\vec{\omega}}  + \epsilon\, \vec{\omega} \times \vec{L} \, ,\ea
\ba \left( M^{-1} \right)^{ab} \to \left( M^{-1} \right)^{ab} - \epsilon\, \left( M^{-1} \right)^{ac} \left( \sum_{i=1}^N  (\vec{\omega} \times \vec{q}_{0i})^c q_{0i}^d + q_{0i}^c (\vec{\omega} \times \vec{q}_{0i})^d  \right) \left( M^{-1} \right)^{db} \, ,  \ea
\end{subequations}
which, when combined, give  $\mathcal{L} \to\mathcal{L}$.

This symmetry becomes even more explicit in the canonical formulation. Indeed, the Legendre transformation to $(\vec{q}_i,\vec{p}_i) \in T^*\cq\simeq\mathbb{R}^{6N}$ is given by
\be \label{Legendre}
p_i^a = \frac{\partial \mathcal{L}}{\partial \dot{q}_i^a} = \dot{q}_{0i}^a + \left(\vec{q}_{0i} \times\left( M^{-1} \cdot \vec{L} \right) \right)^a \, ,
\ee
and those momenta can be checked to verify the constraints (\ref{constraints}). The Hamiltonian then obtained is (\ref{H3D}); indeed, the easiest way to see this is by applying the Legendre transformation (\ref{Legendre}) to (\ref{H3D}) to reconstruct the Lagrangian 
\be \begin{split}
\tilde{\mathcal{L}} &= \sum_{i} \vec{p}_i \cdot \dot{\vec{q}}_i - H \\
&= \sum_{i} \left( \dot{\vec{q}}_{0i} +  \vec{q}_{0i} \times ( M^{-1} \cdot \vec{L} ) \right) \cdot \dot{\vec{q}}_i - \frac{1}{2} \sum_{i} \left( \dot{\vec{q}}_{0i} +  \vec{q}_{0i} \times ( M^{-1} \cdot \vec{L} ) \right)^2 \\
&= \f{1}{2} \sum_i  \dot{q}_i^2 - E_{\rm kin}^{\rm cm} - \f{1}{2} \left( \vec{q}_{0i} \times ( M^{-1} \cdot \vec{L} ) \right)^2 \, .
\end{split}
\ee
Using the formula for the squared norm of a cross product, the last term can be computed to be exactly $E_{\rm kin}^{\rm rot}$, thus proving that $\tilde{\mathcal{L}} = \mathcal{L}$. Note that this reconstruction of the Lagrangian does not mean inverting the Legendre transformation \eqref{Legendre} which defines a map from $T\cq$ to the constraint surface $\cc\subset T^*\cq$ and which thus is not invertible \cite{Henneaux:1992ig}.

\subsection{Dependence of constraints on total collisions and collinearity}\label{app_colcol}

Here, we show that the six constraints (\ref{constraints}) become linearly dependent on total collisions and when all particles are collinear. Moreover, it will follow that the gauge orbits containing these pathological configurations are only three- and five-dimensional, respectively.

The simplest (and most extreme) situation is the $N$ particle collision at the origin $\vec{q}_i=\vec{p}_i=0$, $i=1,\ldots,N$, where the (phase space) gradients of the rotation generators ${R}^a$ vanish and only the translation generators ${P}^a$ define independent flows. The origin does not rotate and this means the gauge orbit in which it lies is only three-dimensional. Indeed, neither the rotation (\ref{constraintR}) nor translation generators (\ref{constraintP}) can change $\vec{p}_i=0$, $\forall\,i$ (see (\ref{gaugeTransfoP}, \ref{gaugeTransfoR})), and only the translation generators can change the location of the total collision, however, not the fact that $\vec{q}_i=\vec{q}$, $\forall\,i$. For $\vec{p}_i=0$ and $\vec{q}_i=\vec{q}$, $\forall\,i$, one has $\mathrm{d}R^a=\epsilon^{abc}q^b\,\mathrm{d}P^c$ and so the gradients of the rotation generators will always be linearly dependent on the gradients of the translation generators. Such a gauge orbit is then three-dimensional.

Similarly, if the $N$ particles are collinear and the momenta are also aligned with the axis of collinearity, only two of the three gradients of ${R}^a$ are linearly independent because the rotation around the axis of collinearity acts trivially on the particles. More precisely, collinearity means that $\vec{q}_i-\vec{q}_j=q_{ij}\,\vec{v}$, $\vec{p}_i=p_i\,\vec{v}$, $i,j=1,\ldots,N$, for some $\vec{v}\in\mathbb{R}^{3}$. On any such configuration
\ba
v^a\,\mathrm{d}R^a&=&v^a\epsilon^{abc}v^b\,\f{1}{N-1}\sum_{i,j}\,q_{ij}\,\mathrm{d}p_i^c-\f{1}{N-1}\,\sum_j\,v^a\epsilon^{abc}q^b_j\,\mathrm{d}P^c+v^a\epsilon^{abc}v^c\,\sum_i\,p_i\,\mathrm{d}q_i^b\nn\\
&=&-\f{1}{N-1}\,\sum_j\,v^a\epsilon^{abc}q^b_j\,\mathrm{d}P^c\,,\nn
\ea
which defines a linear dependence among the six constraints (\ref{constraints}). Note that any such configuration can be mapped to $\vec{q}_i=q_i\,\vec{v}$ and $\vec{p}_i=p_i\,\vec{v}$, where even $v^a\,\mathrm{d}R^a=0$, through a global translation that puts the configuration origin on the axis of collinearity. Since no other special direction exists for such configurations, this is the only linear dependence and so two of the three rotation generator gradients are linearly independent. Neither rotations nor translations can change the collinearity of the particles in both configurations and momenta and so on every point of such a gauge orbit there will always be just five independent constraints.  This means that the gauge orbits in which such collinear scenarios reside are only five-dimensional.

Since the translation generators are always independent and it is not possible that only one rotation generator is independent (it would have to leave two linearly independent vectors invariant), there are no other types of lower dimensional orbits. Hence, the constraint surface is foliated by six-, five- and three-dimensional gauge orbits.

\subsection{Independent Dirac observables on generic $N$ particle configurations}\label{app_generic}

There are various ways to convince oneself geometrically that only $3N-6$ of the $\binom{N}{2}$ relative distances $|\vec{q}_i-\vec{q}_j|$ are independent. Our entire discussion below is for generic configurations, i.e.\ away from pathologies such as total collision or collinearity of all $N$ particles.
 For example, for $N=3$, the particles form a triangle and all three relative distances are independent. Owing to the gauge symmetry, the orientation and localization of this triangle in Newtonian background space is unphysical. Now add a fourth particle; evidently, this particle requires three additional relative distances to be fully localized in relation to the other three particles in the triangle. The physically relevant configurational information of the four particles is thus now contained in the edge lengths of a tetrahedron and again all $\binom{4}{2}=6$ relative distances are independent.  Redundancy now comes in if one adds a fifth particle: in order to fully localize it with respect to the other four, only the relative distances to three of the four particles, forming a triangle, are required, see fig.\ \ref{fig:1}. The fifth particle forms another tetrahedron with the three particles relative to which it has been localized. The relational information in the configurations is thus stored in two tetrahedra that are glued together along a common triangle. There are thus nine independent edge lengths and the tenth relative distance (the one between the fifth and fourth particle) is automatically determined from the others, no matter how the glued tetrahedra are embedded in the ambient three-dimensional Newtonian background space. This construction generalizes to an arbitrary number $N$ of particles: for each further particle, only three independent relative distances are needed to fully localize it relative to the previous ones and all other relative distances between it and the rest are redundant. The independent relational information in the configurations can thus be encoded in a 3D triangulation with $N$ vertices in Euclidean space and this triangulation has $3N-6$ edges. There are thus $3N-6$ independent absolute relative distances for $N\geq3$ particles.

Likewise, it is not hard to see that indeed only $3N-6$ of the $\f{N(N+1)}{2}$ momentum Dirac observables $\vec{p}_i\cdot\vec{p}_j$ are independent. Thanks to (\ref{constraintP}), there are (at most) two linearly independent momenta for $N=3$ and their three inner products are independent. For $N=4$, one has (at most) three linearly independent momenta, thus forming a basis, and their six inner products again are independent. Now for any additional particle $k$: the three inner products of its momentum vector $\vec{p}_k$ with the three basis vectors are independent and clearly determine the inner product of $\vec{p}_k$ with any other momentum vector. Hence, there are $3N-6$ independent Dirac observables among the $\vec{p}_i\cdot\vec{p}_j$.

\subsection{Absence of global canonically conjugate Dirac observables}\label{app_absence}

Recall from sec.\ \ref{sec_mach3d} that any Dirac observable is a function of the basic gauge invariant combinations
\ba
f_{ijkl}:=(\vec{q}_i-\vec{q}_j)\cdot(\vec{q}_k-\vec{q}_l),\q\q\q\q  g_{ijk}:=(\vec{q}_i-\vec{q}_j)\cdot\vec{p}_k\,,\q\q\q\q h_{ij}:=\vec{p}_i\cdot\vec{p}_j\,.\label{appdirac3d}
\ea
It follows from Appendix \ref{app_generic} that, on generic configurations, there will be $3N-6$ independent configuration and $3N-6$ momentum Dirac observables among this set. There cannot be more independent ones because the phase space is $6N$-dimensional and, on generic points of $\cc$, all six first class constraints (\ref{constraints}) are independent, i.e.\ their gradients are linearly independent and tangential to $\cc$.

To analyze what happens for $N$ particle collisions and collinearity, we need to be more precise. (In)dependence of phase space functions is defined in terms of linear (in)dependence of their phase space gradients. Clearly, at the origin of phase space, $\vec{q}_i=\vec{p}_i=0$, $\forall\,i$, which is contained in $\cc$ and lies on a three-dimensional gauge orbit (see Appendix \ref{app_colcol}), all gradients of the basic Dirac observables above vanish, i.e.\ $\mathrm{d}f_{ijkl}=\mathrm{d}g_{ijk}=\mathrm{d}h_{ij}=0$. In this sense, there are no independent Dirac observables among the set (\ref{appdirac3d}) at the origin; the phase space flows they generate (and which are always tangential to $\cc$) vanish at this point on $\cc$. This will already be enough to argue that there can be no global canonically conjugate Dirac observables on $\cc$.

Nevertheless, it is also interesting to firstly consider what happens when all $N$ particles are collinear and also their momenta are aligned along this axis, i.e.\ $\vec{q}_i-\vec{q}_j=q_{ij}\,\vec{v}$, $\vec{p}_i=p_i\,\vec{v}$, $\forall\,i,j$, and some $\vec{v}\in\mathbb{R}^3$. We noted in Appendix \ref{app_colcol} that such configurations reside in five-dimensional gauge orbits. The differentials of the basic Dirac observables (\ref{appdirac3d}) read on such configurations
\ba
\mathrm{d}f_{ijkl}&=&q_{ij}\,\vec{v}\cdot(\mathrm{d}\vec{q}_k-\mathrm{d}\vec{q}_l)+q_{kl}\,\vec{v}\cdot(\mathrm{d}\vec{q}_i-\mathrm{d}\vec{q}_j)\,,\nn\\
\mathrm{d}g_{ijk}&=&p_i\,\vec{v}\cdot(\mathrm{d}\vec{q}_i-\mathrm{d}\vec{q}_j)+q_{ij}\,\vec{v}\cdot\mathrm{d}\vec{p}_k\,,\\
\mathrm{d}h_{ij}&=&p_i\,\vec{v}\cdot\mathrm{d}\vec{p}_j+p_j\,\vec{v}\cdot\mathrm{d}\vec{p}_i\,.\nn
\ea
Hence, the only differentials appearing in these equations are $\vec{v}\cdot(\mathrm{d}\vec{q}_i-\mathrm{d}\vec{q}_j)$ and $\vec{v}\cdot\mathrm{d}\vec{p}_i$. Now there are only $N-1$ linearly independent differences $\vec{v}\cdot(\mathrm{d}\vec{q}_i-\mathrm{d}\vec{q}_j)$ as $\vec{v}\cdot\mathrm{d}\vec{q}_i-\mathrm{d}\vec{q}_j=\vec{v}\cdot(\mathrm{d}\vec{q}_k-\mathrm{d}\vec{q}_j)-\vec{v}\cdot(\mathrm{d}\vec{q}_k-\mathrm{d}\vec{q}_i)$. Similarly, we have $\vec{v}\cdot\mathrm{d}\vec{P}=\sum_i\,\vec{v}\cdot\mathrm{d}\vec{p}_i$. Thus, altogether, on such collinear configurations there will only be $2(N-1)+1$ linearly independent gradients of the observables (\ref{appdirac3d}), of which $2(N-1)$ are non-trivial observables (instead of $6N-12$)) and the additional one corresponds to a linear combination of the momentum constraints $\vec{v}\cdot\mathrm{d}\vec{P}$. That is, $4N-10$ otherwise independent flows generated by the Dirac observables (\ref{appdirac3d}) vanish for such collinear scenarios.

These observations imply that there can be no global canonically conjugate Dirac observable pairs, for if there were, their gradients would have to be non-vanishing on all of $\cc$. In fact, suppose there were a full set of global canonically conjugate Dirac observables $Q^\alpha,\Pi^\beta$, $\alpha,\beta=1,\ldots,3N-6$. They would define a matrix of their gradients with rank equal to $6N-12$ everywhere on $\cc$, thanks to 
\ba
\{Q^\alpha,\Pi^\beta\}=\Omega^{\mu\nu}\,\p_\mu\,Q^\alpha\,\p_\nu\,\Pi^\beta\overset{!}{\approx} \delta^{\alpha\beta},\label{appfullrank}
\ea
where $\Omega$ is an antisymmetric, contravariant tensor whose components in our coordinates read
\ba
\Omega^{\mu\nu}=\{z^\mu,z^\nu\}=\left(\begin{array}{cc}0 & \mathds{1} \\-\mathds{1} & 0\end{array}\right)\,,\nn
\ea
and the $z^\mu$, $\mu=1,\ldots,6N$, label the canonical phase space coordinates $q_1^x,q_1^y,q_1^z,q_2^x,\ldots, \allowbreak p_1^x,p_1^y,p_1^z,p_2^x,\ldots$, while $\p_\mu:=\f{\p}{\p z^\mu}$. Indeed, given that $\Omega^{\mu\nu}$ is invertible, validity of (\ref{appfullrank}) on all of $\cc$ would imply that $\p_\mu Q^\alpha,\p_\nu\Pi^\beta$ constitute $6N-12$ linearly independent gradients everywhere on $\cc$. But this is in conflict with our observation that all Dirac observables are functions of $f_{ijkl}, g_{ijk}, h_{ij}$ in (\ref{appdirac3d}) and that the latter have identically vanishing gradients at the origin of phase space, which lies on $\cc$, or only $2(N-1)$ linearly independent gradients on totally collinear configurations. More precisely, summarizing $f_{ijkl}, g_{ijk}, h_{ij}$ in one label set $O^{\Gamma}$, where $\Gamma$ runs over as many values as there are functions in (\ref{appdirac3d}), we can write $Q^\alpha(O^\Gamma),\Pi^\beta(O^{\Gamma'})$. Note that these will not be unique functional dependences because the basic Dirac observables (\ref{appdirac3d}) have some obvious dependences among them. However, for any consistent such choice of functional dependence, one finds, using the chain rule,
\ba
\{Q^\alpha,\Pi^\beta\}=\Omega^{\mu\nu}\,\p_\mu\,O^\Gamma\,\p_\nu\,O^{\Gamma'}\,\p_\Gamma \,Q^\alpha\,\p_{\Gamma'}\,\Pi^\beta\overset{!}{\approx} \delta^{\alpha\beta}\,,
\ea
where $\p_\Gamma:=\p/\p O^\Gamma$. Given that $\Omega$ is invertible and $\p_\mu\,O^\Gamma$ vanishes at the origin and defines a matrix of rank equal to $2(N-1)$ for total collinearity, the right hand side is impossible to achieve with non-singular $\p_\Gamma\,Q^\alpha,\p_\Gamma\,\Pi^\beta$. But if the latter become singular, then also the gradients $\p_\mu\,Q^\alpha=\p_\mu\,O^\Gamma\,\p_\Gamma \,Q^\alpha$, etc., become ill-defined and so $Q^\alpha,\Pi^\beta$, despite possibly being non-singular as functions (such as, e.g., the relative distances $|\vec{q}_i-\vec{q}_j|$), cannot be conjugate on such pathological configurations.

In summary, there are no global canonically conjugate Dirac observable pairs on $\cc$ and the same, in fact, immediately implies to Dirac observable pairs with affine conjugation relations, where, instead of $\{Q^\alpha,\Pi^\beta\}\approx \delta^{\alpha\beta}$, one would have $\{Q^\alpha,\Pi^\beta\}\approx \delta^{\alpha\beta}\,Q^{\alpha}$ (no sum over $\alpha$). The global absence of either makes reduced quantization a priori challenging as neither canonical nor affine quantization \cite{isham2} can be applied without imposing additional boundary conditions.

\subsection{Proof of the three-body gauge-fixing in the 3D case}\label{proofGaugeFix}

Let us formally prove that the gauge is indeed totally fixed. One way is to consider the Poisson-bracket Matrix $C$ of all our constraints and show that it is invertible. Concatenating all our constraints in a 12-list $(\Lambda_\alpha) = (P^1,P^2,P^3, R^1,R^2,R^3,\chi^x,\chi^y,\chi^z,\phi_1,\phi_2,\phi_3) $, $C$ is a $12\times12$ antisymmetric matrix defined by $C_{\alpha\beta}= \{\Lambda_\alpha, \Lambda_\beta\}$. Computation of $C$ gives (with all 12 constraints imposed):

    \begin{equation}\label{}
C =
\left( \rule{0cm}{4cm}
\begin{array}{c|c}
   \textrm{\huge 0} & 
      \begin{array}{c|c}
        \begin{matrix}
          -1 & 0 & 0 \\
          0 & -1 & 0 \\
          0 & 0 & -1
         \end{matrix} & \textrm{\huge 0}\\ 
      \hline
      \textrm{\huge 0} & 
        \begin{matrix}
          q_B^z & 0 & 0 \\
          0 & -q_B^z & q_C^z \\
          0 & 0 & -q_C^x
         \end{matrix}
     \end{array} \\ 
  \hline
      \begin{array}{c|c}
        \begin{matrix}
          1 & 0 & 0 \\
          0 & 1 & 0 \\
          0 & 0 & 1
         \end{matrix} & \textrm{\huge 0} \\ 
      \hline
      \textrm{\huge 0} & 
        \begin{matrix}
          -q_B^z & 0 & 0 \\
          0 & q_B^z & 0 \\
          0 & -q_C^z & q_C^x
         \end{matrix}
     \end{array}  & \textrm{\huge 0}
 \end{array}
 \right)\,.
\end{equation}

$C$ is then invertible, with its inverse being:

    \begin{equation}\label{inverseConstraintsMatrix}
(C^{-1}) =
\left( \rule{0cm}{4cm}
\begin{array}{c|c}
   \textrm{\huge 0} & 
      \begin{array}{c|c}
        \begin{matrix}
          1 & 0 & 0 \\
          0 & 1 & 0 \\
          0 & 0 & 1
         \end{matrix} & \textrm{\huge 0} \\ 
      \hline
      \textrm{\huge 0} & 
        \begin{matrix}
          -\frac{1}{q_B^z} & 0 & 0 \\
          0 & \frac{1}{q_B^z} & 0 \\
          0 & -\frac{q_C^z }{q_B^z q_C^x} & \frac{1}{q_C^x}
         \end{matrix}
     \end{array} \\ 
  \hline
      \begin{array}{c|c}
        \begin{matrix}
          -1 & 0 & 0 \\
          0 & -1 & 0 \\
          0 & 0 & -1
         \end{matrix} & \textrm{\huge 0} \\ 
      \hline
      \textrm{\huge 0} & 
        \begin{matrix}
          \frac{1}{q_B^z} & 0 & 0 \\
          0 & -\frac{1}{q_B^z} & \frac{q_C^z }{q_B^z q_C^x} \\
          0 & 0 & -\frac{1}{q_C^x}
         \end{matrix}
     \end{array}  & \textrm{\huge 0}
 \end{array}
 \right)\,.
\end{equation}
Note that this matrix is indeed only invertible when also (\ref{constraintDiscrete12}) are imposed.

\subsection{Solving constraints for the redundant momenta} \label{appSolv}

Using the constraints (\ref{constraints}), we can express the momentum degrees of freedom that we are getting rid of in terms of the six surviving phase space degrees of freedom:
\ba \label{redmom}
p_B^y &=& 0\,,\nn\\
p_B^x &=& - \f{1}{q_B^z} R_C^y = - \f{1}{q_B^z} (q_C^z p_C^x - q_C^x p_C^z)\,,\nn\\
p_C^y &=& 0\,,\\
p_A^x& =& - p_B^x - p_C^x = - p_C^x + \f{1}{q_B^z} (q_C^z p_C^x - q_C^x p_C^z)\,,\nn\\
p_A^y&=& - p_B^y - p_C^y = 0\,\nn\\
p_A^z &=& - p_B^z - p_C^z \,.\nn
\ea
Note that this constitutes a (non-global) constraint abelianization \cite{Henneaux:1992ig}.

\subsection{Switching internal perspectives classically}\label{app_gaugetr3d}

Using (\ref{redmom}), the canonical embedding map of the reduced phase space in $A$ perspective into $\cc$ reads 
\ba
\iota_{BC|A}:\cp_{BC|A}&\hookrightarrow&\cc\nn\\
(q_B^z>0,p_B^z,q_C^x>0,p_C^x,q_C^z,p_C^z)&\mapsto& \left(q_B^z>0,p_B^z,q_C^x>0,p_C^x,q_C^z,p_C^z,\vec{q}_A=0,\vec{p}_A=-\vec{p}_B-\vec{p}_C, \right.\nn\\
&&\left.q_B^x=q_B^y=q_C^y=p_B^y=p_C^y=0,p_B^x  = - \f{1}{q_B^z} (q_C^z p_C^x - q_C^x p_C^z) \right)\nn
\ea
and its image is precisely $\cc\cap\cg_{BC|A}$. Conversely, the projection
\ba
\pi_{BC|A}:\cc\cap\cg_{BC|A}&\rightarrow&\cp_{BC|A}\nn
\ea
does precisely the opposite, dropping all redundant information, so that $\pi_{BC|A}\circ\iota_{BC|A}=\text{Id}_{\cp_{BC|A}}$. Clearly, the same structures can be constructed for $C$ perspective. 

Let us construct the gauge transformation $\alpha_{A\to C}$, taking us from $\cc\cap\cg_{BC|A}$ to $\cc\cap\cg_{AB|C}$, where $\cg_{AB|C}$ is defined by (\ref{constraintsChi}--\ref{constraintDiscrete12}), in four steps.
The flow on $\cc$ generated by some constraint $C$ will be denoted by $\alpha_C^s$, where $s$ is the flow parameter. The gauge transformation of some phase space function $F$ simply drags its argument along the flow $\alpha^s_C\cdot F(X)=F(\alpha_C^s(X))$, where $X\in\cc$, Written explicitly,
\begin{equation}\label{Cflow}
\alpha^s_C\cdot F(X)= \sum_{k=0}^\infty \frac{s^k}{k!} \{F,C\}_k(X)\,,
\end{equation}
where  $\{F,C\}_k = \{ \ldots \{\{F,C \}, C \}, \ldots, C \}$ is the $k$-nested Poisson bracket of $F$ with $C$. 

\begin{enumerate}
\item First, we translate $B$ along $z$ to the origin. 
Using (\ref{gaugeTransfoP}), the corresponding gauge transformation is easy to evaluate on the canonical variables, starting at some $X_0\in\cc$
\begin{equation}
\alpha^{s_1}_{P^z}\cdot q^z_i (X_0)= q^z_i(X_0) + s_1\,,
\end{equation}
leaving all other variables invariant. Clearly, $\alpha^{-q_B^z(X_0)}_{P^z}$ does the job, so we flow with parameter distance $s_1=-q_B^z(X_0)$, where $q_B^z(X_0)$ is the actual value of the relative distance of $B$ from $A$ before the translation at $X_0$.

\item Next, we rotate around the $y$-axis until $C$ lies on the $z$-axis. The necessary angle is the one between $A$ and $C$, as seen from $B$ and can be found from gauge-invariant quantities by using:
\begin{equation}\label{thetaC}
\cos \Theta_{AC|B} = \frac{l_{BC}^2 + l_{AB}^2 - l_{AC}^2}{2 l_{AB} l_{BC}}\,,
\end{equation}
where $l_{ij}^2 = \vec{q}_{ij}{}^2$ is the squared distance between particles $i$ and $j$. Using (\ref{gaugeTransfoR}, \ref{Cflow}), this  transformation is easy to evaluate on the canonical variables. Writing $X_1=\alpha^{-q_B^z(X_0)}_{P^z}(X_0)$, we find
\ba
\alpha_{R^y}^{s_2}\cdot q_i^x (X_1) &=& \cos s_2\,q_i^x(X_1)+\sin s_2\,q_i^z(X_1)\,,\\
\alpha_{R^y}^{s_2}\cdot p_i^x (X_1) &=& \cos s_2\,p_i^x(X_1)+\sin s_2\,p_i^z(X_1)\,,\\
\alpha_{R^y}^{s_2}\cdot q_i^z (X_1) &=& \cos s_2\,q_i^z(X_1)-\sin s_2\,q_i^x(X_1)\,,\\
\alpha_{R^y}^{s_2}\cdot p_i^z (X_1) &=& \cos s_2\,p_i^z(X_1)-\sin s_2\,p_i^x(X_1)\,,
\ea
leaving all other variables invariant. Evidently, 
$
\alpha_{R^y}^{-\Theta_{AC|B}}
$
achieves the desired transformation.

\item Now that $C$ lies on the $z$-axis, we translate it to the origin. Writing $X_2:=\alpha_{R^y}^{-\Theta_{AC|B}}(X_1)$, it is clear that 
\ba
\alpha_{P^z}^{-q_C^z(X_2)}
\ea
accomplishes the desired transformation. 

\item However, after this sequence of transformations, we will have $q_A^x<0$ and so we finally rotate once more by an angle $\pi$ around the $z$-axis
$
\alpha_{R^z}^\pi\,,
$
 so that now $q_A^x>0$.

\end{enumerate}

In conjunction, jumping from the reference frame of $A$ to the reference frame of $C$ can be achieved through the sequence of gauge transformations
\begin{equation}\label{}
\alpha_{A \rightarrow C} := \alpha_{R^z}^\pi\circ \alpha_{P^z}^{-q_C^z(X_2)}\circ \alpha_{R^y}^{-\Theta_{AC|B}} \circ \alpha^{-q_B^z(X_0)}_{P^z}\,,
\end{equation}
which completes the map, depicted in the diagram of sec.\ \ref{sec_clswitch3d},
\ba
\cs_{A\to C}:=\pi_{AB|C}\circ\alpha_{A\to C}\circ\iota_{BC|A}:\cp_{BC|A}\rightarrow\cp_{AB|C}\,.
\ea
Recalling that one has to swap some redundant and non-redundant Dirac observables (in line with the $A,C$ label exchange), it reads in coordinates:
\ba
&&(q_B^z>0,p_B^z,q_C^x>0,p_C^x,q_C^z,p_C^z)\q\q\q\q\nn\\ 
&&\q\q\mapsto\q\q\q\q\left(q'^x_A=\f{q_B^z\,q_C^x}{r_{BC}},p'^x_A=\f{r_C^2-q_B^zq_C^z}{r_C\,r_{BC}}(p_B^x+p_C^x)-\f{q_C^x}{r_{BC}}\,(p_B^z+p_C^z),\right.\nn\\
&&\q\q\q\q\q\q\q\left.  q'^z_A=\f{r_C^2-q_B^zq_C^z}{r_{BC}},p'^z_A=-\f{r_C^2-q_B^zq_C^z}{r_C\,r_{BC}}(p_B^z+p_C^z)-\f{q_C^x}{r_{BC}}\,(p_B^x+p_C^x)\right.,\nn\\
&&\q\q\q\q\q\q\q\left.q'^z_B=r_{BC},p'^z_B=\f{r_C^2-q_B^zq_C^z}{r_C\,r_{BC}}p_B^z-\f{q_C^x}{r_{BC}\,q_B^z}\,  (q_C^z p_C^x - q_C^x p_C^z)\right)\,,\nn
\ea
where the primed and unprimed variables are the ones after and before the total transformation $\cs_{A\to C}$, respectively, and $r_C=\sqrt{(q_C^x)^2+(q_C^z)^2}$ and $r_{BC}=\sqrt{(q_C^x)^2+(q_B^z-q_C^z)^2}$.

\section{Using spherical coordinates}\label{appendixSpherical}

\subsection{Spherical coordinates}

In ${L}^2(\mathbb{R}^3)$, we define spherical-coordinates eigenstates by:
\begin{equation}\label{sphericalEigenstates}
 \ket{r, \theta, \varphi} = \ket{x = r \sin \theta \cos \varphi, y = r \sin \theta \sin \varphi, z = r \cos \theta}\,,\,\q \forall \,r > 0, \theta \in [0,\pi], \varphi \in [0, 2 \pi[\,.
 \end{equation}
We have to remember that those states are normalized slightly differently,
\begin{equation}\label{polnorm}
\braket{r', \theta', \varphi'|r, \theta, \varphi} = \frac{1}{r^2 \sin \theta} \,\delta(r-r') \ \delta(\theta - \theta') \ \delta(\varphi-\varphi')\,,
\end{equation}
but they define a basis for ${L}^2(\mathbb{R}^3)$, except for the fact that there are pathological situations at $r=0$ and $\theta=0,\pi$.
One way to see this situation is that the use of spherical coordinates corresponds to an isometry between ${L}^2(\mathbb{R}^3)$ and $\mathcal{H}^{sphe}$, the subspace of ${L}^2(\mathbb{R}_+ \times S^2 )$ defined by the following conditions on a wave function $g$

\begin{subequations}\label{conditionsPhiAngular}
\begin{equation}\label{conditionr0}
g(r=0,\theta,\varphi) = \alpha\,,
\end{equation}
\begin{equation}\label{conditiontheta0}
g(r,\theta = 0,\varphi) = \beta(r)\,,
\end{equation}
\begin{equation}\label{conditionthetapi}
g(r,\theta = \pi,\varphi) = \gamma(r)\,,
\end{equation}
\end{subequations}
and where the integration measure is $\mathrm{d}\mu = r^2 \ \mathrm{d}\Omega = r^2 \ \mathrm{d}\theta \ \sin \theta \ \mathrm{d} \varphi$.

A state $\ket{\phi}$ associated to the density $g$ in spherical coordinates (verifying conditions (\ref{conditionsPhiAngular})) should be written:
\begin{equation}\label{phiAngular}
\ket{\phi} = \int \mathrm{d}r \ r^2 \ \mathrm{d}\Omega \ g(r,\theta,\varphi) \ket{r, \theta, \varphi}\,.
\end{equation}
In this way one has
\begin{equation}\label{}
\braket{\phi_1|\phi_2} = \int \mathrm{d}r \ r^2 \ \mathrm{d}\Omega \ g_1(r,\theta,\varphi)^* g_2(r,\theta,\varphi)
\end{equation}
and
\begin{equation}\label{}
|\braket{r, \theta, \varphi|\phi}|^2 = |g(r,\theta,\varphi)|^2\,,
\end{equation}
so one sees that $g(r,\theta,\phi)$ can physically be considered as \textit{the density of probability that $\ket{\phi}$ is measured at the position $(r, \theta, \varphi)$ in spherical coordinates}, as should be expected.
\medbreak

\subsection{Spherical harmonics states}

Let us restrict ourselves for a moment to $\mathcal{C}_{L}^\infty(\mathbb{R}^3) = \mathcal{C}^\infty(\mathbb{R}^3) \cap {L}^2(\mathbb{R}^3)$, which is dense in ${L}^2(\mathbb{R}^3)$. It is spanned by the:
\begin{equation}\label{sphericalHarmonics}
\ket{r; j,m} =   \int \mathrm{d}\Omega \ Y^{j,m}(\theta,\varphi) \ket{r, \theta, \varphi}\,,
\end{equation}
where the $Y^{j,m}$ are the usual spherical harmonics. Those states are built on the usual basis $\ket{j,m}$ (where $j \geq 0$ and $|m| \leq j$) of common eigenstates of $\hat{R}^z$ and $\hat{R}^2$.

A state in $\mathcal{C}_{L}^\infty(\mathbb{R}^3)$ can then be decomposed as:

\begin{equation}\label{phiAngularHarmonic}
\ket{\phi} = \int \mathrm{d}r \ r^2 \ \sum_{j=0}^\infty \ \sum_{|m| \leq j} \ f(r;j,m) \ket{r;j,m}\,,
\end{equation}
where the only remaining condition on $f$, stemming from (\ref{conditionr0}), is:
\begin{equation}\label{conditionr0bis}
f(r=0;j,m) = \alpha \ \delta_{j,0} \ \delta_{m,0}
\end{equation}

As $\mathcal{C}_{L}^\infty(\mathbb{R}^3)$ is dense in ${L}^2(\mathbb{R}^3)$, decomposition (\ref{phiAngularHarmonic}) (where $f$ verifies (\ref{conditionr0bis})) is also valid for any $\ket{\phi} \in {L}^2(\mathbb{R}^3)$.

\subsection{Spherical coordinates operators}\label{appendixSphericalObservables}

Lastly, let us define operators $\Hat{r}$, $\Hat{\theta}$ and $\Hat{\varphi}$. A natural way to do so would be to define it on the orthogonal basis defined in (\ref{sphericalEigenstates}):
\begin{subequations}\label{sphericalObservables}
\begin{equation}\label{rObservable}
\Hat{r} \ket{r, \theta, \varphi} = r  \ket{r, \theta, \varphi}\,,
\end{equation}
\begin{equation}\label{thetaObservable}
\Hat{\theta} \ket{r, \theta, \varphi} = \theta  \ket{r, \theta, \varphi}\,,
\end{equation}
\begin{equation}\label{phiObservable}
\Hat{\varphi} \ket{r, \theta, \varphi} = \varphi  \ket{r, \theta, \varphi}\,.
\end{equation}
\end{subequations}

Yet, $\Hat{\theta}$ and $\Hat{\varphi}$ defined in this way are not well-defined everywhere, as the action of both of them on states with $r=0$, as well as the action of $\Hat{\varphi}$ on states with $\theta=0$ and $\theta=\pi$, would then map on states which do not verify conditions (\ref{conditionsPhiAngular}). This is why we will say that definition (\ref{sphericalObservables}) is assumed only out of these pathological cases, on which we will complete it by the convention:

\begin{subequations}\label{sphericalObservablesPathological}
\begin{equation}\label{}
\Hat{\theta} \ket{r=0, \theta, \varphi} = 0\,,
\end{equation}
\begin{equation}\label{}
\Hat{\varphi} \ket{r=0, \theta, \varphi} = 0\,,
\end{equation}
\begin{equation}\label{}
\Hat{\varphi} \ket{r, \theta=0, \varphi} = 0\,,
\end{equation}
\begin{equation}\label{}
\Hat{\varphi} \ket{r, \theta=\pi, \varphi} = 0\,.
\end{equation}
\end{subequations}
It is important to note that the choice of definition (\ref{sphericalObservablesPathological}) will have practical consequences only when we consider states comprising a Dirac distribution at $r=0$, $\theta =0$, or $\theta = \pi$; otherwise they do not matter as they only impact states with a zero measure. Note that $r=0$ will correspond to collisions in our model, which we have ruled out dynamically.

\section{Rotational reduction in the quantum theory}

\subsection{The rotation generator trivialization map and its action on states}\label{proofRotational}

We had defined the transformation (\ref{RBC})
\begin{equation}\label{}
{\mathcal{R}}_{B,C} = \exp{ \Big(i\, \Hat{\varphi}_B \hat{R}_C^z \Big) } \exp{ \Big( i\, \Hat{\theta}_B \hat{R}_C^y \Big)} \exp{ \Big( i\,\Hat{\varphi}_B \hat{R}_C^z \Big) }\,.
\end{equation}
It is instructive to understand the construction of ${\mathcal{R}}_{B,C}$ by first having a look at the following unitary transformation on ${L}^2(\mathbb{R}^3)$:
\begin{equation}\label{}
{\mathcal{R}}(\theta,\varphi) = \exp{ \Big(i\, \varphi \hat{R}^z \Big) } \exp{ \Big( i\,\theta \hat{R}^y \Big)} \exp{ \Big( i\,\varphi \hat{R}^z \Big) }
\end{equation}
Let us look at its effect on a state $\ket{\vec{q}_0}$ where $\vec{q}_0 = (r_0 \sin \theta_0 \cos \varphi_0, r_0 \sin \theta_0 \sin \varphi_0, r_0 \cos \theta_0)$. We know that $\hat{R}^z$ and $\hat{R}^y$ are the generators of rotations around the $z$- and $y$-axes, respectively, so:
\begin{equation}
    \begin{split}
        {\mathcal{R}}(\theta,\varphi) \ket{\vec{q}_0} &= \exp{ \Big(i\, \varphi \hat{R}^z \Big) } \exp{ \Big( i\, \theta \hat{R}^y \Big)} \exp{ \Big( i\, \varphi \hat{R}^z \Big) } \ket{\vec{q}_0} \\
        &= \exp{ \Big( i\, \varphi \hat{R}^z \Big) } \exp{ \Big( i\,\theta \hat{R}^y \Big)} \ket{\vec{q}_1}\\
        &= \exp{ \Big( i\,\varphi \hat{R}^z \Big) } \ket{\vec{q}_2}= \ket{\vec{q}_3}\,,
    \end{split}
\end{equation}
where (we only give the values of the important coordinates):
\begin{subequations}
\begin{equation}
    \vec{q}_1 = (r_0 \sin \theta_0 \cos (\varphi_0 - \varphi), r_0 \sin \theta_0 \sin (\varphi_0 - \varphi), r_0 \cos \theta_0)
\end{equation}
\begin{equation}
    q_2^z = q_3^z = r_0 \cos \theta_0 \cos \theta + r_0 \sin \theta_0 \cos (\varphi_0 - \varphi) \sin \theta\,.
\end{equation}
\end{subequations}
$\vec{q}_3$ can then itself be rewritten in spherical coordinates as $\vec{q}_3= (r_0 \sin \gamma \cos \eta, r_0 \sin \gamma \sin \eta, r_0 \cos \gamma)$, where the new angular coordinates $\gamma \in [0,  \pi], \eta \in [0, 2 \pi[$ are defined by:
\begin{subequations}\label{}
\begin{equation}\label{gamma}
\cos (\gamma(\theta, \theta_0, \varphi - \varphi_0)) = q_3^z/r_0 =  \cos \theta \cos \theta_0 + \sin \theta \sin \theta_0 \cos (\varphi-\varphi_0)\,,
\end{equation}
\begin{equation}\label{}
\sin (\eta(\theta, \theta_0, \varphi - \varphi_0)) \sin (\gamma(\theta, \theta_0, \varphi - \varphi_0)) = q_3^y/r_0\,.
\end{equation}
\end{subequations}
The geometrical interpretation of (\ref{gamma}) is that $\gamma(\theta, \theta_0, \varphi - \varphi_0)$ is the angle between the directions $(\theta, \varphi)$ and $(\theta_0, \varphi_0)$, or equivalently the polar coordinate of the direction $(\theta_0, \varphi_0)$ if one takes direction $(\theta, \varphi)$ to be the new $z$-axis; $\eta$ is not relevant to us.

Going back to ${\mathcal{R}}_{B,C}$, one can now see that:
\ba\label{rbc}
{\mathcal{R}}_{B,C} \ket{r_B, \theta_B, \varphi_B}_B \ket{r_C, \theta_C, \varphi_C}_C
&= &\ket{r_B, \theta_B, \varphi_B}_B \otimes ( {\mathcal{R}}(\theta_B, \varphi_B) \ket{r_C, \theta_C, \varphi_C}_C) \\
&=& \ket{r_B, \theta_B, \varphi_B}_B\otimes\nn\\
&&\q\q\ket{r_C, \gamma(\theta_B, \theta_C, \varphi_B - \varphi_C), \eta(\theta_B, \theta_C, \varphi_B - \varphi_C)}_C\,,\nn
\ea
so we see that ${\mathcal{R}}_{B,C}$ acts as a rotation on the $C$-factor of the tensor product, mapping it to a description in spherical coordinates where the polar angle $\gamma$ is now relative to the direction of $B$. The azimuth angle $\eta$ after transformation ${\mathcal{R}}_{B,C}$ is not relevant for our analysis, as we shall see shortly.

$\mathcal{H}^{\rm phys}_{BC|A}$ is spanned by the states $\ket{\Phi(r_B,r_C;j)}$ as defined in (\ref{psiDecomposed}, \ref{Phi}). Let us therefore determine the action of ${\mathcal{R}}_{B,C}$ on the $\ket{\Phi(r_B,r_C;j)}$. Using \ref{rbc}, we find:
\begin{equation}\label{}
\begin{split}
    {\mathcal{R}}_{B,C} \ket{\Phi(r_B,r_C;j)} = (-1)^j &\int \mathrm{d}\Omega_B \ \mathrm{d}\Omega_C \ \sum_{|m| \leq j} \frac{(-1)^{m}}{\sqrt{2j+1}} Y^{j,-m}(\theta_B,\varphi_B) Y^{j,m}(\theta_C,\varphi_C)  \\
    &\ket{r_B,\theta_B,\varphi_B}_B \ket{r_C,\gamma(\theta_B, \theta_C, \varphi_B - \varphi_C), \eta(\theta_B, \theta_C, \varphi_B - \varphi_C)}_C \,.
\end{split}
\end{equation}
The \textit{Legendre addition theorem} states that:
\begin{equation}\label{}
\sum_{|m| \leq j} \frac{(-1)^{m}}{\sqrt{2j+1}} Y^{j,-m}(\theta_B,\varphi_B) Y^{j,m}(\theta_C,\varphi_C) = \frac{\sqrt{2j+1}}{4 \pi} P_j (\cos \gamma)\,,
\end{equation}
where $\gamma$ is defined as in (\ref{gamma}) and $P_j$ is the Legendre polynomial of degree $j$. The change of variables $(\theta_C,\varphi_C) \to (\gamma, \eta)$ leaves the element of integration invariant and yields:
\begin{equation}\label{}
{\mathcal{R}}_{B,C} \ket{\Phi(r_B,r_C;j)} \!=\! (-1)^j\! \int \mathrm{d}\Omega_B  \mathrm{d}\gamma \sin \gamma  \mathrm{d}\eta  \frac{1}{2 \sqrt{\pi}}  \sqrt{\frac{2j+1}{4\pi}} P_j(\cos \gamma) \ket{r_B,\theta_B,\varphi_B}_B \ket{r_C,\gamma,\eta}_C .
\end{equation}
One can recognize the spherical harmonics $Y^{0,0}(\theta_B, \varphi_B) \!= \!\frac{1}{2 \sqrt{\pi}}$ and $Y^{j,0}(\gamma, \eta) \!= \!\sqrt{\frac{2j+1}{4\pi}} P_j(\cos \gamma)$, and thus finally rewrite:
\begin{equation}\label{}
{\mathcal{R}}_{B,C} \ket{\Phi(r_B,r_C;j)} = (-1)^j \ket{r_B;0,0}_B \ket{r_C;j,0}_C 
\end{equation}
Hence, ${\mathcal{R}}_{B,C}$ maps a given $\ket{\psi}^{\rm phys}_{BC|A} \in \mathcal{H}^{\rm phys}_{BC|A}$ as decomposed  in (\ref{psiDecomposed}) to:
\begin{equation}\label{transb,ca}
\ket{\psi}^{\rm phys}_{B,C|A} := {\mathcal{R}}_{B,C} \ket{\psi}^{\rm phys}_{BC|A} = \int \mathrm{d} r_B \ \mathrm{d} r_C \ r_B^2\, r_C^2 \sum_{j=0}^\infty (-1)^j \psi^{\rm phys}_{BC|A}(r_B, r_C; j) \ket{r_B;0,0}_B \ket{r_C;j,0}_C \,.
\end{equation}

Recall that $\ch^{\rm phys}_{BC|A}$ is a proper subspace of $\ch^{\rm TI}_{BC|A}$ and so inherits its inner product from the latter. Now $\calr_{B,C}$  leaves $\ch^{\rm TI}_{BC|A}$ invariant and so just rotates $\ch^{\rm phys}_{BC|A}$ into a new Hilbert subspace $\ch^{\rm phys}_{B,C|A}:=\calr_{B,C}(\ch^{\rm phys}_{BC|A})\subset\ch^{\rm TI}_{BC|A}$ (see fig.\ \ref{commutativeDiagram3D}). It can be easily checked by direct calculation that $\calr_{B,C}$ leaves the inner product invariant and so it defines an isometry from $\ch^{\rm phys}_{BC|A}$ to $\ch^{\rm phys}_{B,C|A}$. Notice also that it is invertible (away from the pathological states \eqref{sphericalObservablesPathological}) with 
\ba
\calr^{-1}_{B,C}= \exp{ \Big(-i\, \Hat{\varphi}_B \hat{R}_C^z \Big) } \exp{ \Big( -i\, \Hat{\theta}_B \hat{R}_C^y \Big)} \exp{ \Big(- i\,\Hat{\varphi}_B \hat{R}_C^z \Big) }\,.
\ea

\subsection{Trivializing the rotation constraints}\label{app_RBCtrivial}

Let us look at how the constraints (\ref{QconstraintR}), which define $\mathcal{H}^\textrm{phys}_{BC|A}$ as a subspace of $\mathcal{H}^\textrm{TI}_{BC|A}$, transform under ${\mathcal{R}}_{B,C}$. 
We will give the results of the calculations, which were done using the usual commutation relations between the $\hat{R}^a$ and using their representation in polar coordinates:
\begin{subequations}
\be \hat{R}^x = i (\sin{\varphi} \, \partial_\theta + \cot{\theta} \cos{\varphi} \, \partial_\varphi)\,, \ee
\be \hat{R}^y = i (- \cos{\varphi} \, \partial_\theta + \cot{\theta} \sin{\varphi} \, \partial_\varphi) \,,\ee
\be \hat{R}^x = - i \partial_\varphi \,.\ee
\end{subequations}
In terms of action on the coefficient function $\psi^{\rm phys}_{B,C|A}$, the constraints get mapped to:
\begin{subequations}\label{mapJ}
\ba \label{mapx} &&{\mathcal{R}}_{B,C} \,(\hat{R}_B^x+\hat{R}^x_C)\, {\mathcal{R}}_{B,C}^{-1} \psi^{\rm phys}_{B,C|A}\\
&&\q\q\q\q\q=( i \sin{\varphi_B} \partial_{\theta_B} + i \cot{\theta_B} \cos{\varphi_B} (\partial_{\varphi_B} - \partial_{\varphi_C}) - i \f{\cos{\varphi_B}}{\sin{\theta_B}} \partial_{\varphi_B}) \psi^{\rm phys}_{B,C|A} = 0  \nn\ea
\ba \label{mapy} &&{\mathcal{R}}_{B,C}\, (\hat{R}_B^y+\hat{R}^y_C)\, {\mathcal{R}}_{B,C}^{-1} \psi^{\rm phys}_{B,C|A} \\
&&\q\q\q\q\q= ( i \cos{\varphi_B} \partial_{\theta_B} + i \cot{\theta_B} \sin{\varphi_B} (\partial_{\varphi_B} - \partial_{\varphi_C}) - i \f{\sin{\varphi_B}}{\sin{\theta_B}} \partial_{\varphi_B} )  \psi^{\rm phys}_{B,C|A} = 0 \nn \ea
\be \label{mapz} {\mathcal{R}}_{B,C}\, (\hat{R}^z_B+\hat{R}^z_C)\, {\mathcal{R}}_{B,C}^{-1} \psi^{\rm phys}_{B,C|A} = -i (\partial_{\varphi_B} - \partial_{\varphi_C})  \psi^{\rm phys}_{B,C|A} = 0 \ee
\end{subequations}

This can also be written in terms of the $\hat{R}$s acting on the state $\ket{\psi}^{\rm phys}_{B,C|A}$:
\begin{subequations}\label{mapJrewritten}
\ba {\mathcal{R}}_{B,C} \,(\hat{R}_B^x+\hat{R}^x_C)\, {\mathcal{R}}_{B,C}^{-1} \ket{\psi}^{\rm phys}_{B,C|A}\!\!\! &&=( \sin{\hat{\varphi}_B} (\sin{\hat{\varphi}_B} \hat{R}_B^x - \cos{\hat{\varphi}_B} \hat{R}_B^y) \\
&& +  \cot{\hat{\theta}_B} \cos{\varphi_B} (\hat{R}_C^z - \hat{R}_B^z) +  \widehat{\f{\cos{\varphi_B}}{\sin{\theta_B}}} \hat{R}_B^z) \ket{\psi}^{\rm phys}_{B,C|A} = 0\,.\nn  \ea
\ba {\mathcal{R}}_{B,C}\, (\hat{R}_B^y+\hat{R}^y_C)\, {\mathcal{R}}_{B,C}^{-1} \ket{\psi}^{\rm phys}_{B,C|A}\!\!\!\!&& = ( \cos{\hat{\varphi}_B} (\sin{\hat{\varphi}_B} \hat{R}_B^x - \cos{\hat{\varphi}_B} \hat{R}_B^y) \\
&& + \cot{\hat{\theta}_B} \sin{\varphi_B} (\hat{R}_C^z - \hat{R}_B^z) + \widehat{\f{\sin{\varphi_B}}{\sin{\theta_B}}} \hat{R}_B^z)  \ket{\psi}^{\rm phys}_{B,C|A} = 0 \,.\nn \ea
\be {\mathcal{R}}_{B,C}\, (\hat{R}^z_B+\hat{R}^z_C)\, {\mathcal{R}}_{B,C}^{-1} \ket{\psi}^{\rm phys}_{B,C|A} = (\hat{R}_C^z - \hat{R}_B^z)  \ket{\psi}^{\rm phys}_{B,C|A} = 0 \ee
\end{subequations}
Inserting (\ref{mapz}) in (\ref{mapx}) and (\ref{mapy}), and then rotating them by angle $\varphi_B$, one finds that (\ref{mapJ}) implies:
\begin{subequations}
\be \partial_{\theta_B} \psi^{\rm phys}_{B,C|A}= 0  \ee
\be \partial_{\varphi_B} \psi^{\rm phys}_{B,C|A}= 0  \ee
\be  \partial_{\varphi_C} \psi^{\rm phys}_{B,C|A}= 0 \ee
\end{subequations}
which is equivalent to the system of constraints:
\ba\label{mapJeq}
\hat{R}_B^x \ket{\psi}^{\rm phys}_{B,C|A}=
 \hat{R}_B^y \ket{\psi}^{\rm phys}_{B,C|A}=
 \hat{R}_B^z \ket{\psi}^{\rm phys}_{B,C|A}=
 \hat{R}_C^z \ket{\psi}^{\rm phys}_{B,C|A}=0\,.
\ea
This is, of course, exactly what one would expect from the shape of (\ref{transb,ca}).
In turn, it is easy to see that (\ref{mapJeq}) implies (\ref{mapJ}); thus (\ref{mapJeq}) can indeed be taken as the set of constraints which defines $\mathcal{H}^\textrm{phys}_{B,C|A}$ as a subset of $\mathcal{H}^\textrm{TI}_{BC|A}$.

\subsection{Dirac observables and rotational reduction}\label{app_RBCobs}

Next, we have to check how our Dirac observables, written in (\ref{Dobserv}) equivalently for $\ch^{\rm phys}$ and $\ch^{\rm TI}$, transform under $\calr_{B,C}$. Since we apply this map to $\ch^{\rm phys}_{BC|A}$, we have to check how the translationally reduced Dirac observables transform under $\calr_{B,C}$. Recall from sec.\ \ref{translationalRed} that the translational reduction of the Dirac observables in (\ref{Dobserv}) amounts to simply dropping the $A$ labels on all sides.

It will also be convenient to write the operator corresponding to the angle $\gamma$ between $B$ and $C$ as seen from $A$ (essentially $\hat{u}$ in (\ref{Dobserv})) on $\ch^{\rm phys}_{BC|A}$ as follows:
\be \begin{split}
    \label{gammaOp} \widehat{\gamma} &= \arccos{ \big( \cos \hat{\theta}_B \cos \hat{\theta}_C + \sin \hat{\theta}_B \sin \hat{\theta}_C \cos (\hat{\varphi}_C-\hat{\varphi}_B) \big)} \\
    &= \arccos \big( \f{\widehat{\vec{q}_{B} \cdot \vec{q}_{C}}}{r_B r_C} \big)
\end{split}  \ee

To compute how the observables transform, one can remember that ${\mathcal{R}}_{B,C}$ acts on the $\Hat{p}_C^a$ and $\Hat{q}_{C}^a$ as a sequence of rotations (whose `parameters' are the operators $\Hat{\theta}_B$ and $\Hat{\varphi}_B$). Indeed, if $s \in \{p,q\}$:
\begin{subequations}
\ba {\mathcal{R}}_{B,C} \,\hat{s}_C^x\, {\mathcal{R}}_{B,C}^{-1} &=& \cos \Hat{\varphi}_B ( \sin{\Hat{\theta}_B} \hat{s}_C^z + \cos{\Hat{\theta}_B} (\cos{\Hat{\varphi}_B} \hat{s}_C^x - \sin{\Hat{\varphi}_B} \hat{s}_C^y)) \nn\\
&&\q\q\q- \sin{\Hat{\varphi}_B} (\sin{\Hat{\varphi}_B} \hat{s}_C^x + \cos{\Hat{\varphi}_B} \hat{s}_C^y)\nn \ea
\ba {\mathcal{R}}_{B,C}\, \hat{s}_C^y \,{\mathcal{R}}_{B,C}^{-1} &=& \sin \Hat{\varphi}_B ( \sin{\Hat{\theta}_B} \hat{s}_C^z + \cos{\Hat{\theta}_B} (\cos{\Hat{\varphi}_B} \hat{s}_C^x - \sin{\Hat{\varphi}_B} \hat{s}_C^y))\nn\\
&&\q\q\q+ \cos{\Hat{\varphi}_B} (\sin{\Hat{\varphi}_B} \hat{s}_C^x + \cos{\Hat{\varphi}_B} \hat{s}_C^y)\nn \ea
\ba {\mathcal{R}}_{B,C} \,\hat{s}_C^z\, {\mathcal{R}}_{B,C}^{-1} &=& \cos{\Hat{\theta}_B} \hat{s}_C^z - \sin{\Hat{\theta}_B} (\cos{\Hat{\varphi}_B} \hat{s}_C^x - \sin{\Hat{\varphi}_B} \hat{s}_C^y))\nn \ea
\end{subequations}

The key point of those transformations is that $\widehat{\gamma}$ in (\ref{gammaOp}), gets mapped to $\Hat{\theta}_C$:
\be {\mathcal{R}}_{B,C} \,\widehat{\gamma} \,{\mathcal{R}}_{B,C}^{-1} = \Hat{\theta}_C\,. \ee
Moreover, the calculations give that the forms of $\Hat{\rho}_B$, $\Hat{\rho}_C$, $\Hat{p}^\rho_B$ and $\Hat{p}^\rho_C$ remain invariant under ${\mathcal{R}}_{B,C}$; only $\Hat{u}_C$ and $\Hat{p}_C^u$ have non trivial transformations (these transformations are only valid on states without support on pathological configuratoins):
\ba
&&{\mathcal{R}}_{B,C} \,\Hat{u}_C\, {\mathcal{R}}_{B,C}^{-1} = - \reallywidehat{\f{\sqrt{(q_C^x)^2 + (q_C^y)^2}}{q_C^z}} = - \widehat{\cot \, \theta_C} \,\nn\\
    &&{\mathcal{R}}_{B,C} \,\Hat{p}_C^u\,{\mathcal{R}}_{B,C}^{-1} \nn\\&&\q=- \f{1}{2} \Big( - \hat{q}_C^z \reallywidehat{\sqrt{(q_C^x)^2 + (q_C^y)^2}}\reallywidehat{((q_C^z)^2 + (q_C^x)^2 + (q_C^y)^2)^{-1}} \,{\vec{\hat q}}_{C} \cdot {\vec{\hat p}}_{C} + \reallywidehat{\sqrt{(q_C^x)^2 + (q_C^y)^2}} \,\Hat{p}_C^z\nn \\ 
    &&\q- {\vec{\hat p}}_{C} \cdot \,{\vec{\hat q}}_{C}\,\hat{q}_C^z \reallywidehat{\sqrt{(q_C^x)^2 + (q_C^y)^2}}\reallywidehat{((q_C^z)^2 + (q_C^x)^2 + (q_C^y)^2)^{-1}} \, + \Hat{p}_C^z\, \reallywidehat{\sqrt{(q_C^x)^2 + (q_C^y)^2}}\Big) \nn \\
    &&\q= - \f{1}{2} \Big( - \widehat{\cos{\theta_C}}\, \widehat{\sin{\theta_C}} \,{\vec{\hat q}}_{C} \cdot {\vec{\hat p}}_{C} + \Hat{r}_C\, \widehat{\sin{\theta_C}} \,\Hat{p}_C^z
    - {\vec{\hat p}}_{C} \cdot \,{\vec{\hat q}}_{C} \,\widehat{\cos{\theta_C}}\, \widehat{\sin{\theta_C}} + \Hat{p}_C^z\, \Hat{r}_C\, \widehat{\sin{\theta_C}} \Big)\,.\nn
\ea

\section{Illustration of the Hamiltonian discrepancy between Dirac and reduced quantization}\label{HamiltonianDiscrepancy}

It can be computed that the Hamiltonians $\hat{H}^\textrm{red}_{BC|A}$ and $\Phi_A'\,\hat H_{\rm tot}\,\Phi_A'^{-1}$, where $\Phi_A'$ is given by \eqref{PHI}, respectively obtained through reduced quantization and through Dirac quantization followed by our quantum reduction method, differ by a term only involving configuration observables:

\be \begin{split}
\hat{H}_{BC|A}^\textrm{red} - \Phi_A'\,\hat H_{\rm tot}\,\Phi_A'^{-1} = \, &\f{1}{16} (1 + \hat{u}_C^2)^\f{3}{2} \Bigg( \Big( 300 \, ( e^{3 \hat{\rho}_B - \hat{\rho}_C} + e^{3 \hat{\rho}_C - \hat{\rho}_B} ) + 567 \, e^{ \hat{\rho}_B + \hat{\rho}_C} \Big) \hat{u}_C (1 + \hat{u}_C^2) \\
&+ \Big( 100 \, ( e^{2 \hat{\rho}_B - 2 \hat{\rho}_C} + e^{2 \hat{\rho}_C - 2 \hat{\rho}_B} ) + 267 \, (e^{2 \hat{\rho}_B} + e^{2 \hat{\rho}_C}) \Big) (1 + \hat{u}_C^2)^\f{3}{2} \\
&+ 133 \, e^{\hat{\rho}_B + \hat{\rho}_C} \hat{u}_C^3 + 333 \, (e^{2 \hat{\rho}_B} + e^{2 \hat{\rho}_C}) u_C^2 (1 + \hat{u}_C^2)^\f{1}{2} \Bigg) \\
&\cdot \reallywidehat{\left( (e^{2 \rho_B} + e^{2 \rho_C}) \, (1 + u_C^2) + e^{ \rho_B + \rho_C} u_C (1 + u_C^2)^\f{1}{2} \right)^{-2}} \, .
\end{split}
\ee
We note that $\Phi_A'\,\hat H_{\rm tot}\,\Phi_A'^{-1}$ is equal to $\Phi_A\,\hat H_{\rm tot}\,\Phi_A^{-1}$, which is the Hamiltonian reduced from the physical Hilbert space to $\ch_{BC|A}$ because the lower left diagram in fig.~\ref{commutativeDiagram3D} commutes and $\hat H_{\rm tot}$ takes the same form on $\ch^{\rm TI}$ as it does on $\ch^{\rm phys}$. However, as argued in the main body, in practice it is easier to construct $\Phi_A'$. The discrepancy between the Hamiltonians thus arises relative to standard choices in Dirac quantization.

There are ways to modify Dirac quantization for systems with constraints linear in momenta, as in the present manuscript, such that it agrees with reduced quantization \cite{Kunstatter:1991ds,Kuchar1986}. While \cite{Kuchar1986} modifies the quantization of gauge degrees of freedom and only constructs a Hilbert space for gauge invariant degrees of freedom, \cite{Kunstatter:1991ds} modifies the quantization of the gauge-invariant degrees of freedom in Dirac quantization. For example, the method in \cite{Kunstatter:1991ds} constructs the Hamiltonian only on the physical Hilbert space $\ch^\textrm{phys}$ in such a way that it is self-adjoint with respect to the measure $\sqrt{|h \gamma|}$, where $h$, given in (\ref{RedVolume}) is the measure on the physical configuration space, and $\gamma$ is the measure on the gauge orbits.\footnote{$\gamma$ is the determinant of $\gamma_{\alpha \beta}$, the induced metric for displacements along gauge orbits, given by $\gamma_{\alpha \beta} = \sum_a K_{\alpha}^a K_{\beta}^a$, where the $K_\alpha$ denote the six constraints (\ref{constraints}).} Applied to our model, it is thus defined not in terms of the momentum operators $\hat{p}_i^a$ on $\ch^\textrm{kin}$, but in terms of the momentum operators $\hat{\pi}^\textrm{phys}_\mu$ (where $\mu \in \{ \rho_{BA}, \rho_{CA}, u \} $) on $\ch^\textrm{phys}$, which are self-adjoint with respect to the measure $\sqrt{|\gamma\,h|}$. The Hamiltonian is then however given by \cite{Kunstatter:1991ds}
\be \label{HKuns}
\hat{H}_{\rm tot}' = \f{1}{2} |h|^{-\f{1}{4}} \, \hat{\pi}^\textrm{phys}_\mu \, |h|^{\f{1}{2}} \, {h}^{\mu \nu} \, \hat{\pi}^\textrm{phys}_\nu \,{|h|^{-\f{1}{4}}} \,  + \, V(\hat{\rho}_{BA}, \hat{\rho}_{CA}, \hat{u}) \, .
\ee
Specifically, the Laplace-Beltrami operator is now constructed with respect to the measure $\sqrt{|h|}$ and no longer with respect to the measure $\sqrt{|\gamma h|}$. Furthermore, the Hamiltonian is already written with respect to a special choice of gauge invariant degrees of freedom, here associated with the reference frame choice $A$. A different reference frame choice, say $C$, would yield a \emph{different} Hamiltonian when written out similarly because it would be based on different canonical pairs. As is well known through the Groenewold-van-Hove phenomenon, classical canonical transformations will in general not translate into unitary transformations in the quantum theory, especially when the variable change is nonlinear \cite{Guillemin:1990ew}.

Applying our quantum reduction procedure to this  modified quantization, the effect of its successive steps will simply be to rotate the physical Hilbert space so as to replace the labels $\rho_{BA}, \rho_{CA}, u$ with, respectively, $\rho_{B}, \rho_{C}, u_C$, and to modify the measure of this space so that the $\hat{\pi}^\textrm{phys}_\mu$ are ultimately mapped to the $\hat{\pi}_\mu$ obtained through reduced quantization. The Hamiltonian (\ref{HKuns}) will, by construction, be mapped precisely to $\hat{H}^\textrm{red}_{BC|A}$ as defined in (\ref{qHred}), thus ensuring that the Dirac and reduced quantized theories are  equivalent under our quantum reduction procedure, as far as the dynamics is concerned. For other composite operators one may similarly have to modify the quantization. While this method is interesting for understanding the relation between Dirac and reduced quantization, the modifications in this construction are not particularly natural from the point of view of the former. Especially the fact that the gauge invariant Hamiltonian \eqref{HKuns} does away with the redundancy inherent in Dirac quantization and is special to a choice of frame is breaking the strength of Dirac quantization that we exploited in the main body for defining a notion of quantum frame covariance. In particular, this means that the modification one has to apply to Dirac quantization in order to obtain equivalence with the reduced quantization of a classical frame perspective will depend on the choice of that frame because the different reduced quantizations are unitarily inequivalent. This is in conflict with quantum frame covariance.

By contrast, in our construction $\hat H_{\rm tot}$ does not face any factor ordering ambiguities on $\ch^{\rm kin}$, which comes with a standard Lebesgue measure, and neither on $\ch^{\rm phys}$, when expressed in the individual kinematical particle variables. Most importantly, our construction admits the redundancy underlying quantum frame covariance and the Hamiltonian $\hat H_{\rm tot}$ treats all degrees of freedom on an equal footing.

\bibliographystyle{utphys}

\bibliography{bibliography}

\end{document}